\documentclass[a4paper,11pt]{article}
\pdfoutput=1 % if your are submitting a pdflatex (i.e. if you have
\usepackage{jheppub} % for details on the use of the package, please
\usepackage[T1]{fontenc} % if needed
\usepackage{amsmath}
\usepackage{amssymb}
\usepackage{amsfonts}
\usepackage{graphicx}

\newcommand{\half}{\frac{1}{2}}

\newcommand{\mymatrix}[1]{\!\begin{pmatrix} #1 \end{pmatrix}\!}

\newcommand{\be}{\begin{equation}}
\newcommand{\ee}{\end{equation}}
\newcommand{\bea}{\begin{eqnarray}}
\newcommand{\eea}{\end{eqnarray}}

\def\l{\left(}
\def\r{\right)}

\newcommand{\MUV}[1]{M_\text{UV}}

\title{\Large{A~small~weak~scale~from~a~small~cosmological~constant}}

\author[a]{Asimina Arvanitaki,}
\author[b]{Savas Dimopoulos,}
\author[b]{Victor Gorbenko,}
\author[b]{Junwu Huang,}
\author[b,c,d]{Ken Van Tilburg}

\affiliation[a]{Perimeter Institute for Theoretical Physics, Waterloo, Ontario, N2L 2Y5, Canada}
\affiliation[b]{Stanford Institute for Theoretical Physics, Stanford University, Stanford, CA 94305, USA}
\affiliation[c]{Center for Cosmology and Particle Physics, Department of Physics, New York University, NY 10003, USA}
\affiliation[d]{School of Natural Sciences, Institute for Advanced Study, Princeton, NJ 08540, USA}

\emailAdd{aarvanitaki@perimeterinstitute.ca}
\emailAdd{savas@stanford.edu}
\emailAdd{vitya@stanford.edu}
\emailAdd{curlyh@stanford.edu}
\emailAdd{kenvt@nyu.edu}

\abstract{
We propose a framework in which Weinberg's anthropic explanation of the cosmological constant problem also solves the hierarchy problem. The weak scale is selected by chiral dynamics that controls the stabilization of an extra dimension. When the Higgs vacuum expectation value is close to a fermion mass scale, the radius of an extra dimension becomes large, and develops an enhanced number of vacua available to scan the cosmological constant down to its observed value. At low energies, the radion necessarily appears as an unnaturally light scalar, in a range of masses and couplings accessible to fifth-force searches as well as scalar dark matter searches with atomic clocks and gravitational-wave detectors. The fermion sector that controls the size of the extra dimension consists of a pair of electroweak doublets and several singlets. These leptons satisfy approximate mass relations related to the weak scale and are accessible to the LHC and future colliders.
}

\begin{document} 
\maketitle
\newpage
\begin{flushright}
~\\
{ \it ``if it was so, it might be;}\\
{\it and if it were so, it would be;}\\
{\it  but as it isn't, it ain't.}\\
{\it That's logic.''}\\
~\\
\hfil{\it {\rm Lewis Carroll,} ``Through the Looking-Glass''.}
~\\
\end{flushright}

\flushbottom

\section{Framework}\label{sec:framework}

\begin{figure}[t]
\begin{center}
\includegraphics[width = 0.99 \textwidth]{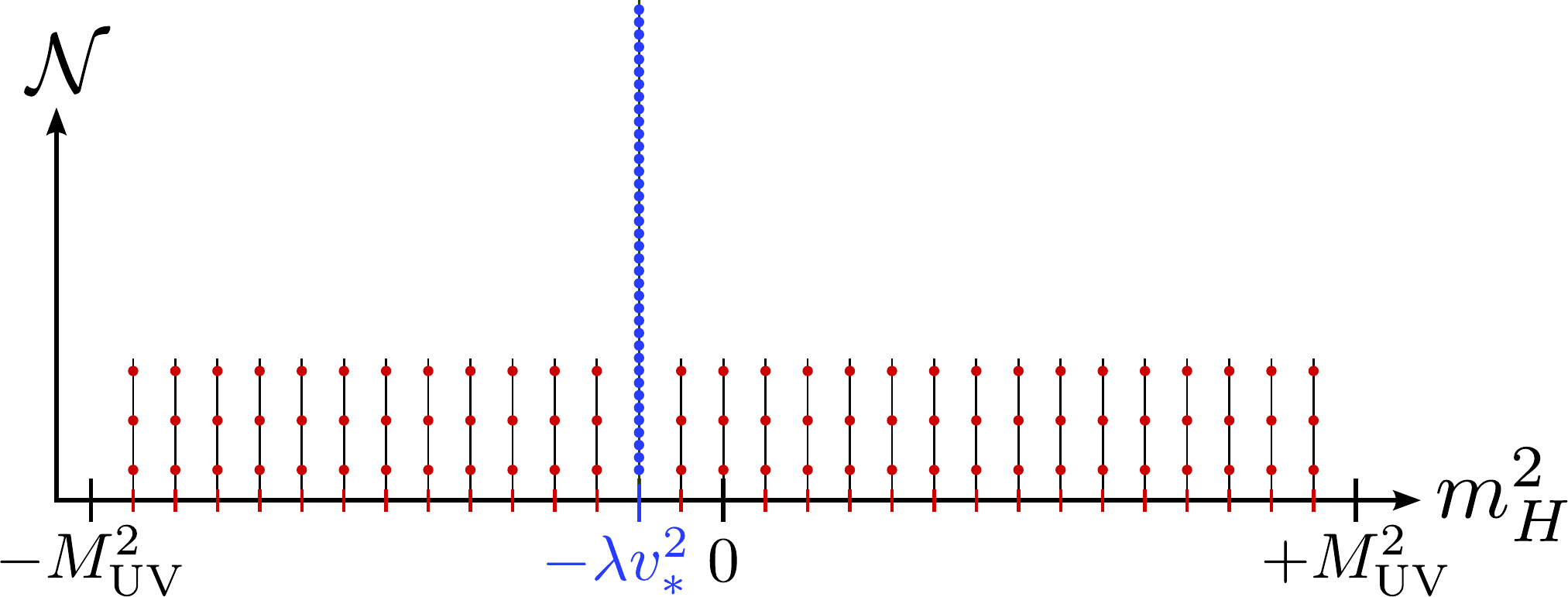}
\end{center}
\caption{Schematic representation of the number $\mathcal{N}$ of vacua (depicted by dots) that can tune the cosmological constant down to a small value as a function of the Higgs mass.}
\label{fig:comb}
\end{figure}

There are two successful approaches for explaining small numbers: dynamics and anthropic selection. For the hierarchy problem, the smallness of the weak scale can be explained by either dynamics, such as supersymmetry or compositeness, or by anthropics---the ``atomic principle'' that postulates the necessity of the existence of atoms~\cite{Agrawal:1997gf}. For the cosmological constant problem, there is only one known approach, using anthropics---the ``galactic principle'' that postulates the necessity of the existence of galaxies~\cite{Weinberg:1987dv,Weinberg:1988cp}. The absence of dynamical solutions to the cosmological constant problem casts doubt on dynamical approaches to the lesser gauge hierarchy problem. 
In this paper, we propose a framework in which the galactic principle can simultaneously solve both the cosmological constant and hierarchy problems. Our strategy involves chiral dynamics that selects the weak scale~$v_*$ by enhancing the number of discrete vacua available to scan the cosmological constant finely enough, down to its observed value. This is shown schematically in figure~\ref{fig:comb}. In our toy-landscape, the only parameters that scan are those that are not protected by symmetries, i.e.~the Higgs mass and the cosmological constant. This assumption is key for our mechanism.

%We construct a model in which the number of vacua with Higgs mass around scale much below the cutoff gets parametrically enhanced. 
%More precisely, we will be using the value of the Higgs vev $v$ that is obviously related to the mass. 
In order for $v$ to be able to influence the vacuum structure of the theory we introduce a pair of $SU(2)$ doublets $L$ and $L^c$ as well as a pair of neutral Majorana fermions $N_1$ and $N_1^c$. Those fermions couple to the Higgs via Yukawa couplings and have vector masses $M_0$ and $M_1$ respectively:
\be
 M_0 L L^c + M_1 N_1 N_1^c + Y H L N_1^c + Y^c H^{\dagger} L^c N_1,
\ee
and have charges listed as in table~\ref{tab:hypercharge}.
The determinant of the fermion mass matrix goes to zero when $Y Y^c v^2 \sim  M_0 M_1$, where one of the fermions becomes massless. This is what singles out a critical electroweak-breaking scale in our model, namely
\be
v_*^2\sim \frac{M_0 M_1}{Y Y^c},
\ee
and relates the Higgs vev---and thus also its mass---to a combination of technically natural quantities. We do not endow the Higgs itself with a new symmetry; indeed, its mass can take on many possible values, most of them near the cutoff $M_\text{UV}^2$. However, in a small, special subset of these Higgs vacua, a chiral symmetry of a new fermion is approximately restored. We know of several ways to turn chiral symmetry restoration into vacuum number enhancement. In this paper, we focus on a model with an extra dimension, although we have also constructed purely four-dimensional versions.

Consider a 5D model with the fifth dimension ending on two branes. The Standard Model fields are localized on one brane, while the second brane has many vacua with different values of its brane tension. The rich vacuum structure of the second brane can help tune the CC to an anthropically allowed value, but only if the extra dimension is dynamically stabilized. 

In our model, the only dynamical field in the bulk is a fermion $\Psi$, which can stabilize the 5D radius at zero effective CC through its Casimir energy if and only if this energy is positive (Section \ref{sec:radiusstabilization}). For generic values of the Higgs vev, the boundary conditions for $\Psi$ are of the type corresponding to negative Casimir energy, so the fifth dimension is not stabilized and the rich vacuum structure is lost. However, if $\Psi$ and $N_1^c$ (one of the brane fermions) have a mass mixing on the SM brane, the boundary conditions for $\Psi$ change their type once chiral symmetry in the fermion sector on the brane gets restored, producing positive Casimir energy. This part of our mechanism is discussed in section~\ref{sec:5dmodels} and appendix~\ref{sec:appCasimir}. 

From the four-dimensional, low-energy point of view the mechanism can be summarized as follows. For Higgs vevs $v \sim v_*$, a restored chiral symmetry changes dramatically the potential for the radion field, which is gravitationally coupled both to these fermions and to a hidden sector with many vacua (the second brane). This modified potential has a huge number of minima so that at least one of them leads to a CC that is anthropically allowed.

Section~\ref{sec:vacua} is dedicated to counting the number of minima with correct ($v \sim v_*$) and wrong (mostly $M_\text{UV}$) values of the Higgs mass. %We assume that only the quantities not protected by symmetries, namely the higgs mass and the cosmological constant, are scanned in our toy-landscape. 
The cutoff of our theory $M_\text{UV}$ can be as high as  $10^{12} \,{\rm GeV}$, and the cosmological constant in the vacua with a Higgs vev different than $v_*$ is $10^{24}$ times larger than the measured value. Only when $v\sim v_*$ can there be vacua with a small enough cosmological constant for galaxies to form.
%We arrive to the conclusion that with an optimal choice of the parameters the cutoff $M_\text{UV}$ can be picked as high as while keeping the smallest cosmological constant among all the vacua with wrong Higgs mass is larger then...

Finally, section~\ref{sec:phenomenology} discusses the phenomenology of the new states near the weak scale and of an ultralight radion. The new electroweak doublets should be below $\sim 4 \pi v_*$ which implies that the fermion sector is accessible at the LHC and future colliders, through searches for direct production of electroweak-charge fermions and measurements of the Higgs invisible width. The radion, which is automatically tuned to be light, is within the reach of equivalence principle tests and fifth-force searches as well as proposed scalar dark matter searches.

\begin{figure*}[t]
\begin{center}\includegraphics[width = 0.45 \textwidth]{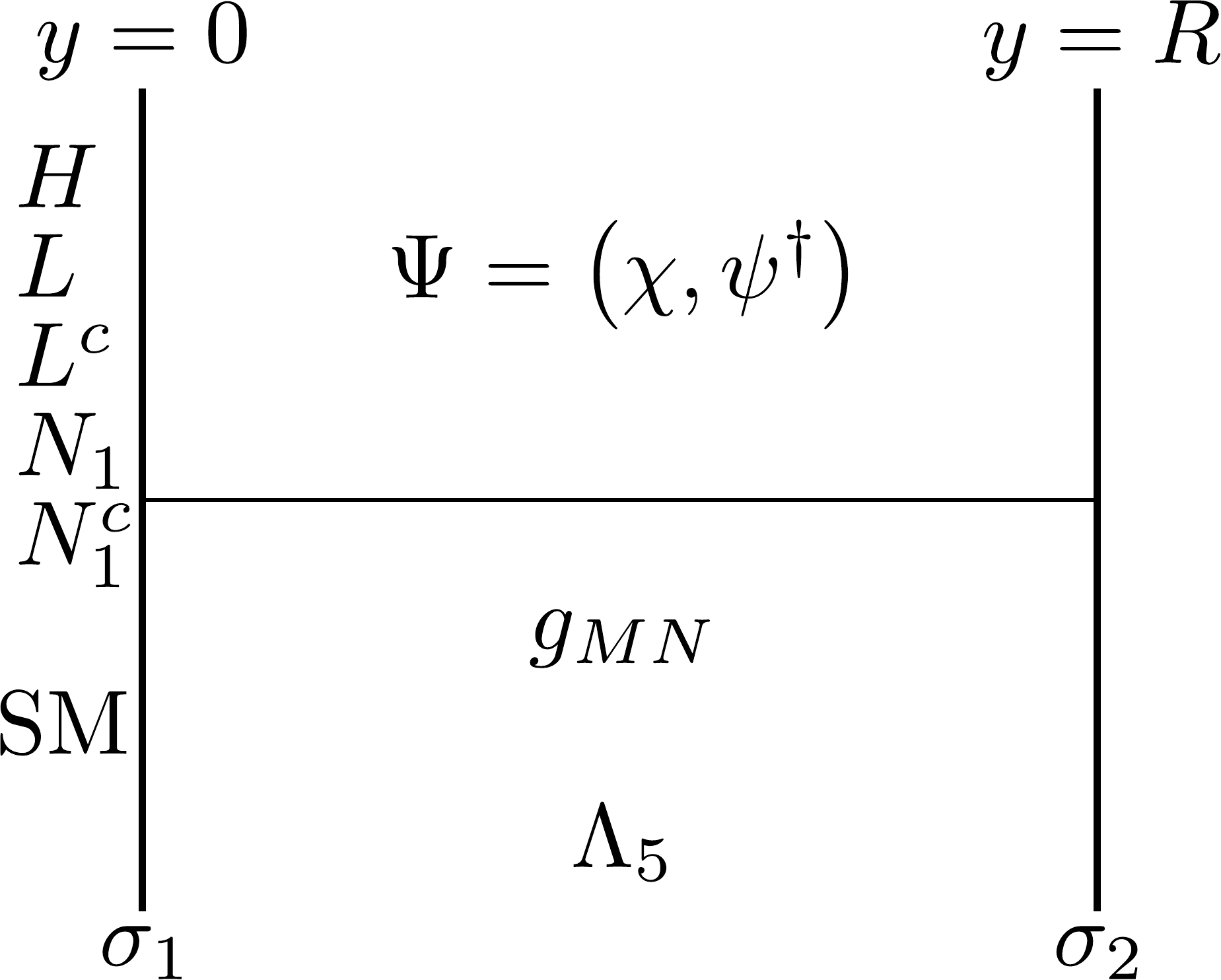}\end{center}
\caption{Geography of the model. The horizontal axis represents the coordinate distance $y$ in the extra dimension, while the vertical scale is representative for one of the usual four spacetime dimensions. Standard Model (SM) fields, including the Higgs field $H$, are confined to a 3-brane (depicted as a thick vertical line) at $y = 0$ with tension $\sigma_1$. The 3-brane is also populated by electroweak-charge fermions $L$ and $L^c$, which couple to $H$ and a pair of neutral fermions $N_1$ and $N_1^c$ through a Yukawa coupling. A bulk fermion $\Psi$ couples to this new fermion sector via a brane-localized mixing term with $N_1^c$, indirectly communicating the vacuum expectation value of $H$ to the bulk via the Casimir stress of $\Psi$. This Casimir stress, along with $\sigma_1$, the tension $\sigma_2$ of the $y=R$ brane, and the bulk cosmological constant $\Lambda_5$, determine the geometry of the space through their effect on the metric $g_{MN}$. We assume the 5D space to be orbifolded around $y=R$, so that there is a second copy of the interval attached to the branes on opposite sides.
}\label{fig:geography}
\end{figure*}

\begin{table}
\centering
\begin{tabular}{c|c|c|c|c}
 & $L = (N_0,E')$ & $L^c = (E^{\prime c},N_0^{c})$ & $N_1$ & $N_1^c$\\ \hline
$SU(3)_{\rm C}$ & $1$ & $1$ & $1$ & $1$ \\
$SU(2)_{\rm L}$ & $2$ & $2$ & $1$ & $1$ \\
$U(1)_{\rm Y}$ & $-1/2$ & $1/2$ & $0$ & $0$
\end{tabular}
\caption{The charges of the new states $L$ $(L^c)$ and $N_1 (N_1^c)$ under the Standard Model $SU(3)_{\rm C} \times SU(2)_{\rm L} \times U(1)_{\rm Y}$ gauge groups.}\label{tab:hypercharge}
\end{table}

\section{Model}\label{sec:5dmodels}

In this section, we present a five-dimensional theory where the Casimir energy density in the bulk depends on the vev of the Higgs field. We introduce couplings of the Higgs field to a fermionic sector with mass parameters that are much below the ultraviolet cutoff of the theory in a technically natural way. In a subset of Higgs-mass vacua, the Higgs vev causes one of the new brane fermions to become much lighter, in turn changing the boundary condition---and thus the \emph{sign} of the Casimir stress-energy---of a bulk fermion that mixes with this state on the SM brane. 

\subsection{Bulk and brane fields}

We consider a 5D theory with the bulk- and brane-localized states as shown in figure~\ref{fig:geography}. We assume that the position of the second brane at $y=R$ is an orbifold fixed point and in particular $\sigma_2$ is allowed to be negative. The 5D action of the theory is
\begin{align}
S = \int_{-R}^{R} \text{d}y\,\int \text{d}^4x  \Bigg \lbrace & \sqrt{-g^{(5)}} \Bigg [ \frac{M_5^3}{2} \mathcal{R}^{(5)} - \Lambda_5 + \frac{i}{2} \bar \Psi \Gamma^M  \overset{\leftrightarrow}{\partial}_M \Psi - M_\Psi \bar \Psi \Psi \Bigg] \label{eq:5action1}\\
 +& \sqrt{-g^{(4)}}\delta(y-0) \big[ - \sigma_1 + \mathcal{L}_\text{SM} + \mathcal{L}_1 \big] + \sqrt{-g^{(4)}}\delta (y-R) \big[- \sigma_2 \big] \Bigg\rbrace \label{eq:5action2}
\end{align}
where $\overset{\leftrightarrow}{\partial} = \overset{\rightarrow}{\partial}-\overset{\leftarrow}{\partial}$, and the signature of the metric $g^{(5)}_{M N}$ is $(-,+,+,+,+)$. The two branes are located at $y = 0$ and $y = R$; the induced metric on them is $g^{(4)}_{\mu\nu}$. 

The fermion $\Psi$ is a neutral 5D Dirac spinor, composed out of two Weyl fermions $\chi$ and $\psi$ as:
\begin{align}
\Psi = \mymatrix{\chi \\ \psi^\dagger }
\end{align}
We will set the bulk mass $M_\Psi = 0$, which can be achieved in a technically natural way with a parity symmetry in the fifth dimension around $y = 0$:
\begin{align}
\chi \to i \chi; \quad \psi^\dagger \to -i \psi^\dagger; \quad \partial_5 \to - \partial_5. \label{eq:5parity}
\end{align}
This forbids the bulk mass term in eq.~\ref{eq:5action1}, which in Weyl components takes the form $M_\Psi (\chi \psi + \text{c.c.})$. The kinetic terms of eq.~\ref{eq:5action1} can be seen to respect the symmetry (see eq.~\ref{eq:5action3} for the expansion into Weyl components). 
This 5D parity is exactly preserved both in the bulk and on the branes, as long as any brane fermions to which $\Psi$ couples also transform appropriately under the parity.

We impose the hard boundary conditions for $\psi$ and $\chi$ at $y = 0$ and $y = R$:
\begin{align}
\psi^\dagger|_{0} = 0; \quad \partial_5 \chi|_{0} = 0; \quad \partial_5 \psi^\dagger|_{R} = 0; \quad \chi|_{R} = 0. \label{eq:hardBC}
\end{align}
These boundary conditions are dynamically modified by brane-localized interactions $\mathcal{L}_1$ at $y = 0^+$, an infinitesimal distance away from $y=0$, so as to avoid treating the values and variations of the fields on the boundary as independent from the bulk values~\cite{Csaki:2003sh}:
\begin{align}
\mathcal{L}_1 = &- i L^\dagger \bar{\sigma}^\mu D_\mu L - i L^{c\dagger} \bar{\sigma}^\mu D_\mu L^c - i N_1^\dagger \bar{\sigma}^\mu D_\mu N_1 - i N_1^{c\dagger} \bar{\sigma}^\mu D_\mu N_1^c \label{eq:5lag1}\\
&+ M_0 L L^c + M_1 N_1 N_1^c + Y H L N_1^c + Y^c H^\dagger L^c N_1 + \text{c.c.}\label{eq:5lag2}\\
&+ \mu^{1/2}N_1^c \chi + \text{c.c.}. \label{eq:5lag3}
\end{align}
The brane-localized states have SM charges shown in table~\ref{tab:hypercharge}. The interactions in $\mathcal{L}_1$ ultimately communicate the Higgs vev to the bulk fermion, whose Casimir stress will affect the stabilization of the extra-dimensional radius.

\begin{figure}[t]
\begin{center}
\includegraphics[width = 0.9 \textwidth]{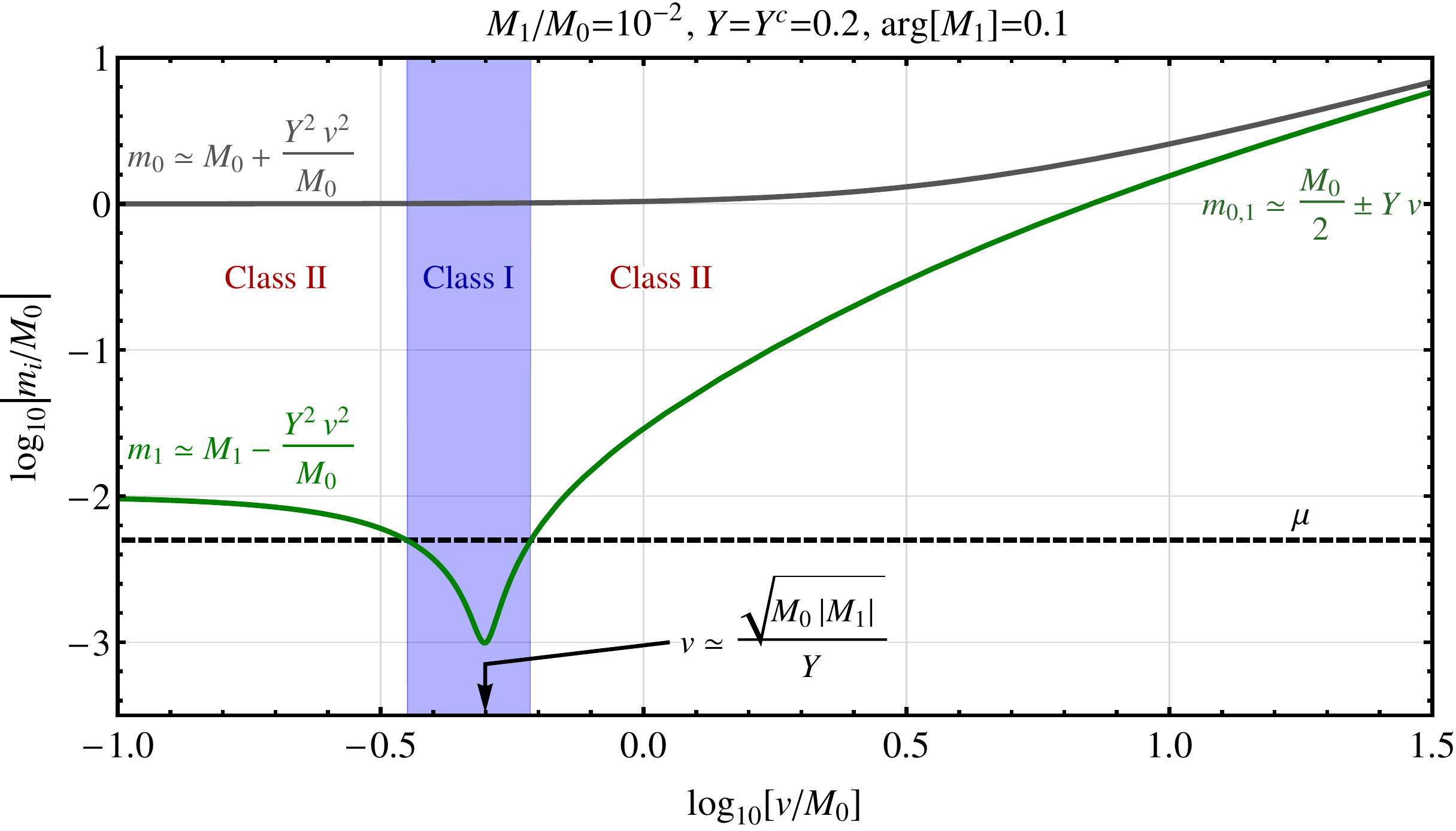}
\end{center}
\caption{
Mass eigenvalues $m_i = \lbrace m_0, m_1 \rbrace$ (gray, green) of neutral brane fermions $N_0$, $N_0^c$, $N_1$, and $N_1^c$, as a function of Higgs vev $\langle |H| \rangle \equiv v$, with quantities on both axes normalized to the bare vector mass $M_0$ on a logarithmic scale. For simplicity, we pick $Y = Y^c$ such that each mass eigenvalue is two-fold degenerate, assume $M_0 > |M_1| > \mu > 0$. For $m_H^2 > 0$ and $v = 0$, there is no mixing and the mass eigenvalues are just the bare masses $M_0$ and $|M_1|$, both larger than the scale of brane-bulk mixing $\mu$. For $Y Y^c v^2 \gtrsim M_0^2$, both mass eigenstates are again much heavier than $\mu$. For $Y Y^c v^2 \lesssim M_0^2$, however, the lighter mass eigenvalue $|m_1|$ can drop below $\mu$ for sufficiently small $\text{arg}[M_1]$ near $v \sim v_* \equiv \sqrt{M_0 |M_1|/Y Y^c}$, at which point mixing of the light state with the bulk fermion becomes important. Vacua with $v$ sufficiently near $v_*$ such that $|m_1(v)|\lesssim \mu$ are categorized in Class I, all others in Class II.
}\label{fig:fermionmass}
\end{figure}

The brane interactions $\mathcal{L}_1$ of the fermions in eqs.~\ref{eq:5lag1},~\ref{eq:5lag2},~and~\ref{eq:5lag3} are engineered such that the brane fermions only mix significantly with the bulk fermion for a select range of Higgs mass vacua. In figure~\ref{fig:fermionmass}, we show the mass eigenvalues $m_0(v)$ and $m_1(v)$ of the mass terms in eq.~\ref{eq:5lag2} as a function of the Higgs vev $\langle | H | \rangle \equiv v$. For Class II vacua, i.e.~all those with positive Higgs mass-squared values (where $v=0$) and most negative Higgs masses (those with $Y Y^c v^2 \gg M_0 |M_1|$), both eigenvalues $m_0(v)$ and $m_1(v)$ are indeed larger than the mixing scale $\mu$, as long as $M_0,|M_1| \gtrsim \mu$. However, for Higgs vevs $v$ near
\be v_* \equiv \sqrt{\frac{M_0 |M_1|}{Y Y^c}}, \label{eq:vstar}\ee
mixing can become important and substantially modify the boundary conditions of the bulk fermion in eq.~\ref{eq:hardBC}. These vacua belong to Class I. The range in $v$ near $v_*$ for which $m_1(v) \lesssim \mu$ and large mixing between brane and bulk fermions occurs, is $\frac{\Delta v^2}{v_*^2}= 2 \mu/M_1$ for $\mu < M_1$. If we take $M_0 \gg |M_1|$ and work in the interesting regime of $Y Y^c v^2 \ll M_0^2$, we can integrate out the heaviest mass eigenstate with mass $m_0(v)$, and arrive at the following effective Lagrangian for $\chi$ and the light states $\tilde{N}_1$ and $\tilde{N}_1^c$---mostly $N_1$ and ${N}_1^c$, with small admixtures of $N_0$ and $N_0^c$:

\begin{align}
\mathcal{L}_1 \supset m_1 (v) \tilde{N}_1 \tilde{N}_1^c + \mu^{1/2} \tilde{N}^c_1 \chi + \text{c.c.}
\end{align}
where 
\be
m_1 (v) =\frac{M_0+M_1-\sqrt{(M_0-M_1)^2+4Y Y^c v^2}}{2} \simeq M_1 - \frac{Y Y^c v^2}{M_0},
\ee
with the latter approximation holding for $v \sim v_*$ as long as $M_1 \ll M_0$.

In appendix~\ref{sec:appCasimir}, we show that this effective brane interaction replaces the first boundary condition at $y=0$ in eq.~\ref{eq:hardBC} with the mixed boundary condition at $y=0^+$:
\begin{align}
\left[-\partial^2 + \left|m_1(v)\right|^2\right] \psi^\dagger|_{0^+} +i \mu \bar{\sigma}^\mu \partial_\mu \chi|_{0^+} = 0.
\label{eq:softBC2}
\end{align}
This new ``soft'' boundary condition contains the essential physics. The second boundary condition in eq.~\ref{eq:hardBC} will similarly be changed but contains no new information since it follows from eq.~\ref{eq:softBC2} and the equations of motion in the bulk.\footnote{For example, a Dirichlet boundary condition for $\psi^\dagger$ automatically implies that its partner $\chi$ has a Neumann boundary condition at the same location, and vice versa.}
Inspecting the limiting behavior of the boundary condition in eq.~\ref{eq:softBC2} at different energy scales $\partial \sim 1/R$, we find for the two Classes of vacua:
\begin{alignat}{7}
&\text{Class I: } && |m_1(v)| \lesssim \mu \quad && \Rightarrow && \quad
\left\{\begin{array}{rl}
        \psi^\dagger|_{0^+} \simeq 0, \quad & \text{for } R \ll 1/\mu  \\
        \chi|_{0^+} \simeq 0 , \quad & \text{for } 1/\mu \ll R \ll \mu/|m_1(v)|^2 \\
        \psi^\dagger|_{0^+} \simeq 0. \quad & \text{for } \mu/|m_1(v)|^2  \ll R .
        \end{array} \right.\label{eq:limitbc1} \\
&\text{Class II: } && |m_1(v)| \gtrsim \mu \quad && \Rightarrow && \quad \psi^\dagger|_{0^+} \simeq 0. \label{eq:limitbc2} 
\end{alignat}
In the deep ultraviolet $1/R \gg \mu$, the brane interactions are never strong enough to change significantly the ``hard'' Dirichlet boundary condition for $\psi^\dagger$ at $y = 0$. For Class II vacua, at all scales below $|m_1(v)|$ the brane fermion $\tilde{N}_1^c$ effectively decouples, so the bulk-brane mixing disappears from the effective theory. Hence the ``soft'' boundary condition in eq.~\ref{eq:softBC2} at $y=0^+$ matches the one at $y=0$ in eq.~\ref{eq:hardBC} at all energy scales in Class II vacua. For Class I vacua, however, there exists a window of extra-dimensional sizes $\mu^{-1} \lesssim R \lesssim \mu/|m_1(v)|^2$ for which the BC flips from Dirichlet for $\psi^\dagger|_0$ to Dirichlet for $\chi|_{0^+}$ and consequently Neumann for $\psi^\dagger|_{0^+}$. Finally, in the far infrared $1/R \ll |m_1(v)|^2/\mu $, the effective frictional term for the $\chi$ field on the brane becomes too diluted relative to the effective brane mass for $\psi^\dagger$, again leaving the hard boundary condition unaffected.\footnote{One can consider a similar model with a single Majorana fermion $N_1$ instead of the Dirac pair $N_1$ and $N_1^c$. In this case, the modification of boundary conditions (Class I) persists for arbitrary $R\gg1/\mu$.}
In the window with flipped boundary conditions for the bulk fermion, we expect the Casimir stress to flip in sign, which we will compute explicitly in section~\ref{sec:casimir}.

\subsection{Technical naturalness}\label{sec:technatural}

Before delving into those machinations, we digress about the naturalness of the fermion sector in our model, since it is a crucial part of the mechanism. Unlike masses for scalar fields, fermion masses can be far below the UV cutoff of the theory, because symmetries can shield them from additive quantum corrections. In our model, the fermionic kinetic terms in eqs.~\ref{eq:5action1}~and~\ref{eq:5lag1} exhibit the symmetries
\be
U(1)_L \times U(1)_{L^c} \times U(1)_{N_1} \times U(1)_{N_1^c} \times U(1)_{\Psi}, \label{eq:fermionsymmetry}
\ee
which follow from invariance of the action under phase rotations of each of the five fermion fields. The Yukawa couplings explicitly break this symmetry down to three factors and associated symmetry rotations:
\begin{align}
U(1)'_{L} \times U(1)'_{L^{c}} \times U(1)_\Psi \quad : \quad \lbrace L,N_1^c \rbrace &\to \lbrace e^{i\alpha} L, e^{-i\alpha} N_1^c \rbrace \\
\lbrace L^c,N_1 \rbrace &\to \lbrace e^{i\beta} L^c, e^{-i\beta} N_1 \rbrace \nonumber \\
\Psi &\to e^{i\gamma} \Psi. \nonumber
\end{align}
The linear combination with $\alpha = -\beta = \gamma$ is exactly invariant, and corresponds to fermion number conservation in this sector.\footnote{Requiring invariance under this symmetry is not necessary, but it does simplify the analysis by forbidding Majorana masses and other Yukawa couplings in the fermion sector.} Another independent linear combination of transformations with $\alpha = \beta = \gamma$ is broken only by (the larger of) $|M_0|$ and $|M_1|$, but not by $\mu$. Finally, a last independent set of rotations, with $\alpha = -\beta = -\gamma$, is only broken by $\mu$, the mass mixing between brane and bulk fermions, and not by the vector masses. Hence the scale of the vector masses $\text{max}\lbrace |M_0|,| M_1| \rbrace$ and the scale of brane-bulk fermion mixing $|\mu|$ are both spurions of two separate symmetries, and will thus not receive additive radiative corrections, provided the theory respects these symmetries in the ultraviolet. A combination of these two scales turns out to determine the scale of electroweak symmetry breaking in our vacuum, with $v \sim |M_0| \gtrsim |M_1| \gtrsim |\mu|$. In this way, we relate the electroweak scale in our vacuum to a symmetry-enhanced point of a fermion sector, one much below the cutoff.

\begin{figure}[t]
\begin{center}
\includegraphics[width = 0.4 \textwidth]{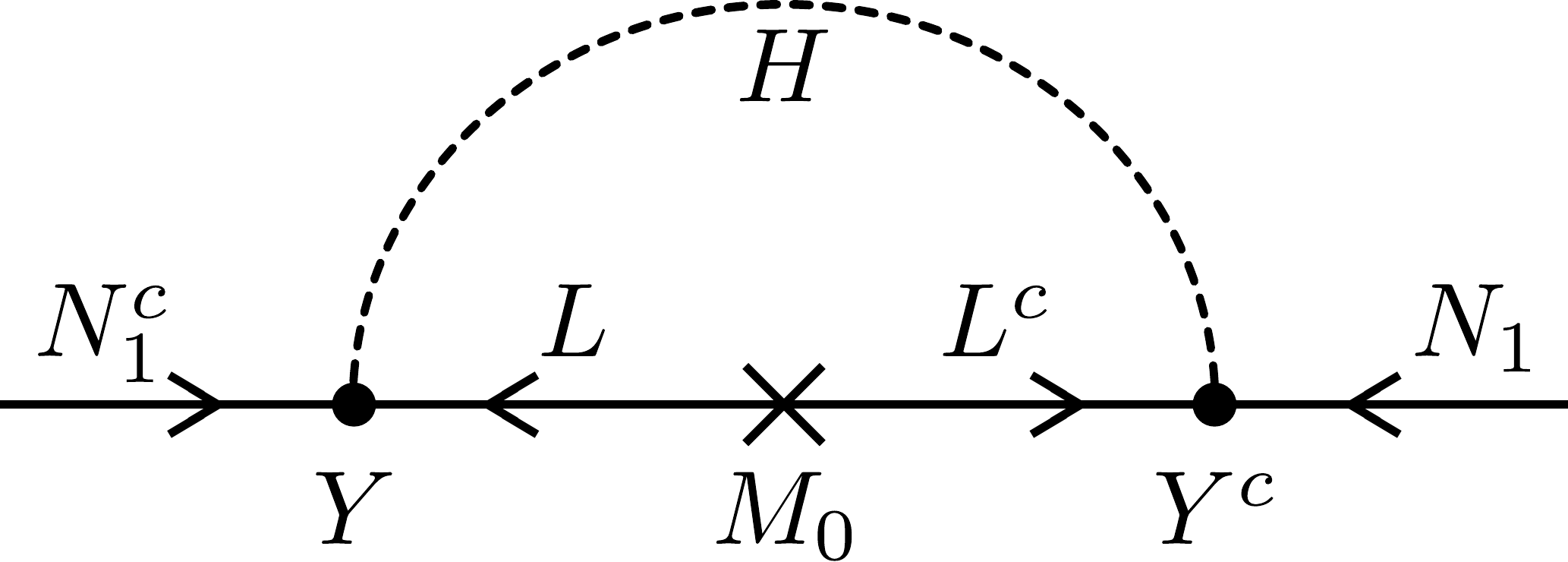}
\end{center}
\caption{
Feynman diagram depicting one-loop radiative corrections to $M_1$ given in eq.~\ref{eq:masscorrection}.
}\label{fig:feynmanloop}
\end{figure}

While on the topic of naturalness, we point out that the masses $M_0$ and $M_1$ are connected in our model, which will lead to important phenomenological consequences. We will often take $|M_0| \gg |M_1|$, but this hierarchy cannot be arbitrarily large; radiative corrections will destabilize this hierarchy. At one-loop level, the diagram of figure~\ref{fig:feynmanloop} leads to the mass correction:
\begin{align}
\delta M_1 \simeq \frac{2 Y Y^c }{(4\pi)^2} M_0^* \log \left( \frac{M_\text{UV}^2}{\max \lbrace |M_0|^2, m_H^2 \rbrace} \right).\label{eq:masscorrection}
\end{align}
Not all phases can be removed from the Lagrangian $\mathcal{L}_1$ in eqs.~\ref{eq:5lag2}~and~\ref{eq:5lag3}. There are five complex parameters, namely $M_0$, $M_1$, $\mu$, $Y$, and $Y^c$, and five possible phase rotations in eq.~\ref{eq:fermionsymmetry}. However, the phase rotation corresponding to lepton number conservation leaves the full Lagrangian invariant, so one physical phase remains. Without loss of generality, we will henceforth take $\lbrace M_0, \mu, Y, Y^c \rbrace$ all to be real and positive, and rotate the physical phase into $M_1 = |M_1| e^{i \arg[M_1]}$. Later, we will require this phase to be somewhat small, specifically $\arg [ M_1 ] \ll \pi$ at energies of order $M_0$, such that $|m_1(v) < \mu|$ can be satisfied for some Higgs vev. Such a small phase is automatically realized when $|M_1| \ll M_0$ in the ultraviolet, and the dominant contributions to $M_1$ at low energies are mediated via the renormalization-group effects in figure~\ref{fig:feynmanloop} and eq.~\ref{eq:masscorrection}.

\subsection{Casimir stress-energy}\label{sec:casimir}

Casimir stress-energy arises due to the radius dependence of vacuum fluctuations.  It can be computed by extracting the finite parts of $\langle 0 | {T^{M}}_N(x,y) | 0 \rangle$ at any point in the bulk and on the branes; divergent terms are absorbed by the local counterterms $\sigma_1$, $\sigma_2$, and $\Lambda_5$. To keep the calculation tractable, we will first restrict ourselves to a flat bulk before generalizing to warped geometries. Because conformal symmetry is approximately preserved in the bulk given that the only dynamical field is the massless spinor $\Psi$,\footnote{The graviton is not conformal, but its Casimir stress can be subdominant to that of $\Psi$. We return to this point at the end of the section.} the Casimir stress must be proportional to ${T^M}_N \propto \text{diag}(1,1,1,1,-4)$.
Poincar\'{e} symmetry in directions parallel to the branes implies ${T^\mu}_\nu \propto {\delta^\mu}_\nu$, Weyl invariance requires tracelessness, which together with stress-energy conservation implies that ${T^M}_N$ is independent of $y$. The Casimir stress in the bulk is thus fixed by symmetries up to an overall coefficient:
\begin{align}
{T^M}_N = \frac{\beta}{R^5}\text{diag}(1,1,1,1,-4). \label{eq:casimirstressflat}
\end{align}
%\begin{align}
%{T^M}_N = \frac{\rho_C}{R}\text{diag}(1,1,1,1,-4) \equiv \frac{\beta}{R^5}\text{diag}(1,1,1,1,-4), \label{eq:casimirstressflat}
%\end{align}
We parametrized the stress in terms of a dimensionless coefficient $\beta$, which depends on $|m_1(v)|$, $\mu$, and $R$.
To determine $\beta$, we first consider
%Finally, we will ignore brane-localized contributions, i.e.~$R$-dependent additions to the brane tensions; they can be checked to be subleading relative to the Casimir energy density integrated over the internal part of the bulk.
the 4D vacuum energy $\rho_C$, the $y$-integral over ${T^0}_0$ above, that is given by the sum over all vacuum bubble diagrams, one for each Kaluza-Klein (KK) state of $\chi$ and $\psi^\dagger$, and equates to:
\begin{align}
\rho_C = -2 \sum_{n=1}^\infty \int \frac{d^4 k}{(2\pi)^4} \log(k^2 + m_n^2)\label{eq:casimirpotentialsum}.
\end{align}
For a real 4D scalar, the prefactor of the sum is $+1/2$ in the effective potential, while for a 5D fermion one gets $-2$.
The minus sign comes from the fermion loop, one extra factor of two from the degeneracy of the KK towers for $\chi$ and $\psi^\dagger$, and another factor of two for the number of helicity states per Weyl spinor. In appendix \ref{sec:appCasimir}, we compute the KK spectrum $\lbrace m_n \rbrace$ given the boundary conditions in eqs.~\ref{eq:hardBC}~and~\ref{eq:softBC2}, and use $\zeta$-function regularization to extract from eq.~\ref{eq:casimirpotentialsum} the finite, $R$-dependent piece, which is:
\begin{align}
\rho_C = -\frac{1}{16 \pi^2} \frac{1}{R^4} \left\lbrace -\frac{3}{2}\zeta(5) + \mathcal{I} \left[\frac{\mu}{|m_1(v)|^2 R} ,\frac{1}{|m_1(v)|^2 R^2}\right] \right\rbrace, \label{eq:casimirpotential}
\end{align}
where $\mathcal{I}(a,b) \equiv 4 \int_0^\infty dx\, x^3 \log \lbrace [ax + (1+ bx^2)\coth(x)]/[a x + (1+ bx^2)]\rbrace$.

\begin{figure}[t]
\begin{center}
\includegraphics[width = 0.8 \textwidth]{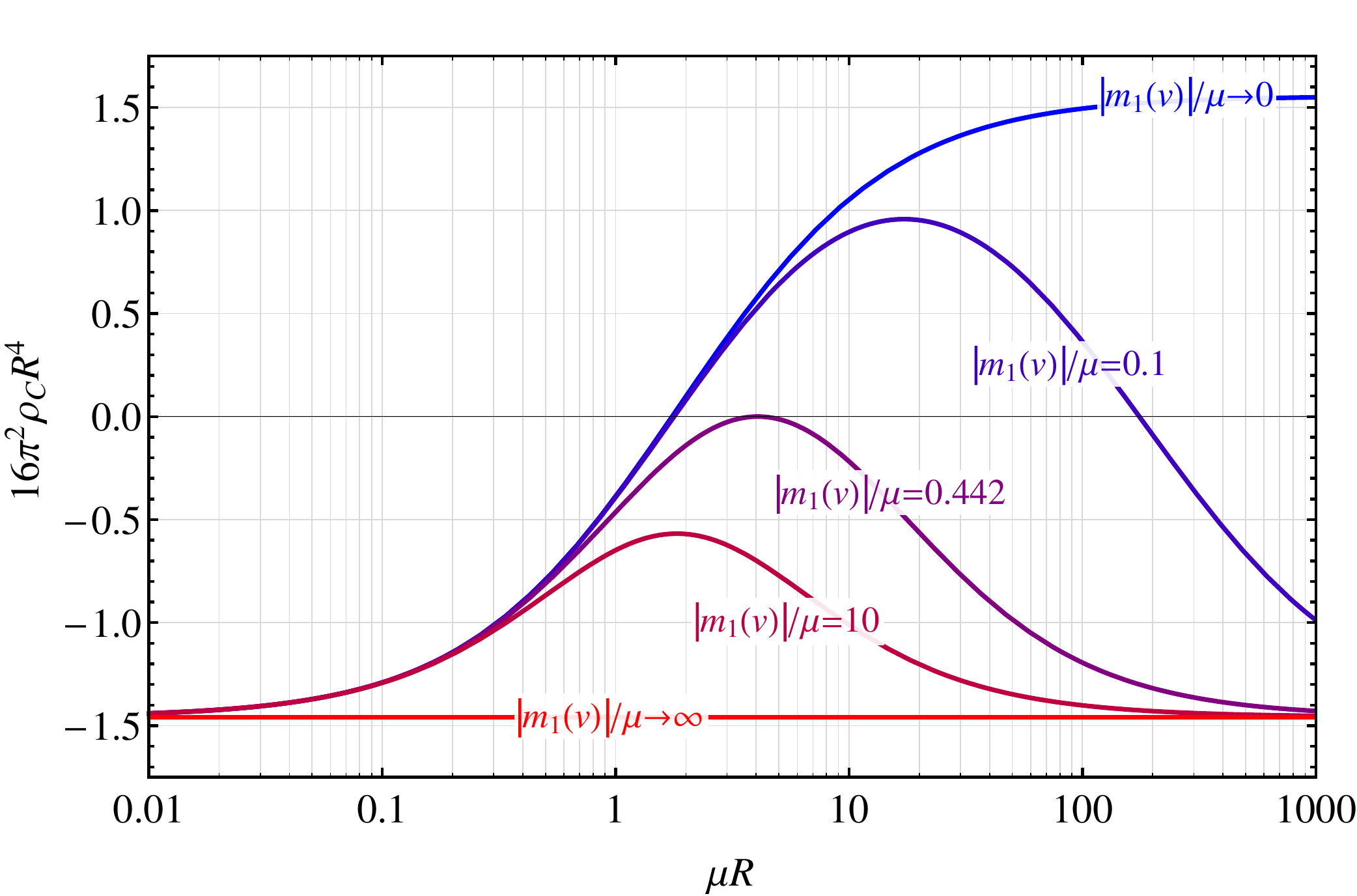}
\end{center}
\caption{Plots of the Casimir energy coefficient $16\pi^2 \rho_C R^4$ as a function of the extra-dimensional radius $R$, in units of (the inverse of) the bulk-brane mixing scale $\mu$. The five curves are for fixed $\mu$ but variable $|m_1(v)|$, increasing from top (blue) to bottom (red). The Casimir energy can be repulsive ($\rho_C,\beta>0$) if and only if $|m_1(v)|/\mu \lesssim 0.442$ in the parametric window $1/\mu \lesssim R \lesssim \mu/|m_1(v)|^2$; otherwise, it is attractive ($\rho_C,\beta < 0$).
}\label{fig:casimirpotential}
\end{figure}

Inspection of eq.~\ref{eq:casimirpotential} confirms the limiting behavior of $\rho_C$ anticipated in eqs.~\ref{eq:limitbc1} and~\ref{eq:limitbc2}. In the deep UV, far IR, and in the window $1/\mu \ll R \ll \mu / |m_1(v)|^2$ (if it exists), we get the standard results for ``pure'' boundary conditions, up to small corrections:
% expressed in terms of the Casimir coefficient $\beta = \rho_C R^4$:
\begin{align}
\rho_C \simeq   \frac{1}{16 \pi^2 R^4} \left\{\begin{array}{lll}
        -\frac{45 \zeta(5)}{32} & \left[1 - \frac{16 \zeta(3)}{15 \zeta(5)} \mu R + \dots\right], \quad & \text{for } R < 1/\mu ;  \vspace{0.1em} \\ \vspace{0.1em}
        +\frac{3\zeta(5)}{2} & \left[1 - \frac{4}{\mu R} + \dots \right], \quad & \text{for } 1/\mu < R < \mu / |m_1(v)|^2 ; \\
        -\frac{45 \zeta(5)}{32} & \left[1 - 4 \frac{\mu}{|m_1(v)|^2 R} + \dots \right], \quad & \text{for } \mu / |m_1(v)|^2  < R .
        \end{array} \right. 
        \label{eq:casimirrho}
\end{align}
In figure~\ref{fig:casimirpotential}, we plot the dimensionless combination $16\pi^2 R^4 \rho_C$ as a function of $\mu R$, for different ratios $|m_1(v)|/\mu$. We conclude that only Class I vacua, with $v \sim v_*$ such that $|m_1(v)| \lesssim \mu$, allow for repulsive Casimir energy density in the case of a flat extra dimension.

When $R$ is far removed from any length thresholds in the functional form of $\rho_C$, ${T^0}_0$ is simply given by $\rho_C/2R$. On the other hand, for $R$ close to $1/\mu$ or $\mu/|m_1(v)|^2$ in  Class I vacua, the boundary conditions are not purely Neumann or Dirichlet. Here, the Casimir stress tensor receives contributions localized on the boundaries, similar to the case of mixed boundary conditions for a scalar~\cite{Romeo:2000wt}. Correspondingly, the coefficient of the bulk ${T^M}_N$ will not be as simply related to $\rho_C$. It is possible to extend our calculation to this case as well, but for simplicity we will assume that there is a mild hierarchy $1/\mu \ll R \ll \mu/|m_1(v)|^2$ so that the boundary contributions can be safely ignored in the region of interest. This subtlety is irrelevant in Class II vacua given the pure boundary conditions of eq.~\ref{eq:limitbc2} at all length scales. In the asymptotic regimes, the coefficient of the Casimir stress in eq.~\ref{eq:casimirstressflat} is $R$-independent and given by
\begin{align}
\beta \simeq   \frac{1}{32 \pi^2 } \left\{\begin{array}{lll}
        -\frac{45 \zeta(5)}{32} , \quad & \text{for } R \ll 1/\mu ;  \vspace{0.1em} \\ \vspace{0.1em}
        +\frac{3\zeta(5)}{2} , \quad & \text{for } 1/\mu \ll R \ll \mu / |m_1(v)|^2 ; \\
        -\frac{45 \zeta(5)}{32} , \quad & \text{for } \mu / |m_1(v)|^2  \ll R .
        \end{array} \right. 
        \label{eq:casimirbeta}
\end{align}

The generalization of eqs.~\ref{eq:casimirstressflat}~and~\ref{eq:casimirbeta} to curved geometries is relatively straightforward. Since we are only interested in approximately static constructions---those with a stable radion and very small Hubble constant $\mathcal{H}$---we can restrict ourselves to spacetimes which are conformally flat to a high degree, with metric
\begin{align}
ds^2=a(z)^2\l dx^2+dz^2\r. \label{eq:conformalmetric}
\end{align}
We denote by $L$ the conformal distance between the two brane locations $z_1$ and $z_2$:
\be
L = z_2 - z_1 = \int_0^{R}\frac{dy}{a(y)}. \label{eq:Lconf}
\ee
The scale $L$ is the relevant one for the Casimir energy of the fermions in the curved bulk, since the masses of their low-lying KK modes are of order $1/L$. 
For conformally flat spacetimes and conformally coupled fields, the Casimir stress takes on a particularly simple form~\cite{birrell1984quantum}. Since the conformal anomaly is absent in odd dimensions,\footnote{In odd dimensions, there are in principle boundary anomalies, but for our purposes their effect will be equivalent to a change in brane tensions.} it differs from the flat space expression eq.~\ref{eq:casimirstressflat} only by the square root of the metric determinant in eq.~\ref{eq:conformalmetric}:
\be
{T^M}_N=\frac{\beta}{a(z)^5 L^5}\text{diag}(1,1,1,1,-4),
\label{eq:casimirconformal}
\ee
where $\beta$ is given by the flat-space expressions of eq.~\ref{eq:casimirbeta} as long as $L$ is far enough from the thesholds where the sign of the Casimir energy flips. %such that conformality holds to a good approximation.

%In Class I vacua, this is the case for $L$ far away from $1/\mu$ and $\mu/|m_1(v)|^2$---or well in between, where the fermion boundary conditions become nearly %pure Neumann or Dirichlet as in eq.~\ref{eq:limitbc1}. In Class II vacua, conformality in the bulk is always a good approximation for the bulk fermion, given the pure %boundary conditions of eq.~\ref{eq:limitbc2} at all length scales.

Besides the fermions in our theory, the 5D gravitons will also contribute to the Casimir stress. They generate attractive, nonconformal Casimir potential contributions, and will therefore tend to destabilize the fifth dimension. If the extra dimension is moderately warped, the gravitational Casimir energy contribution is suppressed to negligible levels by the warping factor. For a flat extra dimension, additional light fermions with even boundary conditions could be added to counter the negative Casimir energy of the gravitons, or $\Psi$ could take on a large multiplicity. As we will show in appendix~\ref{sec:radionmass}, warping and large $\Psi$ multiplicities both increase the radion mass, which is necessary anyway to obtain a phenomenologically viable model.

\section{Radius stabilization}\label{sec:radiusstabilization}

In this section, we discuss the stabilization of the fifth dimension in the context of theory depicted in figure~\ref{fig:geography}. Before presenting the details of our calculation in subsections~\ref{sec:nostab}~and~\ref{sec:Casimirstabilization}, we provide an outline of the main results. After reading this summary, the reader may skip directly to section~\ref{sec:vacua} and come back to the rest of this section at a later time.
%Here, we give an outline and summary of the main results.

\begin{figure}[t]
\begin{center}
\includegraphics[width = 0.99 \textwidth]{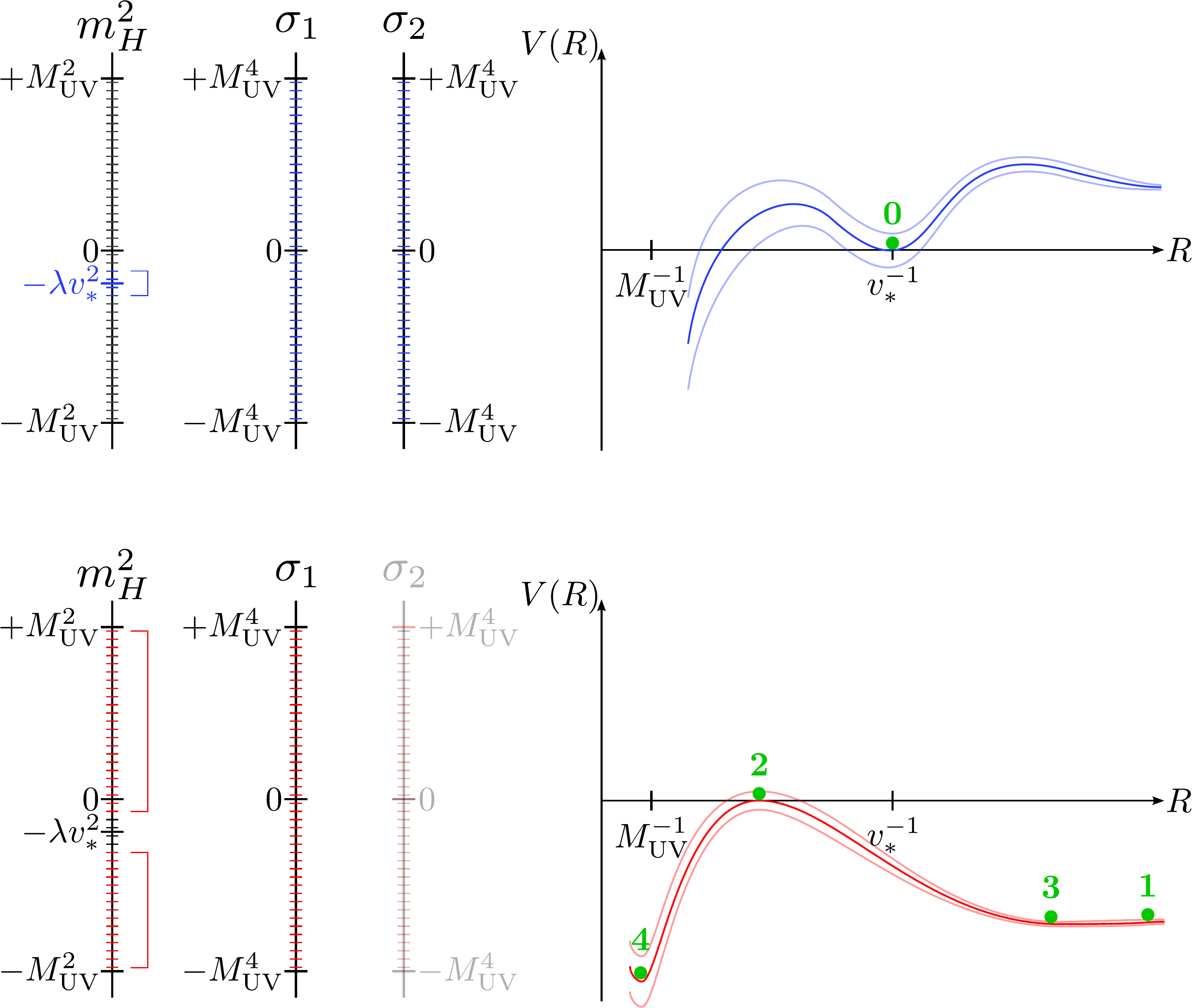}
\end{center}
\caption{Schematic of the classes of vacua. \textbf{Class I vacua} (top): for Higgs masses $m_H^2$ near a technically natural scale  $-\lambda v_*^2$ (indicated in blue), the Casimir stress generated by a bulk fermion is such that the effective radion potential $V(R)$ can be stabilized with near-zero 4D effective cosmological constant $\Lambda_4$, at a physical distance $R$ near the inverse of the critical Higgs vev $v_*$. In such a radion minimum (Point \textbf{0}, green), both brane tensions $\sigma_1$ and $\sigma_2$ can assist in tuning the cosmological constant. \textbf{Class II vacua} (bottom): for all other Higgs masses (indicated in red), the radion potential does not develop stable, static solutions; no configuration exists for which $V(R) \simeq 0$, $V'(R) = 0$, and $V''(R)>0$ for $R \gtrsim M_\text{UV}^{-1}$. Radion extrema are either a runaway direction such that the non-SM brane disappears from the effective theory (Point 1), static but unstable (Point 2), or stable but nonstatic with large $|\Lambda_4|$ (Point 3). The radion could be stabilized in the ultraviolet (Point 4), but with a much sparser distribution of vacua for the brane tensions. Different curves in the radion potential correspond to different values of $\sigma_2$. In this arrangement, the vast majority of vacua with a small $\Lambda_4$ are of Class I, with a Higgs vev near $v_*$ and extra-dimensional radius near $v_*^{-1}$ (Point 0), realizing the vacuum structure of figure~\ref{fig:comb}.}\label{fig:summary}
\end{figure}

In section~\ref{sec:nostab}, we consider a simplified model wherein the only stress-energy in the bulk is that of a bulk cosmological constant $\Lambda_5$, as a warm-up exercise to establish notation and methodology. Physically, this also corresponds to the unstabilized scenario of Point 1 in figure~\ref{fig:summary}, when the distance between the two branes is so large that the Casimir stress, a short-distance effect, is negligible. We will show that the observed cosmological constant $\Lambda_4$ on the SM brane is completely independent of the other brane's tension $\sigma_2$, bearing out the intuition that in absence of a stabilization mechanism, the physics on the two branes is independent by locality in the bulk. Specifically, when the metric in the bulk is pure AdS, a change in $\sigma_2$ can be compensated by a shift in the radius $R$ without any influence on the Hubble curvature. The radion is massless. In this effective single-brane setup, the lower bound on the expected minimum $|\Lambda_4|$ is given by the precision to which the SM brane tension $\sigma_1$ can be tuned, quantified in eq.~\ref{eq:singlebranetuning} and illustrated in the left panel of figure~\ref{fig:tuning}.

In section~\ref{sec:Casimirstabilization}, we include Casimir stress, which breaks AdS symmetry in the bulk and can extremize the radion potential~\cite{Hofmann:2000cj,Kanti:2002zr,Garriga:2000jb} at finite distance (Points 0 and 2 in figure~\ref{fig:summary}). In this case, the effective $\Lambda_4$ on the SM brane depends on both $\sigma_1$ and $\sigma_2$, as summarized by the relations in eqs.~\ref{eq:sigma_mink}~and~\ref{eq:lambda_w} for a flat and warped bulk, respectively. We stress that warping is unnecessary for the functionality of our mechanism; we include it for generality and to obtain a model with a phenomenologically viable radion mass and coupling. Our theory does not rely on warping to lower the Planck scale: the SM is localized on the ``ultraviolet'' brane, i.e.~the one with positive tension, in contrast to RSI~\cite{rsi}. The dependence of $\Lambda_4$ on $\sigma_1$ and $\sigma_2$ is depicted in the right panel of figure~\ref{fig:tuning}. In this two-dimensional $\sigma_1 \otimes \sigma_2$ space, $\Lambda_4$ can be tuned very precisely, contrary to the one-dimensional $\sigma_1$ space in the case without a stabilizing potential.

Next, we analyze the stability of the solutions with small $\Lambda_4$, with the details of the calculation postponed to appendix~\ref{sec:radionmass}. We prove that the radion mass-squared, explicitly given in eqs.~\ref{eq:mrflat}~and~\ref{eq:mrcurved} for flat and warped fifth dimensions, is positive if and only if the Casimir energy is positive ($\beta>0$). Hence static radion extrema in Class II are always unstable maximima (including e.g.~Point 2 in figure~\ref{fig:summary}), as they have $\beta < 0$ due to odd fermionic boundary conditions. Only Class I vacua can attain even boundary conditions for $\Psi$ and $\beta > 0$, and thus stable, static radion minima like Point 0 in figure~\ref{fig:summary}.

Finally, even though Class II vacua do not permit static, stable solutions, they still allow for stable, finite-radius solutions that are not static, with nonzero $\Lambda_4$ like at Point 3 in figure~\ref{fig:summary}. We compute the smallest attainable $|\Lambda_4|$ in these possible radion minima in eqs.~\ref{eq:bound_m}~and~\ref{eq:bound_w}, representing parametrically stronger (for a flat fifth dimension) or similar lower bounds (for a warped fifth dimension) as in the case without a stabilizing potential, which do \emph{not} depend on $\sigma_2$. 

Before proceeding to the technicalities of the calculation, we point out three subtleties. Firstly, the bulk fermion $\Psi$ can only generate a finite amount of \emph{repulsive} Casimir stress ${T^M}_N \sim \beta/L^5 \lesssim \beta \mu^5$ in the window $1/\mu \lesssim L \lesssim \mu/|m_1(v)|^2$. Only for a relatively small range of tensions $\sigma_1$, near a critical tension $\sigma_*$, can the positive Casimir energy density counteract the contributions from the zero-point energies on the SM brane and in the bulk, which are of order 
$|\sigma_1 - \sigma_*|/L$. In observance of decoupling, this requires a tuning of $\sigma_1$ to a precision of $\mu^4$ (near the electroweak scale in our vacuum) but, crucially, not to a precision of the observed cosmological constant. Secondly, throughout we find radion extrema by self-consistently solving Einstein's equations. This procedure automatically includes gravitational backreaction with Ricci tensor of order  $\mathcal{R}_{MN} \sim T_{MN}/M_5^3$, which appears in the radion potential at the same order as the stress-energy that sources it, and can thus never be ignored.\footnote{Some of the literature on extra dimensions does ignore backreaction, including early attempts at Casimir stabilization in refs.~\cite{Appelquist:1983vs,Ponton:2001hq,vonGersdorff:2005ce}.} Thirdly, in our stability analysis, we compute the mass-squared of the lightest mass eigenstate of metric fluctuations $\delta g_{MN}$ on top of the background solution $g_{MN}$ that solves Einstein's equations. This mass eigenstate---the radion---generally has a nontrivial wavefunction in the extra dimension.\footnote{In many works, the radion profile is assumed to coincide with that of the background solution, and simply amounts to a uniform fluctuation in radius $R + \delta R$ with a profile $\delta g_{55}(y) \propto g_{55}(y)$, often leading to qualitatively different results~\cite{Hofmann:2000cj,Kanti:2002zr}.}

\subsection{No stabilization mechanism}\label{sec:nostab}

We consider a five-dimensional geometry with the only contribution to the stress tensor in the bulk given by the cosmological constant $\Lambda_5$ (see figure \ref{fig:geography}). We do not yet include Casimir stress-energy; the analysis in this subsection is thus relevant only for unstabilized brane configurations at very large radius. Solutions of Einstein's equations in this setup have been extensively studied in ref.~\cite{Kaloper:1999sm} where we refer the reader for more details. Throughout this and the next sections, we assume that the 5D cosmological constant is fixed and negative, and that only the brane tensions are responsible for fine tuning the effective 4D cosmological constant $\Lambda_4$.

We look for solutions that are maximally symmetric along directions parallel to the two branes, such that the metric can be put in the form:
\be
d s^2=a(y)^2 ds_\text{dS}^2 + dy^2,
\label{eq:ds2}
\ee
where $ds^2_\text{dS} = -dt^2 + e^{2 \mathcal{H} t} d\mathbf{x}^2$ is the four-dimensional deSitter metric with Hubble constant $\mathcal{H}$ related to $\Lambda_4$ in the usual way: $3 \mathcal{H}^2 M_\text{Pl}^2=\Lambda_4$, with the 4D effective (reduced) Planck mass $M_\text{Pl}$ related to the 5D gravity scale $M_5$ as~\cite{rsi}
\be
M_\text{Pl}^2 = \frac{M_5^3}{k} (1 - e^{- 2 k R}) =
\begin{cases}
2 M_5^3 L,\quad & k L \sim kR \ll 1 \\
M_5^3/k, \quad & kL \gg kR \gg 1 
\end{cases}
\ee
when the fifth dimension is flat or warped, respectively. (The curvature scale $k$ is defined in eq.~\ref{eq:kcurv}.) Hereafter, we will be using both $\mathcal{H}$ and $\Lambda_4$ interchangeably. Anti-deSitter solutions with negative $\Lambda_4$ amount to changing $\mathcal{H}\to i\mathcal{H}$, with otherwise identical conclusions.
In addition to some gauge fixing already done in eq.~\ref{eq:ds2}, we choose the first brane to be located at $y=0$, as in section~\ref{sec:5dmodels}, and normalize the scale factor to $a(0)=1$. The location of the second brane is taken to be at $y=R$, where the extra-dimensional radius $R$ as well as $a(y)$ and $\mathcal{H}$ will be dynamically determined from $\Lambda_5$, $\sigma_1$, and $\sigma_2$.

The five-dimensional Einstein equations resulting from the action in eq.~\ref{eq:5action2} take on a form similar to the classic Friedmann equations, with time replaced by the extra-dimensional, space-like coordinate $y$:
\begin{alignat}{6}
3\frac{a''}{a} && {}+{} 3\l\frac{a'}{a}\r^2  && {}-{} 3 \frac{\mathcal{H}^2}{a^2} && {}+{} \frac{\Lambda_5}{M_5^3} &  = {}-{}\frac{\sigma_1}{M_5^3} \delta(y) - \frac{\sigma_2}{M_5^3}\delta(y-R)
\label{eq:Friedmann1}  \\
 && {}+{} 6\l\frac{a'}{a}\r^2  && {}-{} 6\frac{\mathcal{H}^2}{a^2}  && {}+{}\frac{\Lambda_5}{M_5^3}  & = 0 \label{eq:Friedmann2}
\end{alignat}
In particular, integrating eq.~\ref{eq:Friedmann1} around infinitesimal regions around the branes yields the  jump conditions:
\begin{align}
a'(0)&=-a(0)\frac{\sigma_1}{6 M_5^3}  \label{eq:Jump1},\\
a'(R)&=+a(R)\frac{\sigma_2}{6 M_5^3} \label{eq:Jump2}.
\end{align}
From eqs.~\ref{eq:Friedmann2}~and~\ref{eq:Jump1} evaluated at $y=0$, the Hubble constant is found to be completely independent of $\sigma_2$:
\be 
\mathcal{H}^2=\frac{1}{6 M_5^3}\l\Lambda_5+\frac{\sigma_1^2}{6 M_5^3}\r \equiv \frac{\lambda_1}{6 M_5^3}, \label{eq:Hubble}
\ee
where we have defined $\lambda_1$ for later convenience. A change in the tension $\sigma_2$ of the second brane would result in a change in radius $R$, not a different Hubble constant. In order to obtain a tiny effective four-dimensional constant $\Lambda_4$ close to the observed value, the quantity $\lambda_1$ would have to be tuned very close to zero. For small $|\lambda_1| \lesssim |\Lambda_5|$, the four-dimensional cosmological constant is given by
\begin{equation}
      \Lambda_4 \simeq \sigma_1-\sigma_*, \qquad (|\sigma_1 -\sigma_*| \lesssim \sigma_*) \label{eq:singlebranetuning} 
\end{equation}
where we assumed $L \gg 1/k$, and defined the critical value for the tension which would give rise to an exactly static solution ($\mathcal{H} = 0$):
\be  \sigma_*=\sqrt{-6 M_5^3 \Lambda_5}. \label{eq:sigma*} \ee
For large $\lambda_1 \gtrsim \Lambda_5$, we find that $\mathcal{H}^2 \gtrsim |\sigma_1/6 M_5^3|^2 \gtrsim k^2$, where we employed the usual definition for the curvature scale:
\be k \equiv \sqrt{\frac{-\Lambda_5}{6 M_5^3}}. \label{eq:kcurv}\ee
A Hubble horizon $\mathcal{H}^{-1}$ smaller than the extra-dimensional curvature scale $k^{-1}$ would correspond to an intrinsically five-dimensional world, and would be strongly disfavored anthropically by itself. In the remainder of this work, however, we will take $k$ to be quite large so that the small detuning constraint is the relevant one. 

\begin{figure}
\centering
\includegraphics[width = 0.49\textwidth]{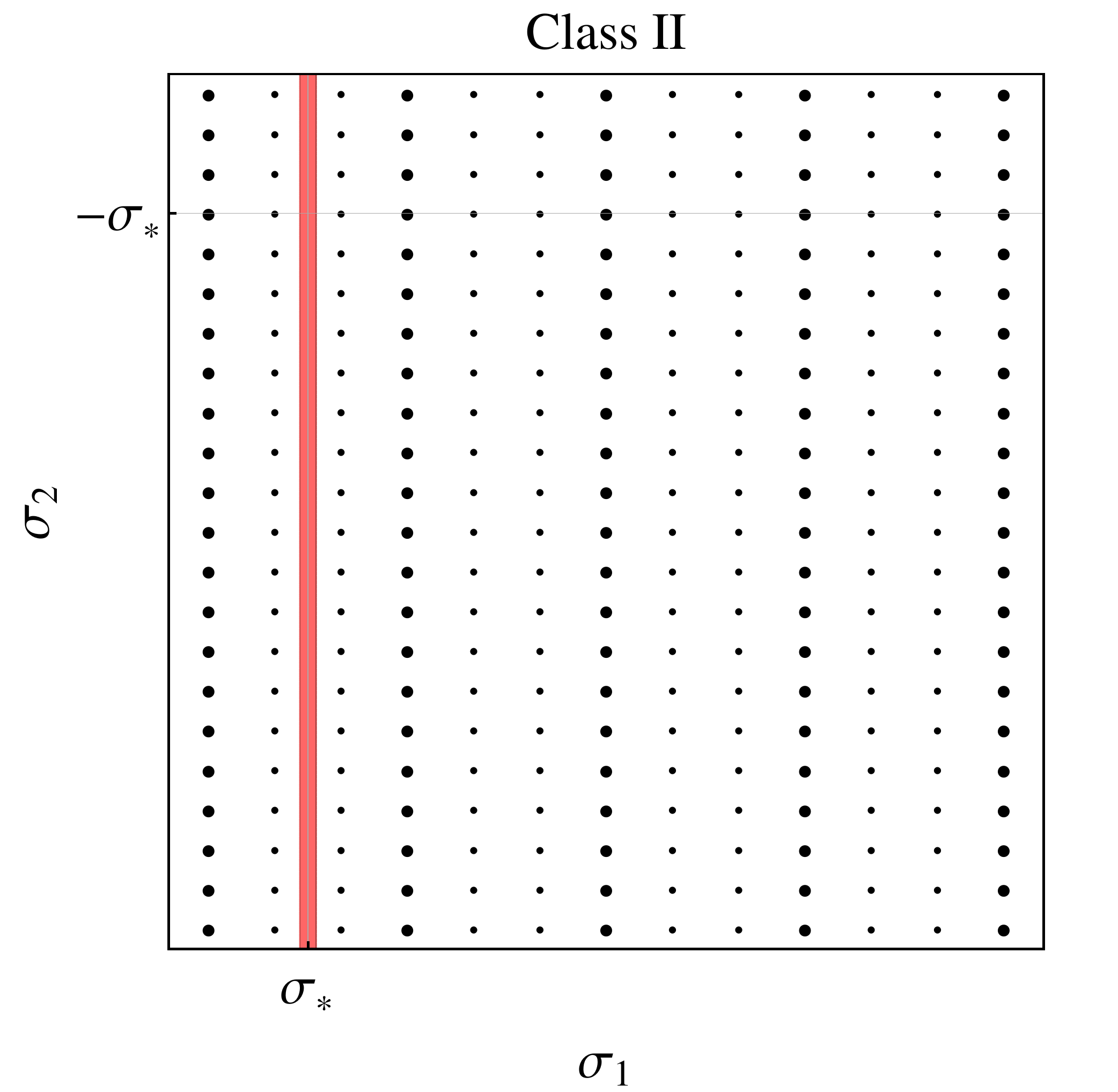}
\includegraphics[width = 0.49\textwidth]{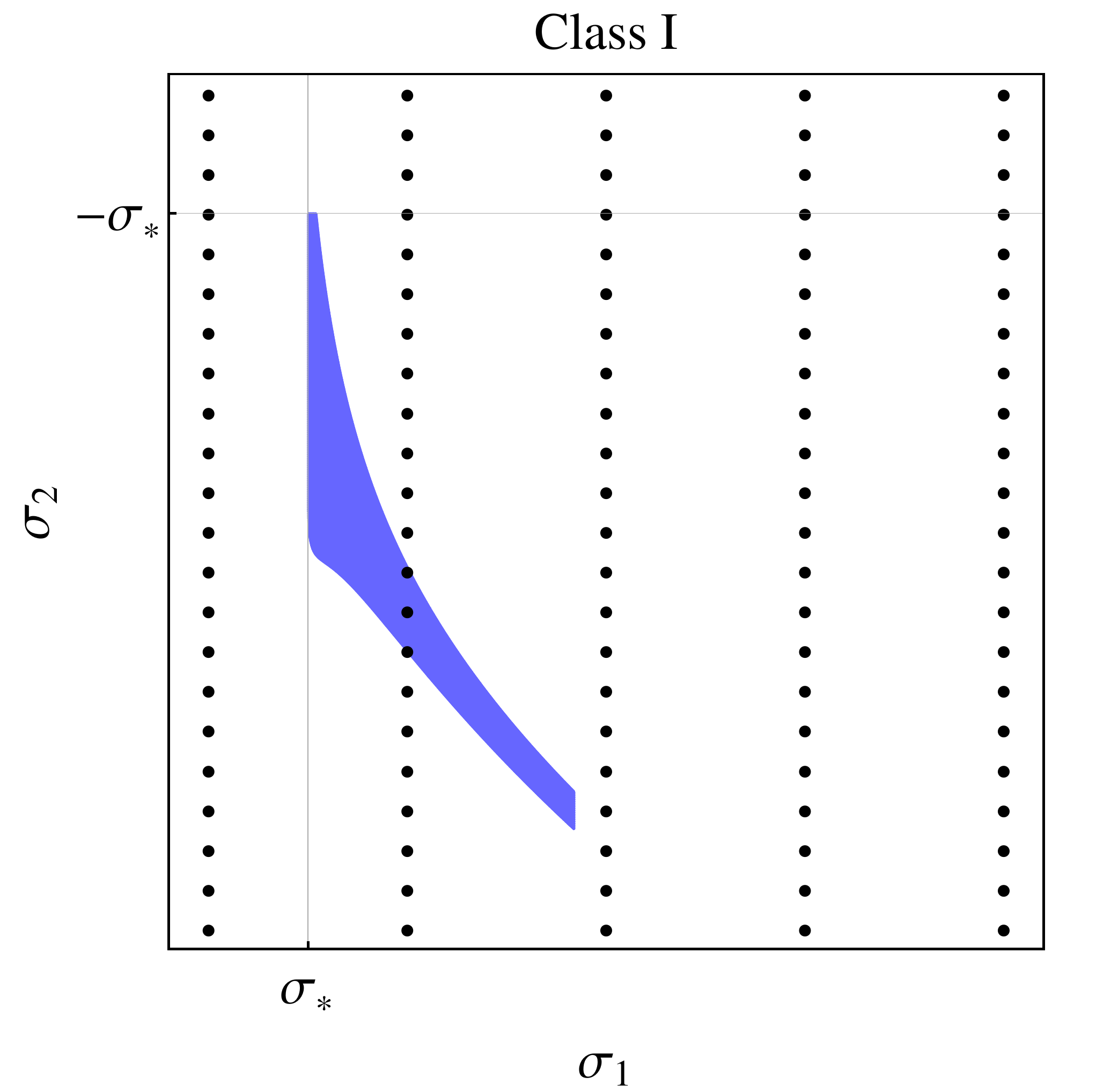}
\caption{Illustration of the scanning of the cosmological constant in the absence (Class II, left) and presence (Class I, right) of a radius stabilization mechanism. Dots correspond to vacua of the brane tensions $\sigma_1$ and $\sigma_2$ on the horizontal and vertical axes, with large dots indicating vacua with a near-critical Higgs mass. The shaded bands represent regions where the four-dimensional cosmological constant is smaller than some value $\Lambda_4$. The shape of the red region is independent of $\sigma_2$ in the unstabilized case, with a width in the $\sigma_1$ direction given by eq.~\ref{eq:singlebranetuning}.
The overall shape of the blue region for a stabilized, warped bulk is given by eq.~\ref{eq:bound_w} (the analogous shape for a flat bulk is described by eq.~\ref{eq:bound_m}). The fattening in the $\sigma_2$ direction as $\sigma_1$ approaches $\sigma_*$ is due to the warping effect of eq.~\ref{eq:sigma2w}, while the finite range in $\sigma_1$ arises because of the need for a large extra dimension to get repulsive Casimir stress (eqs.~\ref{eq:L5}~and~\ref{eq:casimirbeta}). 
}\label{fig:tuning}
\end{figure}

Eq.~\ref{eq:singlebranetuning} quantifies the 4D cosmological constant in a situation when the stabilization mechanism is absent or not effective, such as at very large distances in (our) case of Casimir stress---Point 1 in figure~\ref{fig:summary}. The tuning characteristics of $\Lambda_4$ in terms of the brane tension $\sigma_1$ are illustrated on the left panel of figure~\ref{fig:tuning}. 

\subsection{Casimir energy stabilization}
\label{sec:Casimirstabilization}

We extend the analysis of the previous section by adding Casimir stress-energy in the bulk. For our purposes, it will be sufficient to study the conformal limits with Casimir stress as in eq.~\ref{eq:casimirconformal}, and purely odd BC for $\Psi$ with negative $\beta = - [45\zeta(5)/32]/32\pi^2$, or purely even BC with positive $\beta = + [3 \zeta(5)/2]/32\pi^2$, calculated in eq.~\ref{eq:casimirbeta}. Deviations from conformal symmetry give subleading corrections. 
Including the Casimir stress from eq.~\ref{eq:casimirconformal}, which transforms as a tensor under diffeomorphisms and is thus the same in the gauge of eq.~\ref{eq:ds2} with $a(z)\leftrightarrow a(y)$, the Einstein equations become
\begin{alignat}{8}
3\frac{a''}{a} && {}+{} 3\l\frac{a'}{a}\r^2  && {}-{} 3 \frac{\mathcal{H}^2}{a^2} && {}+{} \frac{\Lambda_5}{M_5^3} && {}+{}\frac{\beta}{M_5^3 L^5 a^5} &  = {}-{}\frac{\sigma_1}{M_5^3} \delta(y) - \frac{\sigma_2}{M_5^3}\delta(y-R)
\label{eq:FriedmannC1}  \\
 && {}+{} 6\l\frac{a'}{a}\r^2  && {}-{} 6\frac{\mathcal{H}^2}{a^2}  && {}+{}\frac{\Lambda_5}{M_5^3} && {}-{}\frac{4\beta}{M_5^3 L^5 a^5} & = 0 \label{eq:FriedmannC2}
\end{alignat}
leaving the jump conditions of eqs.~\ref{eq:Jump1}~and~\ref{eq:Jump2} unchanged.

Since the desired $\mathcal{H}$ is much smaller than all other scales in the problem, we  
first look for solutions with $\mathcal{H}=0$. Later we study perturbations with small nonzero $\mathcal{H}$, or equivalently $\Lambda_4$. A differential equation independent of $\beta$ can be formed from adding four times eq.~\ref{eq:FriedmannC1} to eq.~\ref{eq:FriedmannC2}, which admits the solution~\cite{Hofmann:2000cj}:
\be
a(y)=\left\lbrace\frac{\sinh\left[\frac{5 k}{2}(y_0-y)\right]}{\sinh\left[\frac{5k}{2} y_0\right]}\right\rbrace^{2/5},
\label{eq:sinh}
\ee
where $k$ is the AdS curvature in absence of Casimir stress as in eq.~\ref{eq:kcurv}, and $y_0$ is an integration constant.\footnote{The other $\cosh$-type solution~\cite{Hofmann:2000cj} will be implicitly covered by a perturbative analysis later this section.} The second Einstein equation, eq.~\ref{eq:FriedmannC2}, imposes the constraint
\be
\Lambda_5 L^5=-4\beta\sinh^2\l\frac{5 k}{2} y_0\r \label{eq:lambda1L}
\ee
while the jump condition on the first brane in eq.~\ref{eq:Jump1} implies 
\be
k \coth\l\frac{5k}{2}y_0\r=\frac{\sigma_1}{6 M_5^3}. \label{eq:jumpcoth}
\ee
After combining eqs.~\ref{eq:lambda1L}~and~\ref{eq:jumpcoth}, we arrive at the important result
\be
L^5=\frac{4\beta}{\lambda_1},
\label{eq:L5}
\ee
relating the interbrane distance $L$ in conformal coordinates to $\lambda_1$. Eqs.~\ref{eq:L5}~and~\ref{eq:casimirbeta} together entail that a radion extremum with $\beta > 0$ can only exist for small $\lambda_1 \lesssim \beta \mu^5$, or equivalently for SM brane tensions $\sigma_1$ tuned to a precision of $\beta \mu^5/ k$ close to $\sigma_*$. This is in accordance with the effective field theory consideration that the Higgs vev $v$ cannot significantly affect physical effects at higher energy scales.

Having found the effective size of the extra dimension in terms of the SM brane tension $\sigma_1$ and the bulk CC $\Lambda_5$, the second brane tension $\sigma_2$ is fixed by the second jump condition, which can be thought of as the constraint that $\mathcal{H} = 0$. It will be instructive to investigate the implications of this condition and deformations to nonzero $\mathcal{H}$ perturbatively, in the following two regimes:
\begin{align}
\text{flat bulk: }\begin{cases}
\left|\lambda_1\right| \gg \left|k^5\right| \\
\left|\mathcal{H}\right| \ll \frac{1}{L}
\end{cases};
\qquad \qquad \qquad 
\text{AdS bulk: }
\begin{cases}
\left|\lambda_1\right| \ll \left|\Lambda_5\right|\\
\left|\mathcal{H}\right| \ll \frac{1}{L}
\end{cases}.
\end{align}
We insist on perturbative gravity in the bulk with $|k|^5 \ll |\Lambda_5|$, so that the calculation is under control and no solutions will fall through the cracks; we will show that our results for both regimes agree in the overlapping region. Given eq.~\ref{eq:L5}, $|\lambda_1| \gg |k|^5$ implies a flat bulk with negligible curvature $|k L| \ll 1$, so we will expand around a 5D Minkowski background. In the AdS regime, we will perform an expansion in $\lambda_1 / \Lambda_5$ around the zeroth-order AdS background. We note again that warping is \emph{not} responsible for the workings of our mechanism; to avoid any confusion, we take the SM brane to be the UV brane with \emph{positive} tension $\sigma_1 > 0$, when increased warping in fact reduces the power of our solution to the hierarchy problem.

\subsubsection*{Flat bulk}
In the approximately flat regime, we can Taylor expand the scale factor around $a(0)=1$:
\be
a(y)=1+a_1y+a_2\frac{y^2}{2}+\ldots
\label{eq:a_mink}
\ee
The first jump condition (eq.~\ref{eq:Jump1}) and the bulk Einstein eqs.~\ref{eq:FriedmannC1}~and~\ref{eq:FriedmannC2} respectively determine $a_1$ and $a_2$:
\begin{align}
a_1 & = -\frac{\sigma_1}{6 M_5^3}, \label{eq:a_sol1} \\
a_2 & = -\frac{1}{24}\l\frac{\sigma_1}{M_5^3}\r^2-\frac{5}{12}\frac{\Lambda_5}{M_5^3}. \label{eq:a_sol2}
\end{align}
The second jump condition of eq.~\ref{eq:Jump2} fixes the second tension $\sigma_2$ in terms of $\sigma_1$ to   be, to leading order in $k L \simeq k R$,
\be
\sigma_2^{\mathcal{H}=0}(\sigma_1) = - \sigma_1 - \frac{5}{2^{3/5}} \beta^{1/5}\lambda_1^{4/5},  \label{eq:sigma_mink}
\ee
and is the explicit combination of brane tensions that delivers a static solution with vanishing four-dimensional cosmological constant ($\mathcal{H} = 0$). This extremum exists for both positive and negative $\beta$, though eq.~\ref{eq:L5} requires the positivity of the product $\beta \lambda_1$.

A small $\Lambda_4$ is obtained only when the brane tensions lie within a small neighborhood of eq.~\ref{eq:sigma_mink}. Keeping $\sigma_1$, and therefore $\lambda_1$, fixed and reintroducing a small Hubble constant in the above steps, we find that for small Hubble constant, the solution in eqs.~\ref{eq:a_sol1} and \ref{eq:a_sol2} becomes
\begin{align}
a_1 & = -\frac{\sigma_1}{6 M_5^3}, \\
a_2 & = -\frac{1}{24}\l\frac{\sigma_1}{M_5^3}\r^2-\frac{5}{12}\frac{\Lambda_5}{M_5^3} + \frac{3}{2}\mathcal{H}^2,
\end{align}
and the relation of eq.~\ref{eq:L5} becomes
\begin{equation}
\frac{4\beta}{L^5} = \lambda_1 - 6 \mathcal{H}^2 M_5^3, \label{eq:L5Hubble}
\end{equation}
which changes the expression of eq.~\ref{eq:sigma_mink} to:
\be
\sigma_2 = \sigma_2^{\mathcal{H}=0}(\sigma_1) +6 \mathcal{H}^2 M_5^3 L= \sigma_2^{\mathcal{H}=0}(\sigma_1) + \Lambda_4. \label{eq:deltas2} 
\ee  
As expected, changing the brane tension by a small amount from the solution in the exactly static case looks just like changing the cosmological constant in the effective four-dimensional theory. 

We now turn to the stability of the flat extremum specified by eqs.~\ref{eq:a_sol1},~\ref{eq:a_sol2},~\ref{eq:deltas2},~\ref{eq:L5}. The spectrum of gravitational KK modes is tower of massive spin-2 particles, each with five degrees of freedom, except for the lowest level, which contains a massless graviton and thus a separate scalar---the radion---with a mass-squared that can potentially be negative. The spin-one component of the metric is projected away by parity. 
%In appendix~\ref{sec:radionmass}, we solve Einstein's equations  for linearized scalar perturbations of size $\epsilon$ around the background of eqs.~\ref{eq:ds2}~and~\ref{eq:a_mink}

In general, a scalar perturbation of the metric can be parametrized as:
\be
ds^2=a(y)^2 \left[1 +\epsilon f(y,x^\mu) \right] dx^2+ \left[1+\epsilon g(y,x^\mu)\right] dy^2.
\label{eq:ds2b}
\ee
%with profiles $f(x^\mu,y)$ and $g(x^\mu,y)$ subject to modified jump conditions. (Note that diffeomorphism invariance in the fifth coordinate $y\to y'(y)$ implies a functional ambiguity in the profiles $f$ and $g$. There is, however, generically not enough gauge freedom to make the radion profile constant in the fifth dimension.)
Diffeomorphism invariance in the fifth coordinate $y\to y'(y)$ implies a functional ambiguity in the profiles $f$ and $g$. However, there is not enough gauge freedom to set both $f$ and $g$ to be constant. A nontrivial bulk profile of the radion is expected because all KK modes have to satisfy boundary conditions resulting from a modification of eqs.~\ref{eq:Jump1}~and~\ref{eq:Jump2}. In appendix~\ref{sec:radionmass}, we provide the details of the radion mass calculation. For an approximately flat bulk, the result reads:
\be
m_r^2 = \frac{40\beta}{3 L^5 M_5^3} - 4 \mathcal{H}^2. \label{eq:mrflat}
\ee
Importantly, the sign of the radion mass coincides with that of the Casimir energy density for $\mathcal{H} = 0$: \emph{simultaneously stable and static solutions only exist for $\beta >0$}, a conclusion which will hold true for a warped bulk as well. 

\emph{Nonstatic}, stable minima can exist for negative $\beta$ only beyond the regime of validity of our perturbative expansion in small $\mathcal{H}$, which is transgressed when the Hubble curvature exceeds the size of the Casimir-induced curvature in eqs.~\ref{eq:FriedmannC1}~and~\ref{eq:FriedmannC2}. Taking into account relation~\ref{eq:L5Hubble}, this yields a lower bound on the magnitude of Hubble scale (4D AdS curvature) in this type of radion extremum (cfr.~Point 3 in figure~\ref{fig:summary}):
\be 
\left|\mathcal{H}^2 \right| \gtrsim  \left|\frac{\lambda_1}{M_5^3}\right|, \label{eq:boundH}
\ee
which is parametrically similar to that of eq.~\ref{eq:Hubble}. For negative Casimir energy, the tunability of $\Lambda_4$ is limited by the density of $\sigma_1$ vacua, and does \emph{not} depend on the vacuum structure for $\sigma_2$. Hence the left panel of figure~\ref{fig:tuning} correctly represents this situation as well. Using the definition for the Hubble constant in eq.~\ref{eq:Hubble} and the expression for the 4D Planck mass $M_\text{Pl}^2=2M_5^3L$, eq.~\ref{eq:boundH} can be translated to a bound on the SM brane tension
\be
|\Lambda_4| \gtrsim \left|k (\sigma_1 - \sigma_*) \right|^{4/5}, \label{eq:bound_m}
\ee
a restriction more severe than eq.~\ref{eq:singlebranetuning} for small detuning $|\sigma_1 - \sigma_*| < |k|^4$.

The right panel of figure~\ref{fig:tuning} illustrates the tuning condition for $\beta > 0$. The region in the $\sigma_1$--$\sigma_2$ plane with a 4D cosmological constant less than $\Lambda_4$ is defined by:
\be 
\left|\sigma_2- \sigma_2^{\mathcal{H}=0}(\sigma_1)\right| \leq |\Lambda_4|.  \label{eq:shape_m}
\ee
To leading order, $\sigma_2$ needs to match $\sigma_2^{\mathcal{H}=0}(\sigma_1)$, the value in eq.~\ref{eq:sigma_mink} that gives rise to a static solution, to a precision of $|\Lambda_4|$. However, there can be many values of $\sigma_1$ such that $L \gtrsim \mu^{-1}$ via eq.~\ref{eq:L5}, so to leading order in $kL$, it is the sum $|\sigma_1 + \sigma_2|$ that needs to be tuned to $|\Lambda_4|$ precision.

The calculation above was independent of the sign of the bulk cosmological constant $\Lambda_5$. In what follows, however, we will restrict uniquely to $\Lambda_5<0$. It is easy to see from eq.~\ref{eq:FriedmannC2} that no solutions exist for $\Lambda_5>0$, $\mathcal{H}=0$ and $\beta<0$. Bounds on the minimal $|\Lambda_4|$ for which nonstatic solutions with $\beta < 0$ may exist will be analogous to the AdS case discussed below. If, on the other hand, corresponding static solutions with $\beta>0$ do exist, they can only \emph{increase} the number of Class I vacua, leaving all of our subsequent arguments unchanged.

\subsubsection*{AdS bulk}

The regime of parameters where the small-$\lambda_1/\Lambda_5$ expansion is valid includes highly warped geometries, and in the range $|\Lambda_5| \gg |\lambda_1| \gg k^5$ it also overlaps with the flat-space expansion discussed above. To keep this manifest, we will keep the warping factor $\gamma$
\be
\gamma\equiv e^{k R } \simeq \frac{a(0)}{a(R)}
\ee
arbitrary whenever necessary, and explicitly mention where we utilize large-$\gamma$ simplifications.  %though some formulas simplify significantly in the large $\gamma$ limit. To keep track of the power counting it useful to keep in mind (\ref{eq:L5}) which says that conformal size of the extra dimension $L$ is always of the same order as $\lambda_1^{1/5}$ independently of any approximations done in this section.
% In particular, in the section \ref{sec:radionmass}, we will show that in the $\gamma \to 1$ limit our $\lambda_1/\Lambda_5$ expansion reproduces the flat-space result for the radion mass (see equation \ref{mrflat} and equation \ref{mrcurved}).
First of all, we demonstrate that the regime of small $\lambda_1/\Lambda_5$ indeed corresponds to small corrections to AdS space in the bulk. As long as $\gamma-1 \gtrsim 1$, the conformal distance between the two branes is dominated by the infrared, large-$y$ region of the extra dimension:
\begin{align}
L=\frac{\gamma-1}{k}\sim\frac{\gamma}{k}.
\end{align}
Together with eqs.~\ref{eq:L5}~and~\ref{eq:casimirconformal}, this means that even near the infrared brane, the size of the 5D Casimir stress-energy is only of order $k^5$, less than $\Lambda_5$ as long as gravity itself is perturbative in the bulk. Instead of using the exact but complicated form of eq.~\ref{eq:sinh}, we will expand the scale factor into exponential terms:
\be
a(y)=\frac{1}{1+A_4}\l e^{-k y}+A_4 e^{4 k y}\r+\ldots
\label{eq:a_w}
\ee
which to zeroth order in $\lambda_1/\Lambda_5$ reduces to the AdS solution $a(y) = e^{-ky}$.

We proceed in a way very similar to the approximately flat bulk case discussed above and first restrict ourselves to the $\mathcal{H}=0$ case.
Substituting the ansatz of eq.~\ref{eq:a_w} in the first jump condition of eq.~\ref{eq:Jump1} requires
\be
A_4 = \frac{2\beta}{5 L^5 \Lambda_5} =\frac{\lambda_1}{10 \Lambda_5},
\label{eq:A4}
\ee
whereafter the other jump condition of eq.~\ref{eq:Jump2} implies 
\be
\lambda_2^{\mathcal{H}=0}(\lambda_1) = 10 \gamma^5 \Lambda_5 A_4 = \gamma^5 \lambda_1, \label{eq:lambda2}
\ee
where we defined $\lambda_2 \equiv \Lambda_5+{\sigma_2^2}/{6 M_5^3}$ analogously to $\lambda_1$ in eq.~\ref{eq:Hubble}, with $\lambda_2^{\mathcal{H}=0}(\lambda_1)$ signifying the value corresponding to a static solution. One can check that the above solution indeed satisfies the Einstein eqs.~\ref{eq:FriedmannC1}~and~\ref{eq:FriedmannC2} in the bulk to leading order.
At high warping, eq.~\ref{eq:lambda2} can be massaged to derive the relation between the tensions
\be
\lambda_2^{\mathcal{H}=0}(\lambda_1) = 4\beta k^5 + k^4\lambda_1^{1/5} (4 \beta)^{4/5}, \qquad (\gamma \gg 1)
\label{eq:lambda_w}
\ee
that ensures a vanishing $\Lambda_4$. Eq.~\ref{eq:lambda_w} is the analog of eq.~\ref{eq:sigma_mink} for a flat bulk, and quantifies how the second brane tension is involved in obtaining a static configuration with $\mathcal{H} = 0$.

As for the flat case, we deform the solution by reintroducing a small Hubble constant $\mathcal{H}$. Einstein's equations are now solved by a functionally modified warp factor
\be
a(y)=\frac{e^{-k y}+A_1e^{ky}+A_4 e^{4 ky}}{1+A_1+A_4} + \dots,
\ee
a fact that can be checked upon substitution in eqs.~\ref{eq:FriedmannC1}~and~\ref{eq:FriedmannC2}, which also fixes
\be
A_1 = - \frac{\mathcal{H}^2}{4 k^2},\label{eq:HubbleExp}
\ee
with $A_4$ modified from eq.~\ref{eq:A4} to 
\be
A_4 = \frac{2\beta}{5 L^5 \Lambda_5} =\frac{\lambda_1}{10 \Lambda_5} + \frac{\mathcal{H}^2}{10 k^2},
\ee

The jump condition of eq.~\ref{eq:Jump1} results in the same relation between $\lambda_1$ and $L$ as in the flat case (eq.~\ref{eq:L5Hubble}):
\be
\frac{4 \beta}{L^5} = \lambda_1 - 6 \mathcal{H}^2M_5^3.\label{eq:L5Hubble2}
\ee
For a fixed $\lambda_1$ ($\sigma_1$), we can compute how much $\lambda_2$ ($\sigma_2$) deviates from the value in eq.~\ref{eq:lambda_w} as a function of $\mathcal{H}$:
\begin{align}
\lambda_2 &= \gamma^5 \left(\lambda_1 - 6 \mathcal{H}^2 M_5^3 \right) + 6 \gamma^2 \mathcal{H}^2 M_5^3 = \gamma^5 \left(\frac{4\beta}{L^5}\right) + 6 \gamma^2 \mathcal{H}^2 M_5^3 \nonumber\\
&= \lambda_2^{\mathcal{H}=0}(\lambda_1) - 6 \gamma^4 \mathcal{H}^2 M_5^3, \qquad (\gamma \gg 1), 
\end{align}
where in the second line we used a large-warping approximation ($\gamma \gg 1$). Note that in the large-$\gamma$ limit, the leading term in $\gamma^5 /L^5$ does not depend on $\mathcal{H}$. The leading Hubble dependence is thus only enhanced by $\gamma^4$ for large $\gamma$. This result can be translated in terms of brane tensions:
\be
\sigma_2 = \sigma_2^{\mathcal{H}=0}(\sigma_1) + \gamma^4\Lambda_4. \qquad (\gamma \gg 1) \label{eq:sigma2w}
\ee
The factor of $\gamma^4$ can be understood as the usual change of IR brane energy scales in the effective theory due to warping. 

We present a stability analysis of the warped solutions with a small $\Lambda_4$ parallel to those in a flat extra dimension in appendix~\ref{sec:radionmass}. The radion mass for arbitrary warping factor turns out to be
\be
m_r^2 = \frac{20}{3}\frac{\beta}{L^5 M_5^3}\l\gamma^2+\gamma\r - 4 \mathcal{H}^2 \label{eq:mrcurved}
\ee
up to corrections suppressed by $\lambda_1/\Lambda_5$. The radion mass-squared is again of the same sign as $\beta$ for $\mathcal{H}=0$, so this concludes our proof that positive Casimir energy density is necessary and sufficient for radius stabilization with zero or small $\Lambda_4$. In the limit $\gamma \to 1$, we indeed reproduce the expression in eq.~\ref{eq:mrflat} for $m_r^2$ in a flat geometry. We observe that in the highly warped regime $(\gamma \gg 1)$, the radion mass-squared is enhanced by the warping factor $\gamma$ for a fixed Planck mass $M_\text{Pl}^2 \simeq M_5^3 L /\gamma$. Moreover, in this regime, the radion has an extra-dimensional profile peaked near the IR brane, reducing its couplings to SM states relative to those of the graviton. These two observations are important for radion phenomenology, discussed in section~\ref{sec:radion}.

One may worry that we ignored possible higher-derivative corrections to the Einstein-Hilbert action both in the bulk and on the branes in the preceding calculations. Indeed, even if these corrections are suppressed by powers of $M_5$, naive estimates for the corresponding terms in the action can be parametrically larger than the Casimir stress-energy. However, as we prove in appendix~\ref{sec:lightradion}, none of those terms can qualitatively affect our solution. Their only effect amounts to redefinition of the relation in eq.~\ref{eq:sigma*}. In particular, the radion mass and couplings are quantitatively insensitive to the addition of these operators. 

As before, we address the issue of nonstatic vacua like Point 3 in figure~\ref{fig:summary} that are not small deformations of eq.~\ref{eq:a_w}, where corrections due to Hubble curvature become at least comparable to those from Casimir stress in the Einstein equations. Because eqs.~\ref{eq:L5Hubble}~and~\ref{eq:L5Hubble2} are the same in flat and warped space, eq.~\ref{eq:boundH} persists in the warped case as well. However, since the Planck mass is now $M_\text{Pl}^2 = M_5^3 / k$, in warped radion vacua the minimum cosmological constant is
\be
|\Lambda_4| \geq |\sigma_1-\sigma_*|, \label{eq:bound_w}
\ee
parametrically similar to the case without radius stabilization, eq.~\ref{eq:singlebranetuning}, and thus again captured by the schematic in the left panel of figure~\ref{fig:tuning}.

Besides some quantitative differences, particularly in the size of the radion mass, our conclusions for a warped bulk are qualitatively similar to those for the flat case: only for repulse Casimir stress can both brane tensions contribute to a scanning of the cosmological constant in near-static radion minima. Brane tension deviations by at most $\Delta\sigma_1=\Lambda_4$ or $\Delta\sigma_2=\gamma^4\Lambda_4$ (in the large warping limit) in the neighborhood of eq.~\ref{eq:lambda_w} give rise to a cosmological constant smaller than or equal to $\Lambda_4$, as illustrated in figure~\ref{fig:tuning}.

\section{Vacuum structure}
\label{sec:vacua}

To solve the naturalness problems of the cosmological constant and the Higgs mass, our model requires a large number of vacua; in this section, we explore their statistics. 

%Following our results in Section~\cite{sec:radiusstabilization}, the vacua in our theory are naturally separated into two classes: \textbf{(I)} those with Higgs vevs $v\sim v_*$, where the extra dimension between two 3-branes can be stabilized in the infrared\VG{I don't like the word "infrared" here, let's just say "stabilized" and discuss branes close to each other elsewhere}; and \textbf{(II)} those with Higgs vevs $v \neq v_*$, without a radion stabilization in the infrared.

In Class I vacua, with $v \sim v_* \equiv \sqrt{{M_0 M_1}/{Y Y^c}}$, the radion potential develops minima like Point 0 in figure~\ref{fig:summary}, where the effective 4D cosmological constant depends on \emph{both} brane tensions, $\sigma_1$ and $\sigma_2$, and has scanning density proportional to the \emph{product} of the number of possible tensions on each brane as shown in figure~\ref{fig:tuning}. 
%In Class II vacua, the Higgs vev is not near the critical value ($v \neq v_*$), changing the sign of the Casimir stress relative to that of Class I vacua, as discussed in Section \ref{sec:5dmodels}. Our analysis in Section \ref{sec:radiusstabilization} showed that this sign of the Casimir stress does not allow for a stabilization of the fifth dimension in the infrared within the constraints of our model. Two possibilities remain: \textbf{(Point 1)} the non-SM brane flies away to infinity, or \textbf{(Point 3)} the distance between the two branes gets stabilized at a minimum with a large 4D cosmological constant that cannot be treated as perturbation in the stabilization calculation.
%When the radius gets destabilized to the infrared $L\rightarrow \infty$ (Point 1), 
In Class II vacua, where the Higgs vev is not near the critical value ($v \neq v_*$), the fifth dimension cannot be stabilized in a radion minimum without significant Hubble expansion. The unstable Point~2 can be discounted as a habitable vacuum if any configuration near the maximum of the radion potential is always short lived compared to the Hubble time at which structures would have formed.
In Points~1~and~3, the minimum 4D cosmological constant is set by the scanning density of vacua on the SM brane; non-SM brane physics cannot help tune the effective 4D cosmological constant. 
This can be seen from eq.~\ref{eq:singlebranetuning} for $\Lambda_4$ in Point 1 vacua, and the lower bounds on $|\Lambda_4|$ in eqs.~\ref{eq:bound_m}~and~\ref{eq:bound_w} for Point 3 vacua. These expressions can be combined for a lower bound on the cosmological constant in either situation:
\begin{equation}
|\Lambda_4| \gtrsim \min_{\lbrace\sigma_1\rbrace}|\sigma_1 -\sigma_*|, \label{eq:boundII}
\end{equation}
as long as the fifth dimension has a moderately high curvature scale $k$. The smallest expected value for the quantity on the RHS is determined \emph{solely by the number of vacua on the SM brane}. The main goal of this section is to show that there exist landscapes in which galaxies can form in Class I vacua but not in Class II vacua. For a number of $\sigma_1$ vacua not too large, eq.~\ref{eq:boundII} is simply too stringent to form galactic structures, even after taking into account changes to the micro-physics in different Higgs vacua in our density perturbation analysis.

Point 4 appertains to the possibility that stable 5D vacua may exist near the cutoff. After all, new UV physics may stabilize the radius at distances $1/M_\text{UV}$, beyond the realm of our effective field theory. Supposing that the statistics of vacua on both branes were unchanged in the UV, there would always be enough Point 4 vacua if there are a sufficient number in Class I, spoiling our correlation of the resolution of the CC problem to the Higgs hierarchy problem. However, we will show that the statistics of vacua can change dramatically (i.e.~reduce in number) much above $v_*$ but well before scales of order the cutoff $M_\text{UV}$. In appendix~\ref{sec:branetop}, we present a proof-of-principle module in which the number of possible brane tensions strongly diminishes as the branes approach each other. %Unless the ultraviolet completion of our theory is extremely rich near the 5D Planck scale---antithetical to the logic of growing complexity of the vacuum in the infrared~\cite{Bousso:2000xa}---then again the possible number of values of the effective 4D cosmological constant pale in comparison to the equivalent number in Class I.

\subsection{Vacua with near-critical Higgs masses}\label{sec:classIcount}

To guarantee the existence of a vacuum with a cosmological constant $\Lambda_4$ near the observed value $\Lambda_0 \approx (2~\text{meV})^4$, our theory needs to exhibit large numbers of vacua for the unprotected parameters---$m_H^2$, $\sigma_1$, and $\sigma_2$. Our mechanism requires that each of them individually take on a large number of possible values, denoted by $\mathcal{N}_{m_H^2}$, $\mathcal{N}_{\sigma_1}$, and $\mathcal{N}_{\sigma_2}$, respectively. We will assume these vacua are randomly but uniformly distributed as in figure~\ref{fig:tuning}, a natural choice given the additive renormalization of the corresponding quantities. Since different values for the Higgs mass will give rise to different contributions to the SM brane tension $\sigma_1$ at the quantum level (and also the classical level for $m_H^2 < 0$), by construction $\mathcal{N}_{\sigma_1} \geq \mathcal{N}_{m_H^2}$. We allow for the possibility of $\mathcal{N}_{\sigma_1} > \mathcal{N}_{m_H^2}$; a dark sector with $\mathcal{N}_{1'}$ vacua decoupled from the SM would generically give rise to a total number of tension vacua $\mathcal{N}_{\sigma_1} = \mathcal{N}_{m_H^2} \times \mathcal{N}_{1'}$. For simplicity, we will assume there is no orthogonal sector scanning the bulk cosmological constant $\Lambda_5$ (i.e.~$\mathcal{N}_{\Lambda_5} = 1$), and thus that $k$ is fixed throughout the landscape. This assumption can be relaxed to $\mathcal{N}_{\Lambda_5} > 1$, in which case the maximum UV cutoff for our mechanism would be lower. The bulk curvature can be naturally small if our model is UV completed into a supersymmetric theory with supersymmetry badly broken on the branes but communicated into the bulk only gravitationally~\cite{Randall:1998uk}. 

Statistically, there will be at least one vacuum with a Higgs vev in a $\mathcal{O}(1)$ range around $v_*$ for a sufficient number of different Higgs mass-squared vacua:
\begin{equation}
\mathcal{N}_{m_H^2} \gtrsim \frac{M_{\rm UV}^2}{v_{*}^2}.\label{eq:Nummh}
\end{equation}
For a large hierarchy $M_1 \gg \mu$, the fractional range of Class I vacua would shrink further, since $v$ has to match $v_*$ even more closely. This would increase the lower bound on $\mathcal{N}_{m_H^2}$ by another factor of $M_1/\mu$, but not affect the rest of our discussion.

The majority of vacua with different and potentially small $\Lambda_4$ are generated when the extra dimension is stabilized. This is possible only when $\beta > 0$, which occurs only for a \emph{large extra dimension} $L > 1/\mu$ (see e.g.~eq.~\ref{eq:casimirbeta} and figure~\ref{fig:casimirpotential}). At the location of the Point 0 minimum, the Casimir pressure counteracts a combination of zero-point energies, captured by the relation $4 \beta/L^5 = \lambda_1$ (eq.~\ref{eq:L5}). Stabilization is thus attainable only when $|\sigma_1 -\sigma_*| \lesssim \mu^5/k \sim \mu^4 / \gamma$, demanding a minimal total number of tension vacua:
\begin{align}
\mathcal{N}_{\sigma_1} \gtrsim \underbrace{\frac{M_{\rm UV}^2}{v_{*}^2}}_{v \sim v_*}  ~\times~\underbrace{\frac{M_{\rm UV}^4}{\mu^4/\gamma}}_{L > 1/\mu}. \label{eq:Nums1}
\end{align}
Note that $\mathcal{N}_{\sigma_1}$ need not exceed $M_\text{UV}^4/\Lambda_0$, in which case there would be enough vacua on the SM brane alone to tune the CC to the observed value. 

The existence of a vacuum with a four-dimensional cosmological constant $\Lambda_4$ as large or smaller than the observed one $\Lambda_0$ is assured when the number of possible $\sigma_2$ values is sufficiently large. Following our results for a flat bulk ($\gamma \simeq 1$) in eq.~\ref{eq:deltas2} and a warped bulk ($\gamma \gtrsim 1$) in eq.~\ref{eq:sigma2w}, we observe that the warping down of the IR brane tension reduces the number of required $\sigma_2$ vacua:
\begin{align}
\mathcal{N}_{\sigma_2} \gtrsim \underbrace{\frac{M_{\rm UV}^4}{\gamma^4 \Lambda_0}}_{\Delta \sigma_2/\gamma^4 < \Lambda_0}.\label{eq:Nums2}
\end{align} 
The inequality of eq.~\ref{eq:Nums2} is sufficient but not necessary: the lower bound on $\mathcal{N}_{\sigma_2}$ is reduced if $\mathcal{N}_{\sigma_1}$ comfortably satisfies inequality~\ref{eq:Nums1}, although a larger $\mathcal{N}_{\sigma_1}$ would make a solution to the cosmological constant problem in Class II vacua more likely. 

The total number of vacua in our toy landscape is the product of $\mathcal{N}_{\sigma_1}$ and $\mathcal{N}_{\sigma_2}$, and needs to be at least as large as (from eqs.~\ref{eq:Nums1}~and~\ref{eq:Nums2}):
\begin{align}
\mathcal{N}_{\sigma_1} ~\times ~\mathcal{N}_{\sigma_2} ~\gtrsim  ~\underbrace{\frac{M_{\rm UV}^2}{v_{*}^2}}_{v \sim v_*} ~ \times ~ \underbrace{\frac{M_{\rm UV}^4}{\mu^4/\gamma}}_{L > 1/\mu} ~\times \, \underbrace{\frac{M_{\rm UV}^4}{\gamma^4 \Lambda_0}}_{\frac{\Delta \sigma_2}{\gamma^4} < \Lambda_0}~ = \, \underbrace{\frac{M_{\rm UV}^2}{v_{*}^2}}_{m_H^2 \sim -v_*^2} \times ~ \underbrace{\frac{M_{\rm UV}^4}{\Lambda_0}}_{\Lambda_4 < \Lambda_0} ~ \times ~ \underbrace{\frac{M_{\rm UV}^4}{\gamma^3\mu^4}}_{m_r^2 \approx 0}.\label{eq:totalvacs}
\end{align}
For the equality, we just rearranged the fractions to isolate the naive minimal number of vacua needed to explain the hierarchy problem ($m_H^2 \sim - v_*^2$) and the cosmological constant problem ($\Lambda_4 < \Lambda_0$). Our mechanism requires more. The minimal excess number of vacua can (partly) be traced back to the requirement of a large extra dimension $L > 1/\mu$, which at low energies manifests itself as a \emph{tuned radion mass} $m_r^2$. 
For a radion coupled to the SM with couplings suppressed by $1/\gamma M_\text{Pl}$ (see appendix~\ref{sec:radionmass}), radiative corrections would normally imply a mass-squared of at least $M_\text{UV}^4/\gamma^2 M_\text{Pl}^2$. In our model, the radion mass in Point 0 of a Class II vacuum is parametrically $m_r^2 \sim \gamma \mu^4 / M_\text{Pl}^2$ (eq.~\ref{eq:mrcurved}), much smaller than the naive estimate by a factor equal to the excess number of vacua $M_\text{UV}^4/\gamma^3 \mu^4$ in eq.~\ref{eq:totalvacs}. We discuss the interesting radion phenomenology in detail in section~\ref{sec:radion}.

\subsection{Vacua with generic Higgs masses}\label{sec:classIIcount}

For Higgs vevs not near the critical scale $v_*$ (Class II), the five-dimensional radius cannot be stabilized to a static minimum in which the rich vacuum structure of $\sigma_2$ can aid in tuning the cosmological constant. We confine  most of our analysis to runaway radion vacua like Point~1 and the inherently nonstatic radion extrema like Point~3 in figure~\ref{fig:summary}, for which exactly analagous conclusions can be drawn. We return to the unstable Point 2 vacua, which turn out to set only subleading constraints on our theory, at the end of our discussion.
In all radius configurations like Points 1 and 3, we derived in eqs.~\ref{eq:singlebranetuning},~\ref{eq:bound_m},~\ref{eq:bound_w} that the size of the 4D cosmological constant cannot be smaller than the detuning $|\sigma_1 - \sigma_*|$ of the SM brane tension. Parametrically, the smallest possible cosmological constant $\Lambda_4^{\text{II,min}}$ we can expect to find in stable Class II vacua is thus
\begin{align}
\Lambda_4^{\text{II,min}} \equiv \min_{\text{Class II}} \left\lbrace |\Lambda_4| \right\rbrace &\sim \frac{M_\text{UV}^4}{\mathcal{N}_{\sigma_1}}  \lesssim \frac{\mu^4}{\gamma} \frac{v_{*}^2}{M_{\rm UV}^2} \sim {\rm keV}^4 \left(\frac{\mu}{ {\rm GeV}}\right)^4 \left(\frac{10^4}{\gamma}\right) \left(\frac{10^{12}~{\rm GeV}}{M_{\rm UV}}\right)^2, \label{eq:lambda4minII}
\end{align}
where the inequality follows from the lower bound on $\mathcal{N}_{\sigma_1}$ in eq.~\ref{eq:Nums1} needed to ensure the existence of suitable Class I vacua. The parameters are chosen to satisfy current experimental constraints (see section \ref{sec:phenomenology}). The lowest expected cosmological constant in Class II minima in eq.~\ref{eq:lambda4minII} is easily in gross violation of the anthropic lower bound to form structure in our Universe~\cite{Weinberg:1987dv}, even for $M_\text{UV} \sim 10^{12}~\text{GeV}$ close to the five-dimensional Planck scale $M_5 = ( k /8 \pi G_N)^{1/3} \approx 4 \times 10^{13}~\text{GeV}$ for $k = 10^4~\text{GeV}$. A cosmological constant as high as $\Lambda_4 \sim \text{keV}^4$ would correspond to a Hubble horizon size not much larger than the solar system.

The process of structure formation is sensitive to micro-physics---including the Higgs mass---so to complete our proof that no galaxies can form in Class II vacua, we need to show that vacua with large positive or negative Higgs mass-squared still violate a modified version of Weinberg's lower bound on $|\Lambda_4|$. In other words, there should be choices of $\mathcal{N}_{\sigma_1}$ such that inequality~\ref{eq:Nums1} can be satisfied but the resulting $\Lambda_4^{\text{II,min}}$ of eq.~\ref{eq:lambda4minII} is too large to form galaxies. As outlined in the introduction, we will assume an inflationary sector with fixed reheating temperature $T_\text{reh}$ to the SM and an unspecified dark matter (DM) sector, and density contrast seeded by inflation ${\delta \rho}/{\rho}\big|_\text{I} \sim 10^{-5}$ throughout the landscape, as well as a constant baryon asymmetry $n_B/n_\gamma \sim 10^{-10}$ parametrized by the number density of baryonic matter $n_B$ over that of photons $n_\gamma$.

\begin{figure}
\centering
\includegraphics[width= 0.85\textwidth]{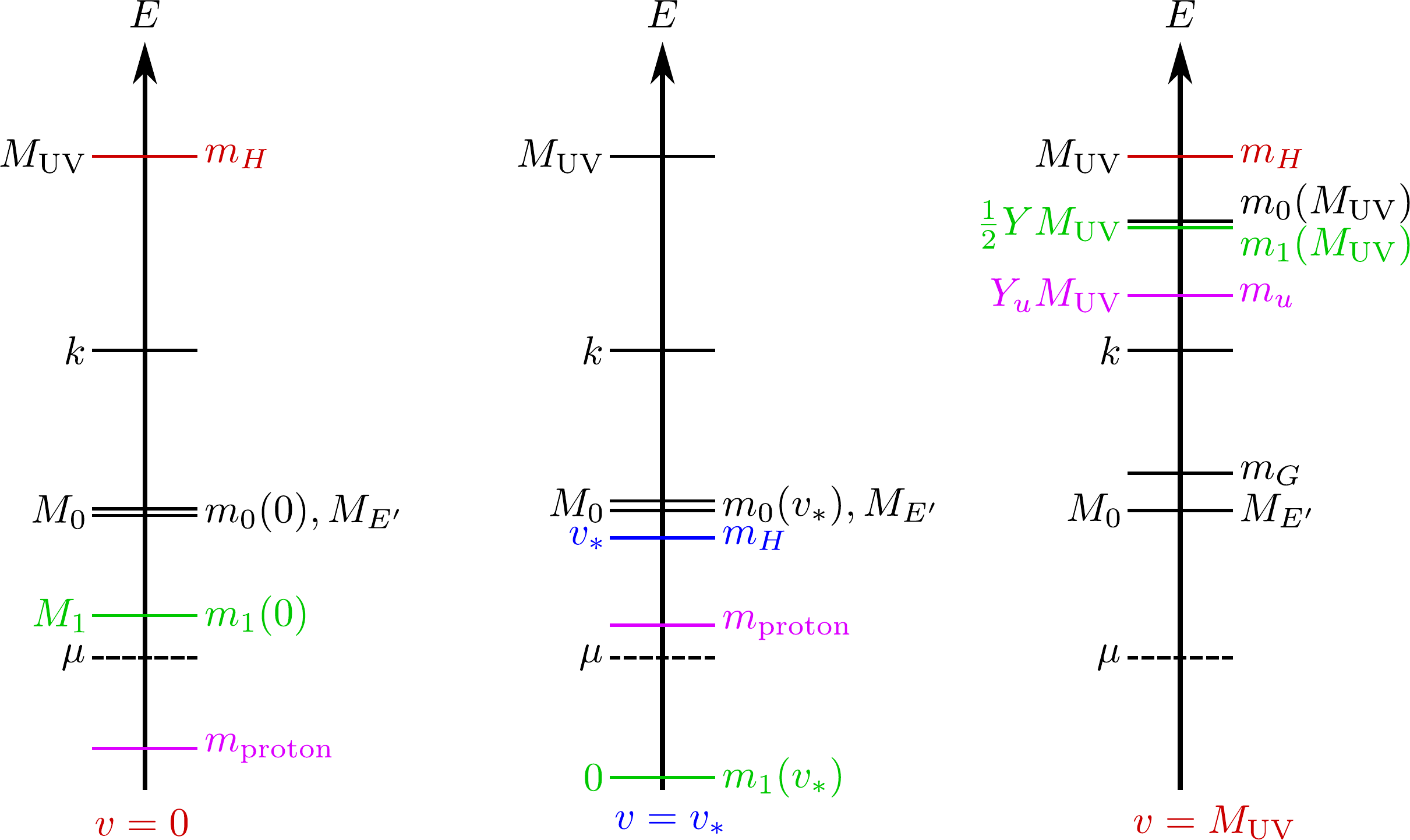}
\caption{Schematic mass spectrum of our theory in three different Higgs mass vacua: $m_H^2 \sim + M_\text{UV}^2$, $m_H^2 \sim -v_*^2$, and $m_H^2 \sim - M_\text{UV}^2$. On the right of each axis, we indicate the masses of the new neutral fermions $m_0(v)$ and $m_1(v)$, the new charged fermion $M_{E'}$, the proton $m_\text{proton}$, the up quark $m_u$, and the glueballs $m_G$. On the left of each axis, we indicate relevant mass scales of the Lagrangian for reference. The lightest, stable baryonic state is indicated in purple. We highlighted in green $m_1(v)$, the mass eigenvalue of the lightest brane fermion \emph{before} mixing with the bulk fermion; if $m_1(v) < \mu$, the lightest mass eigenvalue is of order $\mu$. }\label{fig:spectrum}
\end{figure}

In vacua with a very large positive Higgs mass $m_H^2 \sim + M_\text{UV}^2$, all SM particles would be much lighter, and receive masses only from QCD confinement, which moreover would take place at an even lower scale than in our Universe. We show a spectrum for $m_H^2 \sim + M_\text{UV}^2$ on the left part of figure~\ref{fig:spectrum} contrasted to the spectrum of our vacuum in the middle part; QCD would confine at $\Lambda_\text{QCD} \sim \text{MeV}$, and the electron would be as light as $m_e \sim 10~\text{eV}$. The linear growth of density perturbations requires that matter dominates the cosmological energy density, and that the photon is decoupled from the matter sector.
In a universe with $m_H^2 > 0$, baryons aid even less with structure formation than in our Universe, because the temperature at matter-radiation equality ($T_\text{eq}$) would be slightly lower, and at photon decoupling ($T_\text{dec} \sim \alpha^2 m_e$) would be much lower. In this case, the dark matter overdensities would be solely responsible for forming galaxies. The upper bound on $|\Lambda_4|$ is parametrically the same as in our Universe for structures to form in such a vacuum.
%In a universe with $m_H^2 > 0$, it would be much \emph{harder} to form structure, because both the temperature at matter-radiation equality ($T_\text{eq}$) and at photon decoupling ($T_\text{dec} \sim \alpha^2 m_e$) would be much \emph{lower}, leading to a much more stringent upper bound on $\Lambda_4$ if structures are to form.

\begin{figure}
\centering
\includegraphics[scale=0.5]{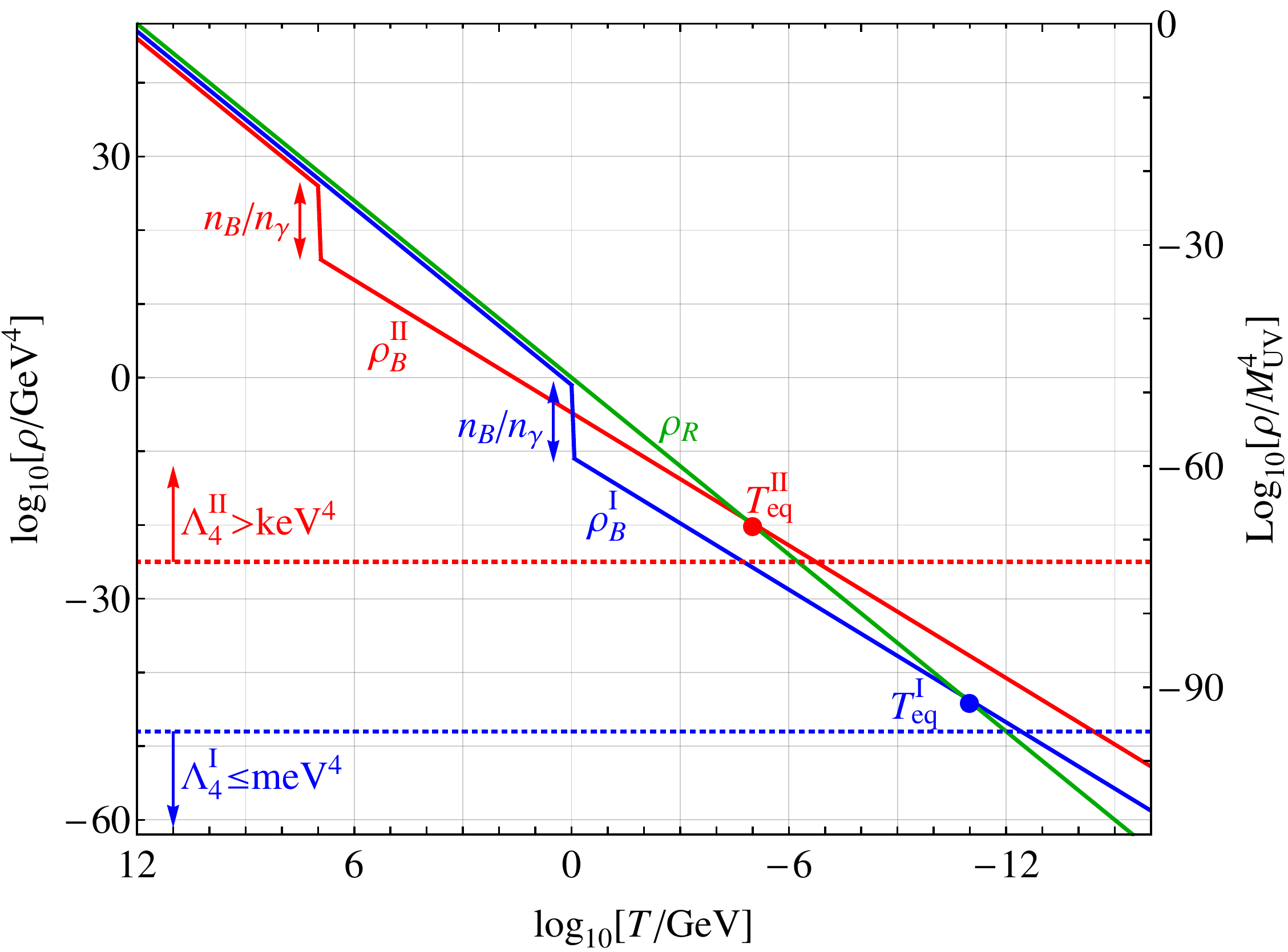}
\caption{The cosmological evolution of energy densities $\rho$ as a function of temperature $T$ with $M_{\rm UV} = 10^{12} ~{\rm GeV}$. The green line shows the radiation energy density decreasing as $T^4$. The blue and red solid lines show the energy densities of the lightest baryonic state $\rho_B^{\text{I,II}}$ in a typical Class~I ($v=v_*$) and Class~II ($v=M_{\rm UV}$) vacuum, respectively. The baryonic energy density drops by a factor of $n_B/n_\gamma$ when the temperature falls below the mass of the lightest baryon. At this point, the anti-baryons annihilate away with most of the baryons until only the asymmetric component remains, which is nonrelativistic and redshifts as $T^3$. The blue and red dots indicate matter-radiation equality and their corresponding temperatures $T_\text{eq}^\text{I,II}$ in both scenarios. The blue and red dotted lines show the critical cosmological constant $\Lambda_4$ for which galaxies can \emph{just} form, and is different in Class~I~and~II. For our mechanism to work, the smallest CC in Class~I vacua $\Lambda_4^\text{I}$ should be smaller than the critical value of $\text{meV}^4$, while $\Lambda_4^\text{II}$ should be larger than the critical value of $\text{keV}^4$ in Class~II.}\label{fig:cosmology}
\end{figure}

A large and negative Higgs mass $m_H^2 \sim - M_{\rm UV}^2$ would lift the lightest fermionic particles in the SM, the up quark and the electron, up to scales of $m_u \sim Y_u M_\text{UV}$ and  $m_e \sim Y_e M_\text{UV}$. The bottom of the spectrum would comprise of (potentially long-lived) glueballs near $\Lambda_\text{QCD}$, which itself would be lifted above its value in our Universe, although not as dramatically as the fermion masses. For sufficiently low reheating temperatures, the fermionic matter density would equal the photon energy density at a temperature of roughly
\begin{equation}
T_{\rm eq} \sim \frac{n_B}{n_{\gamma}} Y_u M_{\rm UV}  \sim  10^{-15} M_{\rm UV} \sim 10^{6}~\text{eV} \left( \frac{M_\text{UV}}{10^{12}~\text{GeV}} \right),\label{eq:TeqII}
\end{equation}
about a million times hotter than the matter-radiation equality temperature $T_\text{eq} \approx 2~{\rm eV}$ in our Universe, for the highest cutoffs $M_\text{UV}$.\footnote{Due to the shallowness of the radion potential, the radion can never dominate the energy density of the universe in Points 1 and 3 vacua. Besides, the radion mass-squared is smaller than $4 |\mathcal{H}^2|$ in a nonstatic vacuum for $\beta<0$. A scalar with sub-Hubble mass does not behave as dark matter and cannot form structure.} In figure~\ref{fig:cosmology}, we outline the cosmological evolution of energy densities in such a vacuum juxtaposed to the evolution in our Universe.  Baryonic density perturbations would start linear growth with the 4D scale factor $a$ much earlier, facilitating the formation of structures. However, these perturbations will \emph{not} have time to grow to galaxies with $\delta \rho / \rho \sim \mathcal{O}(1)$ before a cosmological constant domination phase---when they would be smoothed out again or the universe would crunch---as long as the cosmological constant is sufficiently large:
\begin{align}
\Lambda_4^{\text{II,min}} \gtrsim \rho_{B}\big|_{\delta\sim 1}\equiv \left(\frac{\delta \rho}{\rho}\bigg|_\text{eq} \right)^3 T_{\rm eq}^4, \label{eq:nogalaxies}
\end{align}
where $\left. \delta \rho / \rho \right|_\text{eq}$ is the density contrast at matter radiation equality. (Density perturbations grow logarithmically during radiation domination due to the Meszaros effect~\cite{Meszaros,Weinberg:2002kg}, so $\delta \rho / \rho$ is typically slightly larger at matter-radiation equality than the corresponding primordial density fluctuations at the end of inflation.)
The inequalities of eqs.~\ref{eq:lambda4minII},~\ref{eq:TeqII}~and~\ref{eq:nogalaxies} become mutually exclusive at a UV cutoff of
\begin{align}
M_\text{UV}^\text{max} &\sim \mu^{2/3} v_*^{1/3} \gamma^{-1/6} Y_u^{-2/3} \left({n_B}/{n_\gamma} \right)^{-2/3}\left(\delta \rho /\rho \big|_{\rm eq}\right)^{-1/2}\\
& \sim 10^{12}~{\rm GeV} \left(\frac{10^4}{\gamma}\right)^{1/6} \left(\frac{\mu}{1 \,{\rm GeV}}\right)^{2/3} \left(\frac{10^{-10}}{ n_B/n_{\gamma} }\right)^{2/3} \left(\frac{10^{-4}}{ \delta \rho /\rho \big|_{\rm eq} }\right)^{1/2}, \label{eq:maxmuv}
\end{align}
which can be comparable to the five-dimensional fundamental scale $M_5$. As long as $M_\text{UV} \lesssim M_\text{UV}^\text{max}$, there is always a choice of $\mathcal{N}_{\sigma_1}$ high enough such that a stabilized Class I vacuum can exist, but low enough such that no galaxies can form in stable Class II vacua.

A radion expectation value near Point 2 is unstable. The inverse lifetime~$\Gamma$ of such a configuration is of order the size of the radion mass, which is at least $\Gamma \sim |m_r| \gtrsim \gamma \left( \Lambda_4^{\text{II,min}} / M_\text{Pl}^2\right)^{1/2}$ via eqs.~\ref{eq:mrcurved}, \ref{eq:L5Hubble2}, and~\ref{eq:lambda4minII}. To preclude formation of structures in Point 2 extrema, we require that the lifetime of this type of radion configuration is shorter than the smallest Hubble time at which structures can form. This constraint can be seen to put a lower bound on $\Lambda_4^{\text{II,min}}$ which is weaker by a factor of $\gamma^2 \gg 1$ but otherwise parametrically similar compared to the bound in eq.~\ref{eq:nogalaxies}. We thus conclude that the leading theoretical constraints on the maximum cutoff in our model arise from considerations of the stable vacua (Points~1 and 3), not the unstable vacua of the type in Point 2. 

A large reheating temperature would slightly complicate (but not spoil) the story for negative Higgs mass squared above, as some of the new exotic states could be produced in the thermal bath. Firstly, the new brane fermions $L, L^c, N_1, N_1^c$ would eventually populate the lightest states in this sector, the charged fermions $E'$ and $E^{\prime c}$, which would annihilate rapidly to photons. Secondly, heavy QCD glueballs with mass $m_G \sim 7 \Lambda_\text{QCD}$~\cite{Craig:2015pha,Hook:2014cda} could potentially kick-start a matter-domination phase much before the temperature of eq.~\ref{eq:TeqII} is reached, at $T_\text{eq} \sim m_G$, \emph{if they were long lived}. However, their decay rate $\Gamma \sim m_G^9/m_u^8$ is always faster than the Hubble expansion rate for $M_\text{UV} \lesssim 10^{12}~\text{GeV}$, in which case they could only impede structure formation. Finally, a relic abundance of long-lived KK states of the bulk fermion $\Psi$ would be problematic, but for the UV cutoffs under consideration they are never in thermal contact for $T_\text{reh} < M_\text{UV}$, so they would not populate the universe if the reheating sector only heated up the Standard Model directly.

\section{Phenomenology}\label{sec:phenomenology}

\subsection{Electroweak-scale states}

Our model predicts the existence of two Dirac fermions, a doublet $L = (E^{\prime-}, N_0)$ and a singlet $N_1$. We assume for brevity that the brane fermions are approximately vectorlike with $Y = Y^c$, such that the Weyl spinors can be collected in gauge-eigenstate Dirac spinors $N_0 \equiv (N_0, N_0^{c\dagger})$, $\overline{N}_0 \equiv (N_0^\dagger, N_0^c)$, and similarly for $E^{\prime -}$, $E^{\prime +}$, $N_1$, and $\overline{N}_1$. We reserve tildes for the neutral mass eigenstates $\widetilde{N}_0$ and $\widetilde{N}_1$, the latter of which significantly mixes with the bulk fermion and has a mass of order $\mu$ (see eq.~\ref{eq:5spectrum}).

\begin{figure}[ht]
\centering
\includegraphics[width=0.79\textwidth]{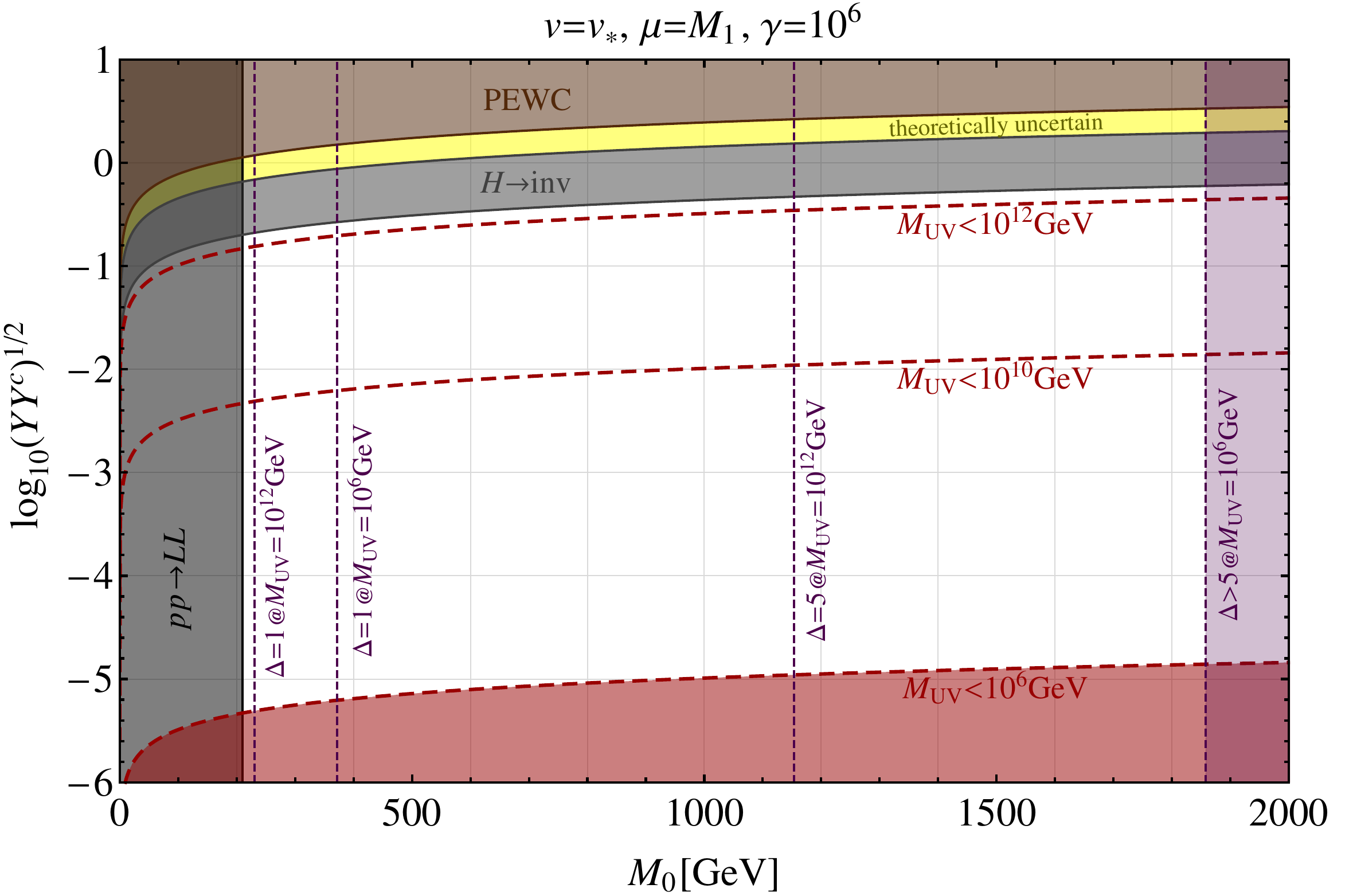}\\
\includegraphics[width=0.79\textwidth]{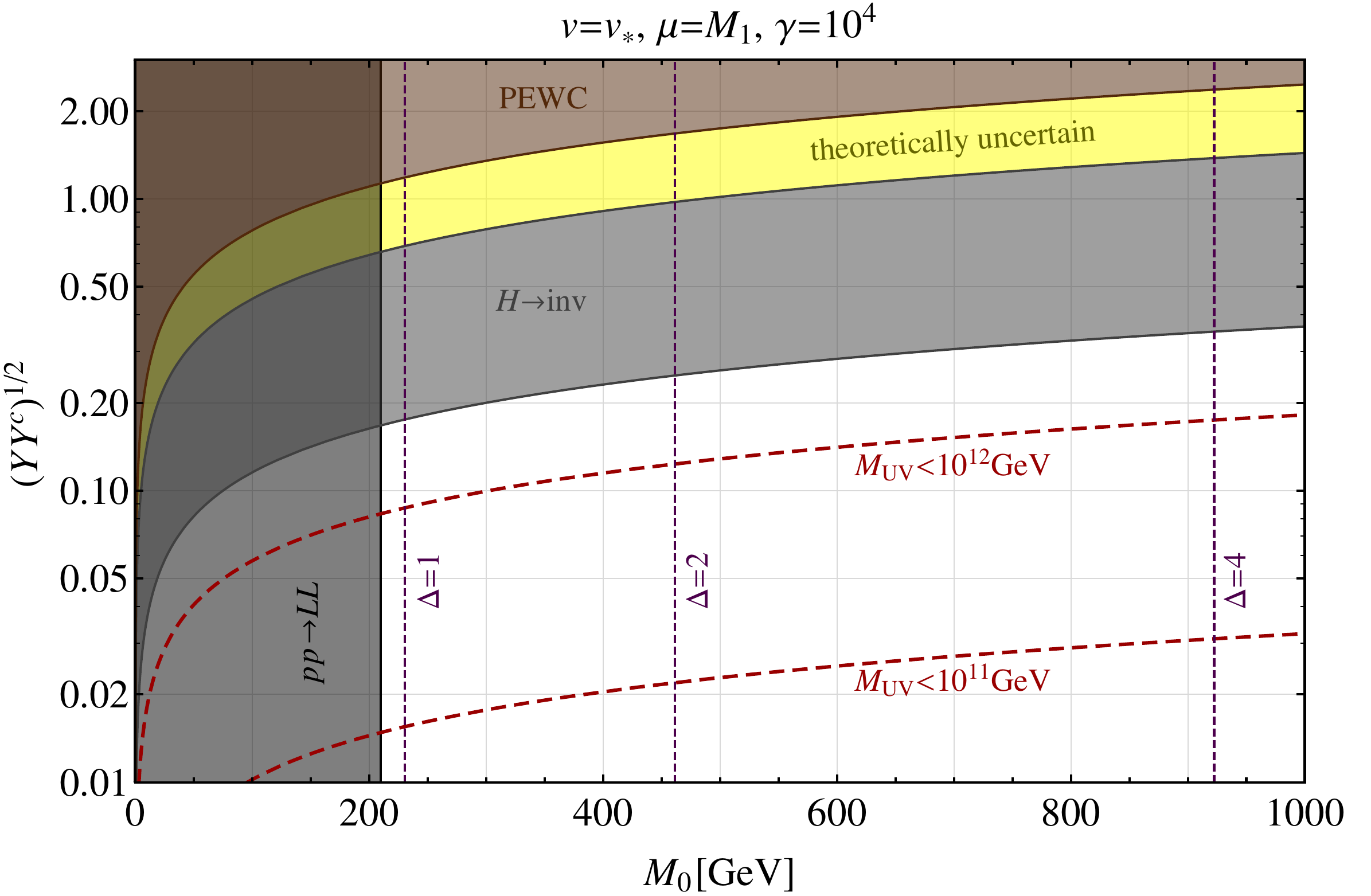}
\caption{\textbf{Top panel:} Experimental and theoretical constraints on our model as a function of $M_0$, assuming a critical Higgs vev $v = v_*$, $\mu = M_1$, and a warping factor $\gamma = 10^6$ throughout. Vertical contours (purple, dashed) show the fine tuning of $M_1$ as a function of $M_0$. Theoretical lower limits on the Yukawa coupling $(YY^c)^{1/2}$ are indicated by the red, dashed contours for $M_\text{UV} = \lbrace 10^{12}, 10^{10}, 10^6\rbrace~\text{GeV}$. The dark gray shaded region on the left is excluded by LHC8 searches for the doublet leptons. The gray shaded band represents the limit on Higgs invisible decays, while the brown region near the top is excluded by precision electroweak constraints (PEWC); between them, there is an experimentally allowed yellow region where the functionality of our mechanism is unclear. \textbf{Bottom panel:} Zoom on a region with larger Yukawa couplings and smaller $M_0$, and for a smaller warping factor of $\gamma = 10^4$ instead. Vertical tuning contours (purple, dashed) of $\Delta = 1,2,4$ are for $M_\text{UV} = 10^{12}~\text{GeV}$, while the theoretical lower limits (red, dashed) on the Yukawa coupling in our model are now shown for $M_\text{UV} = 10^{12}~\text{GeV}$ and $10^{11}~\text{GeV}$.}\label{fig:bounds}
\end{figure}

The doublet has a mass of approximately $M_0$ which must be near the electroweak scale because of naturalness: the radiative mass correction in eq.~\ref{eq:masscorrection} together with the expression for $v_*$ (normalized to $v_* \approx 174\,{\rm GeV}$) in eq.~\ref{eq:vstar} means that $M_0$ cannot be higher than about $4\pi v_* / \sqrt{2 \log [M_{\rm UV}^2/v_*^2]}$ in the absence of a tuning for $M_1$. An ultraviolet cutoff scale of $M_\text{UV} = 10^{12}~\text{GeV}$ requires the electroweak-charge fermions to be lighter than about 250~GeV if the lighter fermion mass scale $M_1$ is to be technically natural. The dashed purple lines in the top panel of figure~\ref{fig:bounds} show tuning contours of $\Delta = 1$ (not tuned) and $\Delta = 5$ (tuned to 20\%) as a function of $M_0$ for a high (low) cutoff of $M_\text{UV} = 10^{12}~\text{GeV}$ ($10^6~\text{GeV}$). Tuning contours of $\Delta = 1, 2, 4$ for $M_\text{UV} = 10^{12}~\text{GeV}$ are also shown in the bottom panel of figure~\ref{fig:bounds}.
Contours of the minimal Yukawa coupling and thus $M_1$ and $\mu$ are depicted by the dashed red lines for different UV cutoffs, according to the relation of eq.~\ref{eq:maxmuv}.

\begin{figure}[htb]
\centering
\includegraphics[trim = 6cm 20cm 6cm 3cm, clip, width=0.45\textwidth]{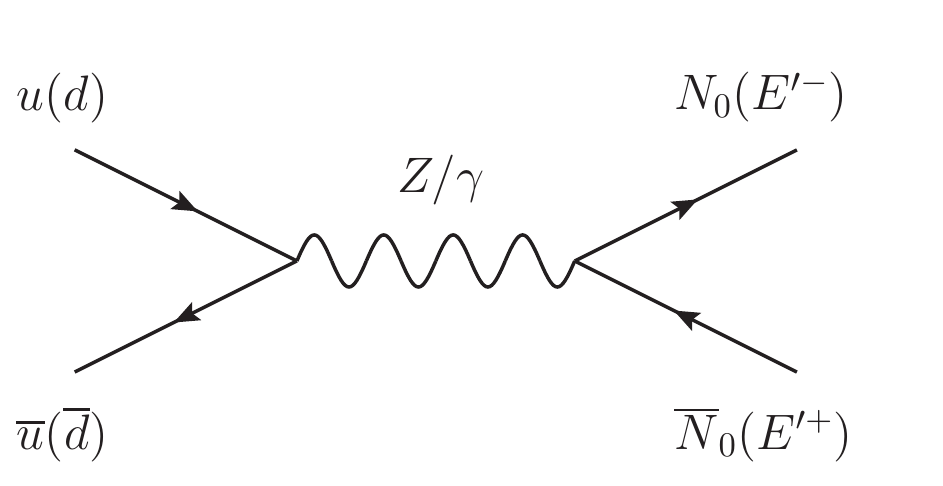}
\includegraphics[trim = 6cm 20cm 6cm 3cm, clip, width=0.45\textwidth]{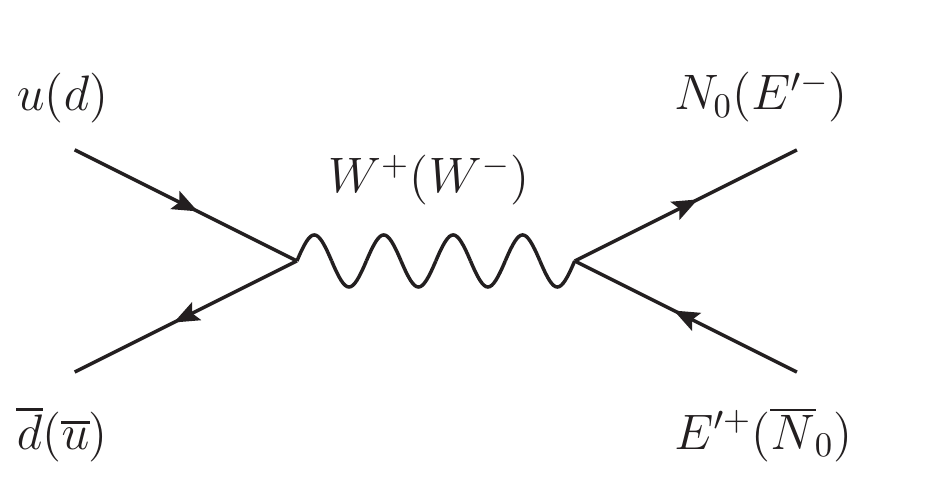}
\caption{Main production channel of the electroweak doublets $L = (E^{\prime -}, N_0)$ at a proton collider. Arrows indicate the flow of the new conserved $U(1)$ fermion number. }\label{fig:prod}
\end{figure}

\begin{figure}[htb]
\centering
\includegraphics[trim = 8cm 21cm 8cm 3cm, clip, width=0.32\textwidth]{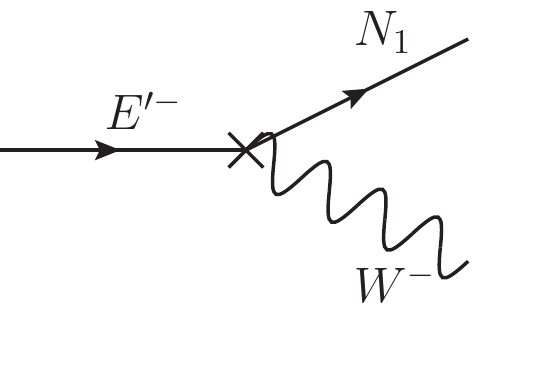}
\includegraphics[trim = 8cm 21cm 8cm 3cm, clip, width=0.32\textwidth]{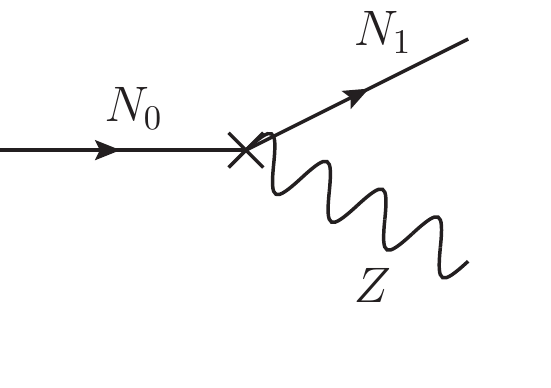}
\includegraphics[trim = 8cm 21cm 8cm 3cm, clip, width=0.32\textwidth]{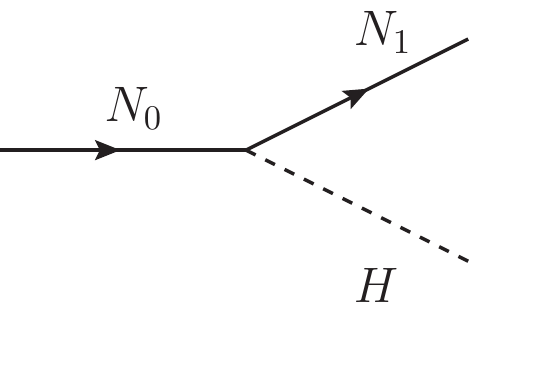}
\caption{Dominant interactions contributing to the decay of the electroweak doublets $L$. Analogous interactions exist for $E^{\prime +}$ and $\overline{N}_0$. The cross is a Higgs vev insertion.}\label{fig:decay}
\end{figure}

\noindent \textbf{Direct searches---}The electroweak doublet $L$ can be produced via electroweak interactions in the processes delineated by the Feynman diagrams in figure~\ref{fig:prod}. The production cross-section of $L$ pairs is similar to that of higgsino pairs; also here, the production channel via a $W$ boson is much larger than Drell-Yan production via $Z/\gamma$~\cite{Beenakker:1999xh,Kramer:2012bx}. The components $E^{\prime -}$ and $N_0$ in the $L$ doublet promptly decay to $N_1$ by emitting a longitudinal component of a gauge boson or a Higgs boson, as drawn in the diagrams of figure~\ref{fig:decay}. A guaranteed signature in this type of model is thus events at the LHC and future colliders with final states containing the decay products of weak gauge or Higgs bosons, and missing energy.

Several searches at LHC8 cover these signatures already~\cite{Aad:2014nua,Khachatryan:2014qwa,Chatrchyan:2013wxa,Chatrchyan:2013mys,Aad:2014vma,Khachatryan:2014mma}. 
The most sensitive searches look for final states from the decay of $WZ$ gauge bosons~\cite{Khachatryan:2014qwa,Khachatryan:2014mma,Aad:2014nua} in channels with three leptons, with two leptons on the $Z$ peak, and missing energy, as well as channels with two leptons on the $Z$ peak, two jets with invariant mass near that of the $W$, and missing energy. The combined 95\%-CL limit on the mass of these states from searches in the LHC 8~TeV run is $180~\text{GeV}$. Recently, CMS updated their analysis~\cite{CMS:2016gvu} with $12.9~\text{fb}^{-1}$ of 13~TeV data, improving this limit to $210~\text{GeV}$.  This bound is indicated by the vertical gray region on the left of figure~\ref{fig:bounds}. (We conservatively extended this bound to the region of large Yukawa couplings and thus large $M_1$ and $\mu$, a squeezed-spectrum limit in which the searches of refs.~\cite{Khachatryan:2014qwa,Khachatryan:2014mma,Aad:2014nua} lose much of their power. See ref.~\cite{CMS:2016gvu} for more detail.)

While the aforementioned bound on $M_0$ already excludes part of the parameter space of our model, it does not yet limit how high the UV cutoff $M_\text{UV}$ can be taken in our framework, with eq.~\ref{eq:maxmuv} being the leading constraint.
With $300~\text{fb}^{-1}$ of LHC collisions at 14~TeV, the projected discovery reach is $450~\text{GeV}$, while $3000~\text{fb}^{-1}$ at a future 100-TeV collider may unveil these states even if they are as heavy as $3~\text{TeV}$~\cite{Han:2013kza,Berggren:2013bua,Gori:2014oua,CMS:2013xfa,ATLAS:2013hta}.
Under the assumption of naturalness in the new fermion sector, this proves that our model is as falsifiable as any other dynamics that can stabilize the electroweak scale, such as supersymmetry or compositeness. Over the lifetime of LHC14, the parameters of our model will be strained at $\Delta > 1$ for all but the lowest cutoffs, while a 100~TeV collider could convincingly rule out---or discover one element of---the construction presented in this work.

If an electroweak doublet is discovered at a collider, the first step towards identifying it with our brane fermion sector and mechanism would be to measure the splitting of the mass eigenstates $E^{\prime -}$ and $\widetilde{N}_0$. The charged lepton $E^{\prime -}$ gets extra radiative mass corrections from gauge loops relative to the neutral fermion $\widetilde{N}_0$, just like in the higgsino sector of split supersymmetry models~\cite{Arvanitaki:2012ps}. The $E^{\prime -}$ mass is approximately $M_0 + 355~\text{MeV}$. The mass of $\widetilde{N}_0$ is $m_0(v) \simeq M_0 + Y Y^c v^2 /M_0$, receiving positive corrections from the Higgs vev. This latter classical correction is one aspect that sets our model apart from others: it would pinpoint the product of Yukawa couplings---and for $v \simeq v_*$, also the Lagranian parameter $M_1$---that can be independently measured via other methods, our next topics of discussion.

For sufficiently large Yukawa couplings and thus $M_1$, lepton colliders such as CEPC~\cite{CEPC-SPPCStudyGroup:2015csa} and the ILC~\cite{Baer:2013cma} may be able to measure the lightest mass eigenvalue(s) in the fermion sector, which will be of order $\mu$, via precision studies of the kinematics in the decays of figure~\ref{fig:decay}. A crucial part of our mechanism is that there is an accidental cancellation in the fermion masses in our vacuum, so a measurement of at least one mass eigenvalue significantly smaller than the bare mass $M_1$ would be a tantalizing hint of the chiral symmetry restoration needed in our model. We leave a careful study on the prospects for measuring these decay kinematics at planned lepton colliders to future work.

\begin{figure}[bt]
\centering
\includegraphics[trim = 8cm 21.5cm 9cm 2.9cm, clip, width=0.32\textwidth]{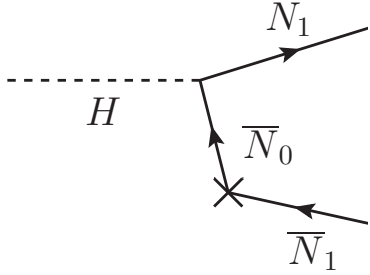}
\caption{Higgs boson decay into a pair of light, invisible fermions. The cross is a Higgs vev insertion.}\label{fig:higgsdecay}
\end{figure}

\noindent \textbf{Higgs invisible width---}The Higgs boson will decay to $\widetilde{N}_1 \widetilde{\overline{N}}_1$ if they are lighter than half the Higgs mass, via the interaction depicted in figure~\ref{fig:higgsdecay}. The LHC 8~TeV run has constrained the Higgs invisible branching ratio to be less than $ \text{BR}( H \rightarrow {\rm inv}) < 0.23$~\cite{Aad:2015pla,CMS:2015naa}, with improvements to come after more integrated luminosity in the 13~TeV run~\cite{CMS:2016rfr,CMS:2016pod,CMS:2016hmx,ATLAS:2016inv}. The limit on the Higgs invisible decay width can be translated into a bound on the effective Yukawa coupling $(Y Y^c)^{1/2}$:
\begin{align}
(Y Y^c)^{1/2} \lesssim 0.24 \left( \frac{M_0}{300 \,{\rm GeV}}\right)^{1/2}.\qquad (95\%~\text{CL})
\end{align}
Assuming the Higgs vev is near the critical one given by $YY^c v_*^2 /M_0 = M_1$, we can interpret this upper bound on the Yukawa coupling as an $M_0$-independent upper bound on the Lagrangian parameter $M_1$ of about 5~GeV, via the relation:
\begin{align}
M_1 \lesssim 5~\text{GeV} \left(\frac{\text{BR}(H\to \text{inv})}{0.23} \right)^{1/2}. \qquad \text{(for $v\simeq v_*$ and $M_1 \lesssim m_H/2$)}
\end{align}
Conversely, a \emph{positive} signal in the invisible decay channel of the Higgs boson can be interpreted as a measurement of the parameter $M_1$ in the context of our model, which should be consistent with the mass splitting of $E^{\prime -}$ and $N_0$.

After $3000~\text{fb}^{-1}$ of LHC14 data, invisible branching ratios of the Higgs may be probed down to the $10\%$ level~\cite{Glaysher:2015bff,CMS:2013xfa}, which corresponds to $M_1 \approx 3~{\rm GeV}$. Future lepton colliders are projected to have a precision on the Higgs invisible branching ratio at the level of $0.14\%$~\cite{CEPC-SPPCStudyGroup:2015csa}, corresponding to $M_1 \approx 0.4~{\rm GeV}$, and putting pressure on our highest-cutoff models in case of a null result. If our mechanism is realized in Nature with a UV cutoff of $10^{12}~{\rm GeV}$, prospects to measure the invisible width of the Higgs boson at a future $e^+ e^-$ machine are hopeful.

\noindent \textbf{Precision electroweak observables---}Virtual effects from the new fermion sector may also be visible, depending on the mass and Yukawa coupling of the doublet fermions~\cite{Martin:2009bg}. The fermionic vector pairs radiatively generate the following corrections to the $S$ and $T$ parameters~\cite{Peskin:1991sw}:
\begin{align}
\Delta T &= 0.06 \left(\frac{Y Y^c}{1} \right)^2 \left(\frac{300~\text{GeV}}{M_0}\right)^2, \\
\Delta S &= 0.014 \left(\frac{Y Y^c}{1} \right) \left(\frac{300~\text{GeV}}{M_0}\right)^2.
\end{align} 
These observables have been constrained experimentally to $\Delta T \lesssim 0.20$ and $\Delta S \lesssim 0.14$, both at 95\%~CL~\cite{Agashe:2014kda}. The resulting constraint on the Yukawa coupling is the relatively mild $(Y Y^c)^{1/2} \lesssim 1.1$ for $M_0 = 200~{\rm GeV}$ and worse at higher doublet masses. The most relevant constraint comes from the bound on the $T$ parameter, and is interpreted in the context of our model as the brown exclusion regions in figure~\ref{fig:bounds}.

There is a small region of parameter space depicted by the yellow region in figure~\ref{fig:bounds} between the precision electroweak constraints (PEWC) and $H \to \text{inv}$ bounds that is currently still experimentally allowed for not too large a hierarchy between $M_1$ and $\mu$, each around 100~GeV.
This sliver of parameter space may lead to interesting LHC signatures~\cite{CMS:2016com}, but the absence of mass hierarchies puts our Casimir energy calculation at the edge of its region of validity, so we postpone a detailed analysis of this region to future work.
Nevertheless, future lepton colliders will likely have the capability to determine the $T$ parameter to a precision of $0.02$ for $S=0$~\cite{CEPC-SPPCStudyGroup:2015csa}, sufficient to close up this gap. In addition, a pair of light charged fermions with considerable Yukawa couplings to the Higgs will also change the Higgs decay rate into two photons~\cite{ArkaniHamed:2012kq,Kearney:2012zi,Carena:2012xa}. Though current measurements~\cite{Khachatryan:2016vau} do not place additional constraints on our parameter place, a combined effort of the high-luminosity LHC and future lepton colliders will measure the diphoton rate to percent level precision~\cite{Fan:2014vta}, probing some of the yellow band at low $M_0$.

\subsection{Radion signatures}\label{sec:radion}

The radion in our model is extremely light, and may produce exciting signatures in fifth-force experiments, equivalence-principle tests, black-hole superradiance and, if abundant in the Universe, scalar dark matter searches. We will first summarize the radion's main properties derived in detail in appendix~\ref{sec:radionmass}, and then discuss the signatures.

\begin{figure}
\centering
\includegraphics[width=0.9\textwidth]{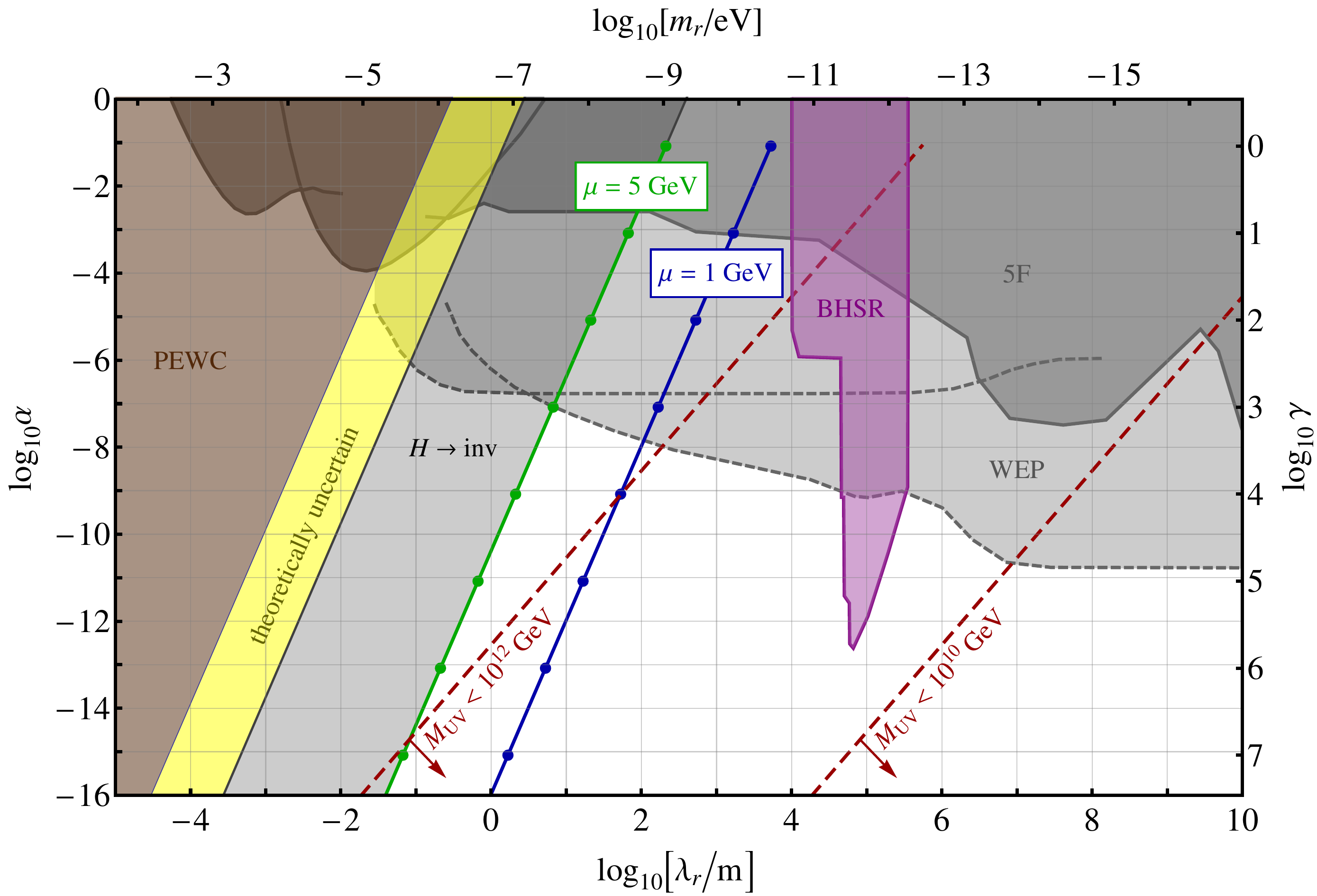}
\caption{Plot of the fractional change $\alpha$ in Newton's potential due to the radion Yukawa force as a function of its range set by its Compton wavelength $\lambda_r = 1/m_r$. The green (red) solid line shows the radion mass-coupling relation for fixed $\mu = 2/L = 5~\text{GeV}$ ($1~\text{GeV}$) and variable $\gamma$, with dots indicating $\gamma = 1, 10, 10^2, \dots$ from top to bottom. The darker gray region above the solid gray line is excluded by searches for a fifth force (5F)~\cite{Hoskins:1985tn,Kapner:2006si,Geraci:2008hb,Fischbach:1996eq}. The lighter gray region above the dashed gray line is excluded if the radion couplings violate the weak equivalence principle (WEP) by an $\mathcal{O}(1)$ amount~\cite{Wagner:2012ui,Smith:1999cr}, which can occur if SM states have different profiles in the extra dimension. The brown region is excluded by precision electroweak constraints (PEWC), while the lighter gray region is excluded by null observations of the Higgs invisible width. The yellow band is theoretically uncertain. The purple shaded region is excluded by null observations of the black hole superradiance effect (BHSR)~\cite{Arvanitaki:2014wva}. %The radion is a dark matter candidate through misalignment mechanism if it is heavier or more strongly coupled than indicated by the dotted blue line. 
The dashed red contours indicate maximum UV cutoffs.}\label{fig:radion}
\end{figure}

For a flat fifth dimension, the radion couples with roughly the same strength as the graviton.
The radion mass is determined by the equilibrium size $L$ of the extra-dimension, which is expected to be of order $1/\mu$:
\begin{align}
m_r = \frac{1}{\lambda_r} \simeq \sqrt{\frac{40 \beta}{3}} \frac{1}{L^2 M_\text{Pl}} \approx \frac{1}{300~\text{m}} \left(\frac{\beta}{\frac{1}{32\pi^2}\frac{3 \zeta(5)}{2}}\right)^{1/2} \left(\frac{\mu}{5~\text{GeV}} \right)^2 \left(\frac{2}{\mu L}\right)^2.
\end{align}
A gravitationally coupled scalar this light is already excluded, as we show in figure~\ref{fig:radion}. For a warped extra dimension, the radion mass-squared is enhanced by the warping factor $\gamma$ for a fixed 4D Planck mass $M_\text{Pl}$, as we calculated in eq.~\ref{eq:mrcurved}:
\begin{align}
m_r = \frac{1}{\lambda_r} \simeq \sqrt{(\gamma+1) \frac{20 \beta}{3}} \frac{1}{L^2 M_\text{Pl}} \approx \frac{1}{4.2~\text{m}} \left(\frac{\gamma}{10^4}\right)^{1/2} \left(\frac{\beta}{\frac{1}{32\pi^2}\frac{3 \zeta(5)}{2}}\right)^{1/2} \left(\frac{\mu}{5~\text{GeV}} \right)^2 \left(\frac{2}{\mu L}\right)^2.
\end{align}
Moreover, for a warped bulk, the radion has a profile that peaks near the IR brane, and therefore has suppressed couplings of $\frac{B_r}{\sqrt{6}\gamma M_\text{Pl}} T^{\mu}_{\mu}$ to the Standard Model states on the UV brane, where $T^{\mu}_{\mu}$ is the trace of the energy momentum tensor of the brane-localized SM states~\cite{Garriga:1999yh,Csaki:1999mp}.

The radion has a mass that appears ``unnatural'' to a 4D low-energy observer. This is a feature that stems from the tuning of the brane tensions against the bulk cosmological constant, which guarantees a large but finite fifth dimension, and a correspondingly small radion mass due to nonlocal effects, including Casimir energy (see appendix~\ref{sec:lightradion} for further details). This tuning is necessary for getting a fifth dimension whose stabilization can be controlled by the Higgs vev, and is therefore a prerequisite for getting a small four-dimensional cosmological constant. The excess number of vacua $\frac{M_{\rm UV}^4}{\gamma^3 \mu^4}$ found in eq.~\ref{eq:totalvacs} is parametrically the amount one needs to tune the radion mass-squared from the ``natural'' value $\frac{M_{\rm UV}^4}{\gamma^2 M_\text{Pl}^2}$ to the ``tuned'' value $\frac{\gamma \mu^4}{M_\text{Pl}^2}$. The lightness of the radion brings into play a variety of precision instruments that can look for this new degree of freedom.

The radion generates a Yukawa force between massive objects on the UV brane to which the SM is localized, and modifies the gravitational potential between objects with mass $m_1$ and $m_2$ spaced a distance $d_{12}$ apart:
\begin{equation}
V_r (d_{12}) = \frac{G m_1 m_2}{d_{12}} \left(1+ \alpha e^{-m_r d_{12}}\right)\label{eq:radionforce}
\end{equation}
where $m_{r}$ is the radion mass, and $\alpha \simeq 1/12\gamma^2$ parametrizes the strength of the new Yukawa interaction compared with gravity. The force mediated by the radion is only relevant at distances $d_{12}$ shorter than its Compton wavelength $\lambda_r \equiv 1/m_r$. In figure~\ref{fig:radion}, we show that the radion is sufficiently weakly coupled to evade all current fifth-force constraints for $\gamma > 10^{5}$~\cite{Hoskins:1985tn,Kapner:2006si,Geraci:2008hb,Fischbach:1996eq}. The exact value of the coefficient $\alpha$ might depend on the chemistry of the objects involved---violating the equivalence principle---if the SM brane has a finite thickness, with different SM fields localized at different places along the extra dimension~\cite{ArkaniHamed:1999za}.  %(In figure~\ref{fig:radionDM}, we assume that the relative radion couplings to different SM fields violate the weak equivalence principle by an $\mathcal{O}(1)$ amount only for the curves and regions labeled ``WEP'', ``Rb/Cs'', ``Dy'', ``O/O clocks'', and ``N/O clocks''.)
In this case, there are additional model-dependent constraints from searches for new forces that violate the weak equivalence principle~\cite{Wagner:2012ui,Smith:1999cr}. Future equivalence principle tests could shed more light on the mass and coupling of the radion~\cite{Dimopoulos:2006nk}. The radion interactions also violate the strong equivalence principle; these effects are the dominant model-independent constraints on the radion in the mass regime below $10^{-20}~\text{eV}$~\cite{bertotti2003test,Blas:2016ddr}. The KK modes of the graviton are at most gravitationally coupled to the Standard Model, and their masses are typically too large for detection in short-distance gravity experiments.

\begin{figure}
\centering
\includegraphics[width=0.99\textwidth]{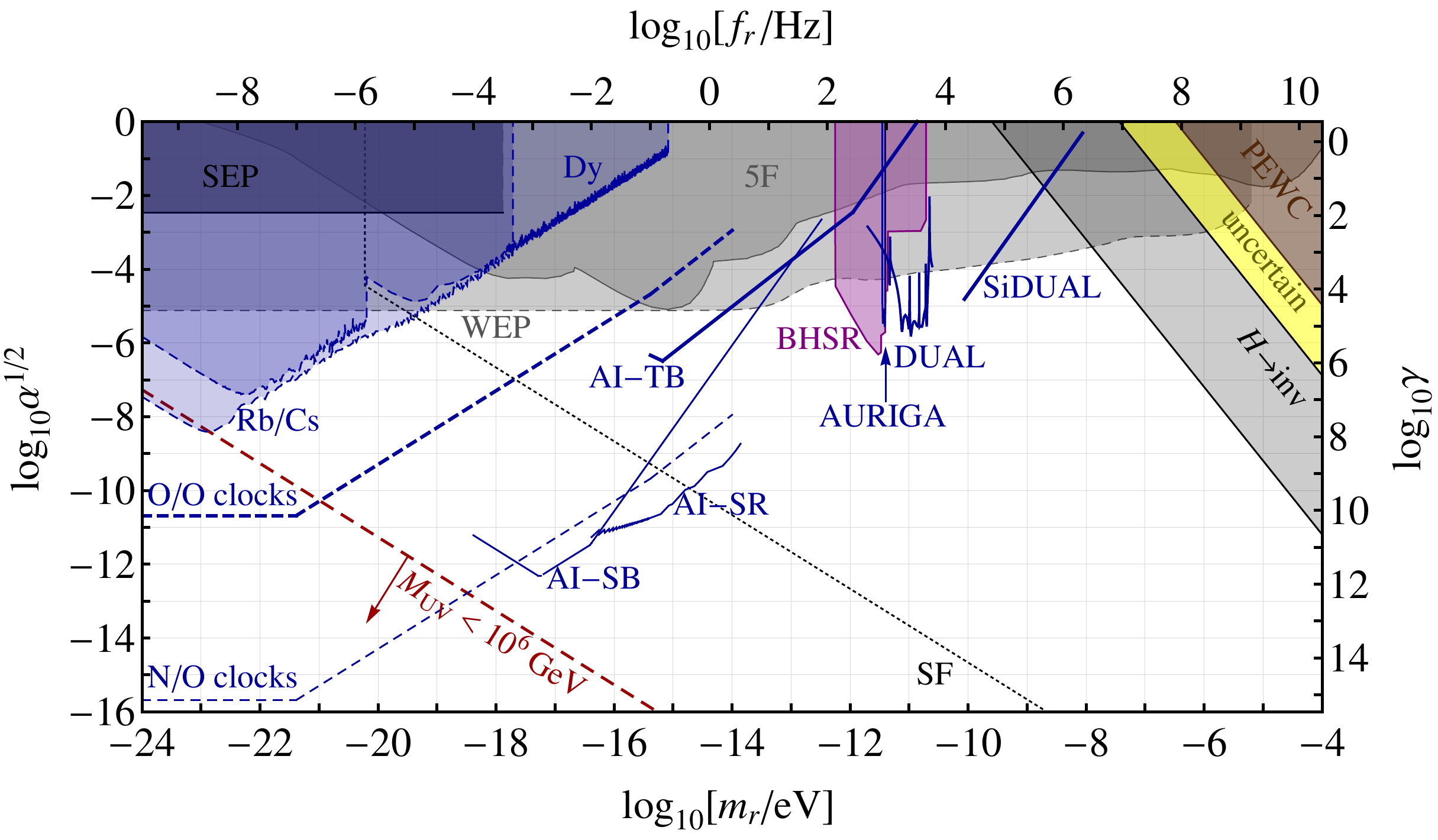}
\caption{Parameter space of the radion coupling $\alpha^{1/2}$ relative to gravity as a function of mass $m_r$ and frequency $f_r = m_r/2\pi$. Under the assumption that the radion constitutes all of the dark matter, the discovery reach of scalar dark matter searches is shown by the blue curves. At sub-Hz frequencies, current optical clocks (O/O clocks) and a future nuclear-optical clock pair comparison (N/O clocks) have prime sensitivity~\cite{Arvanitaki:2014faa}. Atom-interferometric gravitational wave detectors on earth (AI-TB) and in space (AI-SB) can cover frequencies between $10^{-4}~\text{Hz}$ and $10~\text{Hz}$ in broadband mode, while a space-based resonant-mode detector (AI-SR) could perform a deep scan above $10^{-2}~\text{Hz}$~\cite{Arvanitaki:2016fyj}. Experimental proposals (DUAL~\cite{Leaci:2008zza}, SiDUAL) based on resonant-mass gravitational wave detector technology can probe the radion above kHz frequencies~\cite{Arvanitaki:2015iga}. 
95\%-CL limits from existing scalar dark matter searches by atomic clocks (Dy~\cite{VanTilburg:2015oza}, Rb/Cs~\cite{Hees:2016gop}) and the AURIGA resonant-mass detector~\cite{Branca:2016rez} are indicated by blue regions. The radion can only be a subcomponent of DM left and below of the black dotted line due to disruption of structure formation (SF)~\cite{Arvanitaki:2014faa}. Virtual radion exchange also leads to violations of the strong equivalence principle (SEP), which have been most strongly constrained by Doppler tracking of the Cassini spacecraft~\cite{bertotti2003test}. Curves and regions labeled ``WEP'', ``Rb/Cs'', ``Dy'', ``O/O clocks'', and ``N/O clocks'' assume that the relative radion couplings to different SM fields violate the weak equivalence principle by an $\mathcal{O}(1)$ amount.
Other abundance-independent exclusion regions and $M_\text{UV}$ bounds are similar to those in figure~\ref{fig:radion}.  }\label{fig:radionDM}
\end{figure}

There is a corner in our parameter space with small $\mu$ and large $\gamma$ where the radion Compton wavelength $\lambda_r$ matches the Schwarzschild radius of astrophysical black holes. Through the superradiance effect~\cite{Arvanitaki:2010sy}, angular momentum and energy can be extracted via the production of radion particles. This process constrains models with small warping factor $\gamma$ since the radion cloud can only build up to an appreciable size if the effective radion quartic coupling of $\gamma^2 m_r^2/M_\text{Pl}^2$ is sufficiently small~\cite{Arvanitaki:2014wva}. Note that the radion self-coupling grows with the warping factor $\gamma$, while its coupling with matter decreases as $\alpha^{1/2} \sim 1/\gamma$. Black-hole superradiance thus constrains a radion with large couplings to matter. The horizontal edge of the superradiance constraints in figures~\ref{fig:radion}~and~\ref{fig:radionDM} is uncertain due to nonlinear effects discussed in ref.~\cite{Gruzinov:2016hcq}.

The radion in our model can also be a dark matter candidate through the misalignment mechanism (see section~\ref{sec:cos} for more details). If abundant in the Universe today, searches for dilaton-like light scalar dark matter~\cite{Arvanitaki:2014faa,Graham:2015ifn, Arvanitaki:2015iga,Arvanitaki:2016fyj,Blas:2016ddr} have a promising discovery reach in the radion mass-coupling parameter space as shown in figure~\ref{fig:radionDM}. For masses lighter than $10^{-14}$~eV, broadband searches for oscillatory signals in atomic clocks can already probe a large part of the parameter space~\cite{Arvanitaki:2014faa} if the radion couplings to the SM violate the weak equivalence principle. Two data analyses on isotopes of Dy~\cite{VanTilburg:2015oza} and on hyperfine Rb and Cs clocks~\cite{Hees:2016gop} have already set the best limits in this part of the dark matter parameter space. 

In the future,  comparisons between different optical clocks~\cite{2013Sci...341.1215H,2015NatCo...6E6896N, 2016NaPho..10..258N,PhysRevLett.116.063001, 2016arXiv160706867S}, and especially between a nuclear thorium clock~\cite{peik2003nuclear,campbell2012single,tkalya2015radiative,von2016direct} and an optical clock will greatly extend the discovery potential of atomic clock pair comparisons. In figure~\ref{fig:radionDM}, we estimate the $\text{SNR}=1$ reach after $10^7\text{ s}$ integration for an optical clock pair with a fractional frequency instability and thus sensitivity to variations in the fine structure constant $\alpha_\text{EM}$ of $\delta f / f \sim \delta \alpha_\text{EM} / \alpha_\text{EM} \sim 10^{-16} ~\text{Hz}^{-1/2}$. We show a similar estimate for a comparison between an optical clock and a future $^{229\text{m}}$Th clock with $\delta f / f \sim 10^{-15} ~\text{Hz}^{-1/2}$ stability, boosted by an enhancement factor in the coupling to nucleons of about $10^6$~\cite{flambaum2006enhanced}. Terrestrial and space-based atom-interferometric gravitational wave sensors will complement and expand even further the reach at the higher end of this frequency band~\cite{Arvanitaki:2016fyj}. These gravitational wave detectors rely on the time-domain response of the setup to the DM wave, and have sensitivity even if the radion couplings do respect the weak equivalence principle.

For radion masses above $10^{-12}$~eV, current resonant-mass detectors are already probing new parameter space~\cite{Arvanitaki:2015iga}, as evidenced by a limit from the AURIGA experiment in a narrow band around $3 \times 10^{-12}~\text{eV}$~\cite{Branca:2016rez}. Smaller and more sensitive devices offer the possibility to expand the reach of this acoustic signature to higher frequencies in the near future~\cite{Arvanitaki:2015iga,Leaci:2008zza,Goryachev:2014yra}.
The SiDUAL curve in figure~\ref{fig:radionDM} shows the ultimate reach after $10^7~\text{s}$ integration as limited by thermal noise on the lowest breathing-mode resonance of a silicon sphere with a quality factor $Q \sim 10^6$ cooled to $20~\text{mK}$, for variable sphere radii smaller than $20~\text{cm}$. Like the interferometric gravitational wave sensors, the resonant-mass detectors retain their sensitivity even if the radion couplings obey the weak equivalence principle.

Finally, both at very low masses and at large self-interactions (i.e.~high warping), the radion ceases to be a good DM candidate. For too light or too strongly self-coupled scalar DM, the associated Jeans length can become larger than the size of observed structures in our Universe~\cite{Arvanitaki:2014faa}. Structure formation will essentially be unaffected to the right and above the black dotted line in figure~\ref{fig:radionDM}; to the left and below the dotted line, the radion can likely only constitute a fraction of the dark matter. We note that all the sensitivity curves and exclusion bounds depicted in blue in figure~\ref{fig:radionDM} only weaken by a factor $(\rho_r/\rho_0)^{1/2}$ for radion matter densities $\rho_r$ below the total DM density $\rho_0$.

\subsection{Cosmology}\label{sec:cos}

\noindent \textbf{Light fermions---}The lightest of the new fermion states $\widetilde{N}_1$ and $\widetilde{\overline{N}}_1$ will be stable due to the unbroken new fermion number discussed in section~\ref{sec:technatural}, and are therefore a potential dark matter candidates. If once abundant in the Universe, they can only annihilate into pairs of Standard Model particles through the Higgs portal with a small annhilation cross section. If ever in thermal equilibrium, the freeze-out abundance of these fermions would overclose our Universe. Thermal equilibrium with the SM can be avoided if the reheating temperature $T_\text{reh}$ is below the mass of the lightest mass eigenstate: $T_\text{reh} \lesssim \mu$. Requiring that a ``freeze-in'' abundance from the tail of the Maxwell distribution of the Standard Model particles~\cite{Hall:2009bx} does not produce too much matter content, yields a slightly stronger bound of $T_\text{reh} \lesssim \mu /15$. It is also possible that $YY^c$, $M_1$, and $\mu$ are all small enough---that is, $M_1 \sim \mu \lesssim 10^{-4}~\text{GeV}$---such that $\widetilde{N}_1$ and $\widetilde{\overline{N}}_1$ are never in thermal contact with the Standard Model for reheating temperatures compatible with Big-Bang nucleosynthesis ($T_\text{reh} \gtrsim 1~\text{MeV}$).
Higher reheating temperatures together with larger couplings could be accommodated if the unbroken fermion number is weakly gauged, in which case the lightest states can annihilate into two dark photons of the gauged $U(1)$ symmetry.

\noindent \textbf{Radion---}The radion is an extremely light scalar and as such it will be displaced from its minimum during inflation. A large misalignment of the radion remains fixed after the end of inflation, until the Hubble rate becomes comparable to the radion mass. At this point, the misaligned radion field starts oscillating and redshifting like matter. The radion can thus be the dark matter of our Universe, or a small component thereof, with an abundance:
\begin{align}
\rho_{r,\text{mis}}^{0} \simeq 
\frac{1}{\gamma^2} \left(\frac{m_r}{\mathcal{H}_\text{eq}}\right)^{1/2} \left(\frac{B_r^\text{mis}}{M_\text{Pl}/\gamma}\right)^2 \rho^0.
\end{align}
Here, $\rho^0$ is the DM abundance in today's Universe, $B_r^\text{mis}$ is the primordial radion field misalignment amplitude that has natural size $\sim M_{Pl}/\gamma$ (see appendix~\ref{sec:radionmass}), and $\mathcal{H}_\text{eq}$ is Hubble at matter-radiation equality.
The mass for which the radion makes up all the dark matter of the Universe ($\rho_{r,\text{mis}}^{0} = \rho^0$) is
\begin{equation}
m_r \approx 10^{-8}~\text{eV} \left(\frac{\gamma}{10^5}\right)^4 \left(\frac{M_\text{Pl}/\gamma}{B_r^\text{mis}}\right)^4.
\end{equation}
The primordial misalignment amplitude is a priori unknown: we expect inflation to scan all possible values. This uncertainty implies that the radion is an excellent dark matter candidate in a wide range of the mass vs.~coupling plane as shown in figure~\ref{fig:radionDM}.

In principle, the radion field can take any value, but for displacements larger than $M_\text{Pl}/\gamma$, perturbative control of the radion potential is lost. This means that the extra dimension can possibly (but not necessarily) be destabilized. For high reheating temperatures, one might worry that the radion is driven into this regime in the early Universe by its coupling to the large matter energy density.
In particular, eq.~\ref{eq:L5Hubble} suggests that the approximately static solution is not valid when the brane-localized energy density exceeds $\mu^4/\gamma$. 
To see how the radion evolves after reheating, it is necessary that we restore the time dependence of the radion equation of motion. If the 4D Hubble scale is much smaller than the inverse size of the fifth dimension $\mathcal{H}^2/\mu^2 \ll 1$, the evolution of the radion can be calculated with the radion potential and couplings found in appendix~\ref{sec:radionmass}. The radion equation of motion has the following form:
\begin{equation}
\ddot{B_r} + 3 \mathcal{H} \dot{B_r} + \frac{\mathrm{d} V(B_r)}{\mathrm{d} B_r} + \frac{T^\mu{}_\mu}{\sqrt{6}\gamma M_\text{Pl}} = 0
\end{equation}
For a radion \emph{without} any primordial misalignment, the field $B_r$ will first increase as $\sqrt{t}$ during radiation domination until $\mathcal{H}\simeq m_r$, at which point it reaches
\begin{align}
\overline{B}_r \simeq \frac{M_\text{Pl}}{\gamma} \left(\frac{\mathcal{H}_\text{eq}}{m_r}\right)^{1/2}.\label{eq:radionamp}
\end{align}
When the Hubble rate falls below the mass of the radion, the field starts oscillating, with its energy density redshifting like matter. The radion energy density sourced from this effect is thus at least
\begin{align}
\overline{\rho}_\text{r}^0 \simeq \frac{1}{\gamma^2}\left(\frac{\mathcal{H}_\text{eq}}{m_r}\right)^{1/2}\rho^0. \label{eq:sourcedabundance}
\end{align}
One can see from eq.~\ref{eq:radionamp} that for radion masses smaller than the $\mathcal{H}_\text{eq}$, the radion can develop an amplitude comparable to $M_\text{Pl}/\gamma$, where significant change of the geometry may occur. We restrict to cases in which
$m_r > \mathcal{H}_\text{eq}$ only, where on top it is heavy enough to be a good DM candidate. Even without any primordial misalignment from inflation, eq.~\ref{eq:sourcedabundance} shows that the radion is generically expected to make up some fraction of the dark matter.

\section{Implications of Weinberg's no-go theorem}

We wish to take a step back, and contextualize our framework against other ways to address the naturalness problems of the Higgs mass and the cosmological constant. Approaches toward solving either of them fall in three broad categories: symmetry, adjustment, and landscape (in order of decreasing elegance). Our mechanism to explain both the small Higgs mass and cosmological constant is squarely categorized in the last. We now clarify why we did not pursue more aesthetic approaches.

\noindent \textbf{Symmetry---}The Higgs mass and the cosmological constant are positive-mass-dimension quantities that can be protected with conformal symmetry, supersymmetry, or compositeness, as long as the dynamics associated with breaking the symmetry occurs near the respective observed energy scales. However, there is strong experimental evidence signaling that Nature does not work in this way, overwhelmingly so for the cosmological constant, and increasingly so for the Higgs mass.

\noindent \textbf{Adjustment---}Weinberg's ``no-go theorem''~\cite{Weinberg:1988cp} laid bare the main challenge for adjustment mechanisms of the cosmological constant. If fields---here parametrized by the placeholder $\phi$---are to adjust the cosmological constant down to its observed value, both the value \emph{and} the derivative of the potential must vanish:
\begin{align}
\label{eq:weinbergCC1}
\partial_\phi V(\phi)&=0,\\
\label{eq:weinbergCC2}
V(\phi)&\approx 0.
\end{align}
This system of equations is overdetermined. It allows no solutions in absence of a symmetry enforcing the second equation at the scale of the observed cosmological constant, $\Lambda_0^{1/4} \sim \text{meV}$, because zero of the overall potential is not special. %\footnote{One possible loophole would be to couple $\phi$ directly to a gravitational operator $\mathcal{O}(\mathcal{R})$, though radiative quantum corrections from such a coupling dwarf the contributions from the classical expectation value $\langle \mathcal{O}(\mathcal{R}) \rangle_0$.}  
The only other way eqs.~\ref{eq:weinbergCC1}~and~\ref{eq:weinbergCC2} can be reconciled is to have an enormous number---a  discretuum---of vacua; this defines a ``landscape''. If these minima exhibit different values of the cosmological constant $\Lambda_4$, and their number exceeds $\mathcal{N}_{\Lambda_4} \sim M_\text{UV}^4/\Lambda_0\sim 10^{120} \left( \frac{M_\text{UV}}{M_\text{Pl}}\right)^4$, there can be a minimum with an accidentally small value $\Lambda_4 \sim \Lambda_0 \approx 0$---the vacuum of our Universe.

The same logic can be applied to the hierarchy problem, with the conditions:
\begin{align}
\partial_\phi V(\phi,H)&=0,\label{eq:weinbergH1}\\
\partial_H V(\phi,H)&=0,\label{eq:weinbergH2}\\
\partial_H^2 V(\phi,H)&\approx 0.\label{eq:weinbergH3}
\end{align} 
In the absence of new dynamics or symmetries such as supersymmetry at the weak scale, this system of equations does not admit a solution either, unless there is a discretuum of at least $\mathcal{N}_{m_H^2} \sim M_\text{UV}^2/m_0^2\sim 10^{32} \left( \frac{M_\text{UV}}{M_\text{Pl}}\right)^2$ vacua with $m_0 \sim 100~\text{GeV}$. This also defines a landscape.

There is an innate coupling of the cosmological constant problem to adjustment scenarios for the Higgs mass like those of ref.~\cite{Graham:2015cka}. Suppose classical evolution leads to a preferred final vacuum with $m_H^2 \approx 0$. This vacuum, in addition to eqs.~\ref{eq:weinbergH1},~\ref{eq:weinbergH2},~\ref{eq:weinbergH3}, must also satisfy:
\begin{align}
V(\phi,H)&\approx 0.\label{eq:weinbergH4}
\end{align}
With this addition, the system of equations becomes doubly overdetermined. The zero of energy has to be tuned to coincide with the preferred final state for which eq.~\ref{eq:weinbergH3} holds. This vacuum is just one out of $\sim10^{32} \left( \frac{M_\text{UV}}{M_\text{Pl}}\right)^2$ possible vacua, so requiring eqs.~\ref{eq:weinbergH3}~and~\ref{eq:weinbergH4} are simultaneously satisfied reintroduces the hierarchy problem. To avoid this, one needs a sector that relaxes both $m_H^2$ and $\Lambda_4$ concurrently. Unfortunately, there is no known adjustment mechanism for the cosmological constant~\cite{Abbott:1984qf}, although there has been some recent progress in this direction~\cite{Graham:2017abc,Alberte:2016izw}.

Furthermore, a discretuum inherently defines a landscape. In a landscape of vacua, an approach to the hierarchy or the cosmological constant problem using classical evolution is misleading. Relying on classical equations involves starting with some specific initial conditions and time evolving them until a preferred local minimum is reached. In the presence of such a vast number of vacua, any specific initial conditions have measure zero and constitute a fine tuning. Arguing that some initial conditions are preferred or generic is tantamount to solving the measure problem in this landscape (see for example~\cite{Susskind:2012xf,Linde:2006nw} and references therein).

%Generically, all possible vacua are populated along the global history of the Universe, so any rationale for a specific initial, intermediate, or final state would be tantamount to a solution of the measure problem. Firstly, adjustment mechanisms typically call for a particular set of \emph{initial} conditions, e.g.~positive $\Lambda_4$ or $m_H^2$. While not terribly fine tuned, their measure is nevertheless ill defined.

In addition, reaching a minimum as the endpoint of classical evolution is misleading. There is a huge number of possible quantum tunnelings from this minimum to other vacua that will eventually populate the landscape. Arguing that tunneling takes a long time is no consolation, since there is no preferred unit of time in this eternally-inflating multiverse of vacua. Postulating that we can ignore tunneling because the multiverse is young is another extreme fine tuning~\cite{Susskind:2012xf,Linde:2006nw}.

\noindent \textbf{Landscape---}In view of the above difficulties, we adopt a pragmatic approach. We postulate the only known solution to the cosmological constant problem---the galactic principle in tandem with a landscape of vacua~\cite{Weinberg:1987dv,Weinberg:1988cp,Polchinski:2006gy,Bousso:2007gp}---and investigate theories for which doing so also solves the hierarchy problem.\footnote{As with any anthropic argument, we cannot exclude the possibility that an extremely unlikely fluctuation would result in, for example, a habitable solar system in a universe with an enormous cosmological constant---the so-called Boltzmann brain problem.} Specifically, we construct an extra-dimensional model with a discretuum of vacua,  where eq.~\ref{eq:weinbergH4} automatically implies eq.~\ref{eq:weinbergH3}:\footnote{The idea that the number of vacua with a low Higgs vev can be enhanced in the landscape appeared in~\cite{Dvali:2003br}, albeit without implications for the cosmological constant problem.}
\begin{align}
V(\phi,H)&\approx 0 \qquad \Rightarrow \qquad \partial_H^2 V(\phi,H) \approx 0.\label{eq:implication}
\end{align}
From a low-energy perspective, $\phi$ can be loosely viewed as the radion in our model. In fact, the implication is even stronger than in eq.~\ref{eq:implication}: in our five-dimensional construction, a vanishing cosmological constant not only implies a light Higgs, but also an ultralight radion: $\partial_\phi^2 V(\phi,H) \approx 0$.\footnote{This is reminiscent of arguments in ref.~\cite{Polyakov:1993tp,Damour:1994zq}, connecting the smallness of the cosmological constant to the lightness of the dilaton. However, we have also constructed strictly four-dimensional models accomplishing the relation in eq.~\ref{eq:implication} without the appearance of an additional ultralight scalar~\cite{Arvanitaki:4D}.} In this sense, the galactic principle dictates we live in a triply special place in the landscape: a vacuum with a small cosmological constant, a light Higgs, and an ultralight radion, all of which appear tuned and unnatural to a low-energy observer.

To realize our framework, we have carefully engineered a landscape in which only unprotected quantities, i.e.~additively renormalized Lagrangian parameters, are assumed to vary wildly over scales of order the cutoff. The emergent electroweak scale $v_*^2 \sim M_0^2$ is determined by a fermion mass scale $M_0$ that \emph{is} protected by symmetry. We have tacitly assumed this scale to be fixed and much below the cutoff; if this assumption were relaxed, the resulting electroweak scale would likely be at the largest possible value of $M_0$ in the landscape.\footnote{A similar issue also plagues the marriage of an anthropic explanation of the CC problem with dynamical solutions to the hierarchy problem. If e.g.~the SUSY breaking scale varied throughout the landscape, the galactic principle would naturally tend to select vacua with the largest possible SUSY breaking scale, since heavier matter is more favorable for structure formation. Large SUSY breaking scale is also favored in the Bousso Polchinski framework \cite{Bousso:2000xa}.}
 Because $M_0$ (and also $M_1$ and $\mu$) are technically natural parameters, ultraviolet completions exist in which these fermion masses remain parametrically below $M_\text{UV}$.
 
Anthropic arguments are often abhorred, as they usually make no experimental predictions. This is not necessarily so. Our model is both falsifiable and verifiable, and has phenomenological implications for a wide range of experiments, including particle colliders, fifth force experiments, and light scalar dark matter searches.

\acknowledgments
We thank Sergei Dubovsky, John March-Russell, and Giovanni Villadoro for critical conversations at several stages throughout the completion of this work. We are also grateful to Masha Baryakhtar, Tony Gherghetta, Peter Graham, Anson Hook, Kiel Howe, Nemanja Kaloper, David Kaplan, Robert Lasenby, Maxim Pospelov, Surjeet Rajendran, and Timothy Wiser for helpful discussions. We thank the CERN Theory Group for hospitality during part of this work. SD, VG, JH, and KVT acknowledge the support of NSF grant PHYS-1316699. KVT is supported by a Schmidt Fellowship funded by the generosity of Eric and Wendy Schmidt. AA acknowledges the support of NSERC and the Stavros Niarchos Foundation. Research at Perimeter Institute is supported by the Government of Canada through Industry Canada and by the Province of Ontario through the Ministry of Economic Development \& Innovation.

\appendix

\section{Casimir energy calculation}\label{sec:appCasimir}
We specialize to the case of an approximately flat fifth dimension for the purpose of the explicit calculation of the Casimir energy; the generalization to a warped bulk follows straightforwardly from eq.~\ref{eq:casimirconformal}. We use methods similar to those of ref.~\cite{Csaki:2003sh} to determine the boundary conditions and fermionic spectrum, and evaluate the Casimir energy with techniques of ref.~\cite{Ponton:2001hq}. 

\subsection*{Boundary conditions}
Expanding eqs.~\ref{eq:5action1}~and~\ref{eq:5action2} around five static, flat dimensions ($g^{(5)}_{MN} = \eta_{MN}$), with a fifth dimension compactified on an interval of length $R$,\footnote{To compute mass eigenvalues, it is enough to consider one interval since the fermions' wave functions are (anti-)symmetric around $y=R$.} we get
\begin{align}
S = \int \mathrm{d}^4 x\, \int_{0}^{R} \mathrm{d}y\,  \Bigg\lbrace  & -i \chi^\dagger \bar{\sigma}^\mu \partial_\mu \chi - i  \psi \sigma^\mu \partial_\mu \psi^\dagger + \half \left(\psi \overset{\leftrightarrow}{\partial}_5 \chi - \chi^\dagger \overset{\leftrightarrow}{\partial}_5 \psi^\dagger \right) \label{eq:5action3}\\
&+ \delta(y-0^+)\Big[ - i \tilde{N}_1^\dagger \bar{\sigma}^\mu D_\mu \tilde{N}_1 - i \tilde{N}_1^c \sigma^\mu D_\mu \tilde{N}_1^{c\dagger} \\
&\left. \hspace{6em} + \, m_1 (v) \tilde{N}_1 \tilde{N}_1^c + \mu^{1/2} \tilde{N}^c_1 \chi + \text{c.c.} \right. \Big] \Bigg\rbrace.
\end{align}
after integrating out the heavy fermions with mass $m_0 \simeq M_0$. The equations of motion for the light fermions are:
\begin{alignat}{8}
&-i \bar{\sigma}^\mu \partial_\mu \chi && - \partial_5 \psi^\dagger && + \mu^{1/2}  \tilde{N}_1^{c\dagger} \delta(y-0^+) && && = 0 \label{eq:5eom1}\\
&-i\sigma^{\mu}\partial_\mu \psi^\dagger && + \partial_5 \chi && && && = 0 \label{eq:5eom2}\\
&-i \sigma^\mu \partial_\mu \tilde{N}_1^{c\dagger} && && + \mu^{1/2} \chi|_{0^+} &&+ m_1(v) \tilde{N}_1 && = 0 \label{eq:5eom3}\\
&-i \bar{\sigma}^\mu \partial_\mu \tilde{N}_1 && && && + m_1(v)^* \tilde{N}_1^{c\dagger} && = 0. \label{eq:5eom4}
\end{alignat}
eq.~\ref{eq:5eom1} can be integrated around $y = 0^+$ to yield a jump condition for $\psi^\dagger$, which together with the hard BC for $\psi^\dagger$ in eq.~\ref{eq:hardBC}, gives the following constraint on the SM brane:
\begin{align}
\psi^\dagger|_{0^+} = \mu^{1/2} \tilde{N}_1^{c\dagger}. \label{eq:softBC1}
\end{align}
Similarly, eqs.~\ref{eq:5eom3}~and~\ref{eq:5eom4} can be combined for a Klein-Gordon-type equation for $\tilde{N}_1^{c\dagger}$ which, after combining with eq.~\ref{eq:softBC1}, turns into the boundary condition of eq.~\ref{eq:softBC2}:
\begin{align}
\left[-\partial^2 + \left|m_1(v)\right|^2\right] \psi^\dagger|_{0^+} + i \mu \bar{\sigma}^\mu \partial_\mu \chi|_{0^+} = 0. 
\label{eq:appsoftBC2}
\end{align}

\subsection*{Kaluza-Klein spectrum}
The 5D fields $\chi$ and $\psi^\dagger$ are quantized into massive 4D Kaluza-Klein mode pairs with Dirac masses $m_n$, which solve the Dirac equations
\begin{align}
-i \bar{\sigma}^\mu \partial_\mu \chi_n(x) - m_n \psi^\dagger_n(x) = 0; \quad -i {\sigma}^\mu \partial_\mu \psi_n^\dagger(x) - m_n \chi_n(x) = 0.
\end{align}
The decomposition that solves the above Dirac equations as well as the bulk equations of motion, eqs.~\ref{eq:5eom1}~and~\ref{eq:5eom2}, is
\begin{align}
\chi(x,y) & = \sum_{n=1}^\infty \chi_n(x) \left[A_n \cos(m_n y) + B_n \sin(m_n y) \right]\\
\psi^\dagger(x,y) & = \sum_{n=1}^\infty \psi^\dagger_n(x) \left[-B_n \cos(m_n y) + A_n \sin(m_n y) \right].
\end{align} 
The quantities $A_n$, $B_n$, $m_n$ are now uniquely determined by the BC at $y=0^+,R$ and a canonical normalization condition.  The latter is not needed to derive the spectrum $m_n$, given implicitly as
\begin{align}
\cot(x_n) = + \frac{a x_n}{1 - b x_n^2}, \label{eq:5spectrum}
\end{align}
where we have defined $x_n \equiv m_n R$, $a \equiv \mu / |m_1(v)|^2 R$, and $b \equiv 1/|m_1(v)|^2 R^2$. The LHS and RHS each come from solving for the ratio $-B_n/A_n$ from the BC at $y=R$ (see eq.~\ref{eq:hardBC}) and $y = 0^+$ (see eq.~\ref{eq:appsoftBC2}), respectively.

\subsection*{Evaluation of Casimir sums}
To set up the regularization procedure, we rewrite eq.~\ref{eq:casimirpotentialsum}
\begin{align}
\rho_C = -2 \sum_{n=1}^\infty \int \frac{\mathrm{d}^4 k}{(2\pi)^4} \log(k^2 + m_n^2) \equiv \left. - \frac{d}{ds}\zeta_f(s)\right|_{s \to 0}
\end{align}
in terms of the special $\zeta$-function:
\begin{align}
\zeta_f(s) = -\frac{1}{8\pi^2}\frac{1}{(2-s)(1-s)}\frac{1}{R^{4-2s}} F(2s-4); \quad F(s) \equiv \sum_{n=1}^\infty x_n^{-s},
\end{align}
where $x_n$ are the positive roots of eq.~\ref{eq:5spectrum}. The sum $F(s)$ can be computed via the contour integral:
\begin{align}
I(s) = \frac{1}{2\pi i} \oint_C \mathrm{d} z\, \frac{1}{z^s}\frac{f'(z)}{f(z)}; \quad f(z) \equiv a z - (1 - bz^2) \cot z.
\end{align}
When evaluated over a counterclockwise, infinite, semicircle contour in the ${\rm Re}(z)>0$ half-plane (infinitesimally avoiding the origin, by a semicircle of radius $\varepsilon$), the contour integration picks up all positive roots of eq.~\ref{eq:5spectrum}. It also picks up the poles at $z = n \pi$ with residue $-(n\pi)^{-s}$, but those can be easily subtracted:
\begin{align}
F(s) = I(s) + \frac{1}{\pi^s} \sum_{n=1}^\infty \frac{1}{n^s} = I(s) + \frac{\zeta(s)}{\pi^s},
\end{align}
with $\zeta(s)$ the standard Riemann $\zeta$-function. Evaluating $I(s)$ explicitly for sufficiently large ${\rm Re}(s)>0$ gives:
\begin{align}
I(s) = & \frac{s}{\pi} \sin\left(\frac{\pi s}{2} \right) \int_\varepsilon^\infty \mathrm{d}y\, y^{-s-1} \log\left[\frac{a y + (1+by^2)\coth(y)}{a y + (1+by^2)}\right] \\
&+\frac{s}{\pi} \sin\left(\frac{\pi s}{2} \right) \int_\varepsilon^\infty \mathrm{d} y\, y^{-s-1} \log[a y + (1+by^2)]\\ 
&+\frac{s \varepsilon^{-s}}{2 \pi} \int_{-\pi/2}^{\pi/2} \mathrm{d}\theta\, e^{-is\theta} \log f(\varepsilon e^{i\theta}) - \frac{\varepsilon^{-s}}{2} \cos\left(\frac{\pi s}{2} \right).
\end{align}
The terms in the second and third lines give vanishing contributions (for $a>0$) when analytically continued to $s<0$ in the limit $\varepsilon \to 0$. Hence only the first line remains in this regime, where it is finite and $\varepsilon$-independent. Ultimately, we are interested in the behavior near $s = -4$; one can easily check that $F(-4) = 0$, so the Casimir energy is proportional to:
\begin{align}
-\frac{d}{ds}\left[F(2s-4) \right]_{s \to 0} & = - 2 \pi^4 \zeta'(-4) + \mathcal{I}(a,b),\\
\mathcal{I}(a,b) & \equiv 4 \int_0^\infty \mathrm{d} y\, y^3 \log\left\lbrace \frac{a y + (1+by^2)\coth(y)}{ay+(1+by^2)}\right\rbrace. \label{eq:rhofunction}
\end{align} 
Finally, we get to a manageable expression for the Casimir energy, quoted in eq.~\ref{eq:casimirpotential}:
\begin{align}
\rho_C = -\frac{1}{16 \pi^2} \frac{1}{R^4} \left\lbrace -\frac{3}{2}\zeta(5) + \mathcal{I}(a,b) \right\rbrace. \label{eq:casimirpotentialA}
\end{align}
As a cross-check, one can see that for $\mu = 0$, we have $\mathcal{I}(0,b) = 93\zeta(5)/32$ independent of $b$ as it should for zero brane-bulk mixing, so that the terms within curly brackets evaluate to $45\zeta(5)/32 \approx 1.46$. 

\section{Radion mass and coupling}\label{sec:radionmass}

In this appendix, we calculate the radion mass and coupling for the background solutions given in section~\ref{sec:Casimirstabilization}. As discussed around eq.~\ref{eq:ds2b}, there is a gauge ambiguity in parametrization of the scalar modes of the metric.
To isolate the radion from the unphysical components of the graviton, it is convenient to pick the gauge~\cite{Csaki:2004ay,Ponton:2001hq,Csaki:2000zn,Charmousis:1999rg}:
\be
f(y,x^\mu)=-\frac{1}{2}g(y,x^\mu)= b(y) B_0 (x^{\mu}), \label{eq:gauge}
\ee
where $B_0 (x^{\mu})$ is the radion field in the 4D effective theory (not yet canonically normalized). To find the radion mass, we assume a 4D plane wave ansatz, $B_0 (x^{\mu}) \propto e^{i p x}$ with momentum $p$ parallel to the branes and $p^2=-m_r^2$, factorized from an extra-dimensional profile $b(y)$.
Plugging in the metric from eq.~\ref{eq:ds2b} in the gauge of eq.~\ref{eq:gauge} into the bulk Einstein equations gives a deviation from eqs.~\ref{eq:FriedmannC1}~and~\ref{eq:FriedmannC2} to leading order in $\epsilon$:
\begin{alignat}{14}
+ \frac{3}{2}b'' + 6\frac{a''}{a}b & {}+{} 6\l\frac{a'}{a} \r^2 b && {}+{} 9 \frac{a'}{a}b'  && {}+{} 3 \frac{\mathcal{H}^2}{a^2} b && {}-{}\frac{5}{2} \frac{\beta}{M_5^3 L^5 a^5}b  && {}+{}\frac{15}{2}\frac{\beta}{ M_5^3 L^6 a^5} \int_0^R \frac{b}{a} dy &  {}={} 0,
\label{eq:Friedmannb1} \\
+ \frac{3}{2} \frac{1}{a^2} m_r^2 b & {}+{} 12\l\frac{a'}{a} \r^2 b && {}+{} 6 \frac{a'}{a}b' &&   {}+{} 6 \frac{\mathcal{H}^2}{a^2} b && {}+{} 10 \frac{\beta}{ M_5^3 L^5 a^5}b  && {}-{} 30\frac{\beta}{M_5^3 L^6 a^5} \int_0^R \frac{b}{a} dy & {}={} 0.
\label{eq:Friedmannb2} 
\end{alignat}
These are the leading corrections to the Einstein equations away from the branes ($0<y<R$), with one upper and one lower index. In eq.~\ref{eq:Friedmannb2}, we extracted the radion mass via the equation of motion $\ddot{B}_0 + 3\mathcal{H} \dot{B}_0 = -m_r^2 B_0$ for e.g.~a purely time-like momentum $p$.
The integrals in the Casimir terms arise due to the radion dependence of the conformal distance
$L[\epsilon b B_0]\simeq\int_0^{R}dy (1-\frac{3}{2}\epsilon b B_0)/a$.
The jump conditions of eqs.~\ref{eq:Jump1}~and~\ref{eq:Jump2} impose the boundary conditions for the radion profile $b'/b = -2 a'/a$ at both brane locations:
\be
b'(0)= + \frac{\sigma_1}{3 M_5^3} b(0); \qquad \qquad b'(R)= - \frac{\sigma_2}{3 M_5^3}b(R).
\label{eq:Jumpb}
\ee
Eqs.~\ref{eq:Friedmannb1},~\ref{eq:Friedmannb2},~and~\ref{eq:Jumpb} fully determine the radion profile as well as its mass. For brevity, we will proceed with the rest of the calculation in the flat-bulk and Ads-bulk regimes as in section~\ref{sec:Casimirstabilization}.

\subsubsection*{Flat bulk} 
Inserting the expansion of eq.~\ref{eq:a_mink} and a similar one for $b(y)=1+b_1y +b_2\frac{y^2}{2}+ \dots$ into eqs.~\ref{eq:Friedmannb1}, \ref{eq:Friedmannb2}, and~\ref{eq:Jumpb}, and substituting the background solution coefficients of eqs.~\ref{eq:a_sol1} and~\ref{eq:a_sol2}, we can constrain the radion profile to be
\be
b_1=\frac{\sigma_1}{3 M_5^3}, \qquad b_2=  \frac{\sigma_1^2}{4M_5^6}+  \frac{5\Lambda_5}{6M_5^3} - 3 \mathcal{H}^2.
\ee
while the mass of the radion is
\be
m_r^2 = \frac{10}{3}\frac{\lambda_1}{M_5^3} - 24 \mathcal{H}^2 =\frac{40\beta}{3 L^5 M_{5}^3} - 4\mathcal{H}^2.
\label{eq:Amrflat}
\ee
Note that there are more algebraic equations than unknown coefficients (three, namely $b_1$, $b_2$, and $m_r^2$) at this order. For example, one could solve for the profile $\lbrace b_1, b_2\rbrace$ via the two jump conditions of eq.~\ref{eq:Jumpb}, and subsequently for $m_r^2$ via eq.~\ref{eq:Friedmannb2} evaluated at $y = 0$. As a consistency check, one can then see that the same profile then also solves eq.~\ref{eq:Friedmannb1} to leading order at any location in the bulk $0 < y < R$.

When the fifth dimension is almost flat, the radion has an approximately uniform profile in the bulk, and is therefore near-gravitationally coupled to matter on the SM brane up to $\mathcal{O}(1)$ factors (namely $1/\sqrt{6}$ in the coupling, $\alpha \simeq 1/12$ in the force).

\subsubsection*{AdS bulk}
We parametrize the scalar background as in eq.~\ref{eq:a_w} and the scalar perturbation as in eq.~\ref{eq:ds2b}. We expand the radion profile in three different exponents:
\be
b(y)=e^{2 k y}+B_4 e^{4 k y }+B_5 e^{5 k y} + B_7 e^{7 k y}+\ldots
\ee
Substitution into eq.~\ref{eq:Friedmannb1} and also into the two jump conditions of eq.~\ref{eq:Jumpb} determines the $B_i$ coefficients:
\begin{align}
B_4 &= -\frac{5\beta}{L^5 \Lambda_5}\left(\gamma^2+\gamma \right) - \frac{ \mathcal{H}^2}{2 k^2},\\
B_5 &=  \frac{10\beta}{3 L^5 \Lambda_5}\left(\gamma^2+\gamma+1 \right), \\
B_7 &=  - 7 A_4 = - \frac{14\beta}{5 L^5 \Lambda_5}.
\end{align}
With this profile, the radion mass can then be computed via eq.~\ref{eq:Friedmannb2}:
\be
m_r^2= -\frac{4}{3}\frac{B_4 \Lambda_5}{M_5^3} =\frac{20}{3}\frac{\beta}{L^5 M_5^3}\l\gamma^2+\gamma\r-4\mathcal{H}^2, \label{eq:Amrcurved}
\ee
in agreement with eq.~\ref{eq:Amrflat} in the limit $\gamma \to 1$. As a consistency check, one can see that these values also solve the Einstein equations at any location in the bulk, to leading order in $\lambda_1/\Lambda_5$ and $\mathcal{H}^2/k^2$.

To determine the couplings of the radion with matter on the SM brane, or self-couplings of the radion, we need to go beyond the 4D plane-wave approximation. Plugging in the metric from eq.~\ref{eq:ds2b} in the gauge of eq.~\ref{eq:gauge} into the Einstein-Hilbert action, we find the radion kinetic term
\begin{equation}
\mathcal{L}_r \supset -\int^R_{-R} \mathrm{d} y \frac{3 M_5^3}{4} e^{2 k |y|} \left(\partial_{\mu} B_0 \right) \left(\partial^{\mu} B_0 \right).
\end{equation}
Integrating over the fifth dimension and canonically normalizing, we find the radion couples to the trace of energy momentum tensor as~\cite{Charmousis:1999rg}:
\begin{equation}
\mathcal{L}_r \supset \frac{B_r}{\sqrt{6} \gamma M_{\rm Pl}} T^{\mu}_{\mu},
\end{equation}
where $B_r (x^{\mu}) \equiv \frac{\sqrt{6} \gamma M_{\rm Pl}}{2} B_0 (x^{\mu})$ is the canonically-normalized radion field. The suppression of the radion coupling compared to that of the graviton by the warping factor $\gamma$ arises because of the small overlap of the radion wave function with the SM brane. 
In a similar way, we find the form of the cubic and quartic coupling of the radion to be $\frac{ m_r^2}{M_{\rm Pl}/\gamma} B_r^3$ and $\frac{m_r^2}{\left(M_{\rm Pl}/\gamma \right)^2} B_r^4$, respectively. At the scale of $M_{\rm Pl}/\gamma$, our linear expansion of the radion in the Friedmann equation~\ref{eq:Friedmannb1}~and~\ref{eq:Friedmannb2} starts to fail.

\section{The unbearable lightness of the radion}\label{sec:lightradion}
In this appendix, we will show that the mass of the radion is always protected, even in the presence of (local) higher-dimensional operators. Under certain assumptions, the radion mass can be arbitrarily small when the brane tensions are tuned to achieve a static geometry as well as a large inter-brane separation. The main take-away is that these higher-dimensional operators modify the necessary $k$ and $\sigma_*$ to achieve the desired geometry from the values found in section~\ref{sec:radiusstabilization} by small fractional amounts, but do not change the conclusion that the radion mass is light when the extra dimension is large.

The sufficient conditions are {\bf (1)} 5D Poincar\'{e} invariance of vacuum stress-energy sources and currents in the bulk, {\bf (2)} 4D Poincar\'{e} invariance of vacuum stress-energy sources and currents on the brane(s), and {\bf (3)} diffeomorphism invariance. We will prove that with these assumptions the radion is massless provided that {\bf (4)} we restrict to metric solutions that are constant (static, isotropic, homogeneous) along the brane directions, and {\bf (5)} we ``turn off'' nonlocal sources such as Casimir energy.

Provision {\bf (5)} is not possible---indeed we found in section~\ref{sec:radiusstabilization} that Casimir stress is responsible for giving a mass to the radion---but we imagine there is a brane tension tuning to achieve a large brane separation such that nonlocal effects are small, as in the main text. If we succeed in our proof of $m^2_r = 0$ with assumptions {\bf (1)--(5)}, then it follows that the radion mass must be proportional \emph{only} to the Casimir stress-energy, the leading nonlocal contribution to ${T^M}_N$. Higher-dimensional operators only give small fractional corrections to the mass.
Consequently, the radion mass will be automatically tuned to be light \emph{if and only if} the extra dimension is tuned to be large. 
\subsection*{Local bulk operators}
Away from the branes, the gravitational equations of motion are
\begin{align}
^{(5)}{\mathbb{R}^M}_N = -\frac{\Lambda_5}{M_5^3 } \delta^M_N,\label{eq:eomgrav5}
\end{align}
where on the RHS we have allowed only Poincar\'{e}-invariant stress-energy in the vacuum $\langle {T^M}_N \rangle = \Lambda_5 \delta^M_N$. The LHS can be any diffeomorphism-covariant operator involving 5D gravity; in pure Einstein-Hilbert gravity, it would just be the Einstein tensor: $^{(5)}{\mathbb{R}^M}_N = {G^M}_N = {\mathcal{R}^M}_N - \frac{1}{2} \mathcal{R} \, \delta^M_N$. The LHS can  be an arbitrary function of the 5D Riemann tensor $^{(5)}{\mathcal{R}_{MN}}^{OP}$ and other currents ${J^M}_N$. We ignore additional  geometric invariants with extra covariant derivatives since they do not change the argument.
Under assumption {\bf (1)}, ${J^M}_N(x,y) = \text{constant} \times \delta^M_N$, while the Riemann tensor will be maximally symmetric:
\begin{align}
^{(5)}{\mathcal{R}_{MN}}^{OP} = \frac{1}{20} {}^{(5)}\mathcal{R} \left( \delta_M^O \delta_N^P - \delta_M^P \delta_N^O \right). \label{eq:maxsym}
\end{align}
Hence the LHS of eq.~\ref{eq:eomgrav5} must be some function $f_{(5)}$ of ${}^{(5)}\mathcal{R}$ times $\delta^M_N$, so the equation of motion reduces to
\begin{align}
f_{(5)}\left({}^{(5)}\mathcal{R}\right)= -\frac{\Lambda_5}{M_5^3}. \label{eq:eomgravtrace}
\end{align}

We take the form of the metric to be:
\begin{align}
g_{MN} (x,y) = \text{diag} \Big\lbrace
\begin{array}{ccccc}
\frac{-a(y)^2}{1-\epsilon b(y) B_0(x)}, & \frac{+a(y)^2}{1-\epsilon b(y) B_0(x)}, & \frac{+a(y)^2}{1-\epsilon b(y) B_0(x)} & \frac{+a(y)^2}{1-\epsilon b(y) B_0(x)}, & \left[1-\epsilon b(y) B_0(x)\right]^2
\end{array}
\Big\rbrace, \label{eq:metric5}
\end{align}
and make the Ansatzes:
\begin{align}
a(y) &= e^{- k y}, \label{eq:solution1}\\
b(y) &= e^{+ 2 k y}.
\label{eq:solution2}
\end{align}
The metric with $\epsilon = 0$ is the maximally-symmetric background solution which solves eq.~\ref{eq:eomgravtrace} for some $k$. This curvature scale will be different from the relation in eq.~\ref{eq:kcurv}, but only by small fractional corrections of order $k^2/M_5^2$ if we assume that higher-order terms beyond the Einstein-Hilbert action are suppressed by powers of $M_5$, not a lower scale. We are only interested in background solutions that are constant along brane directions {\bf (4)}, so we identify the first four coordinates $x^\mu$ with the brane coordinates.

On top of the background solution of eq.~\ref{eq:solution1}, we perturb by the radion fluctuation $\epsilon b(y) B_0(x)$. The amplitude $\epsilon$ is taken to be infinitesimal, while its appearance in the metric of eq.~\ref{eq:metric5} is such that $B_0(x)$ does not mix with the graviton at leading order in derivatives (like in eq.~\ref{eq:gauge}), and the profile $b(y)$ of eq.~\ref{eq:solution2} is such that $B_0(x)$ is the lowest 4D mass eigenstate.

If $B_0(x)$ were constant, $\epsilon$ would parametrize a diffeomorphism  of the maximally symmetric solution, leaving curvature invariants unchanged. Hence we must have that to leading order in $\epsilon$, they can only depend on derivatives of $B_0(x)$ in the brane directions. The LHS of eq.~\ref{eq:eomgrav5} must thus take the form:
\begin{align}
^{(5)}{\mathbb{R}^M}_N = f_{(5)}(k^2)\delta^M_N + \mathcal{O} \left[\epsilon \partial_\mu \partial^\mu B_0(x) \right].
\end{align}
The expansion of the equation of motion in eq.~\ref{eq:eomgrav5} to leading order in $\epsilon$ around the background solution for $a(y)$ contains only derivatives of $B_0(x)$---the radion is massless.

In the presence of two branes, two brane-tension tunings would need to occur to set up a geometry in this regime: one to get a static geometry, the other to get a very large extra dimension to suppress nonlocal effects like Casimir energy. Equivalently, these two tunings can be viewed as tuning both $\Lambda_4$ and $m_r^2$. In our model, a stable extra dimension requires a second tension tuning for the extra dimension to be large enough so that the ``soft'' boundary condition of eq.~\ref{eq:softBC2} becomes relevant. This automatically also tunes the radion to be correspondingly light.

This concludes the bulk of our proof. For completeness, we will show that, with the inclusion of assumption {\bf (2)}, higher-dimensional curvature operators localized on the brane also do not lift the radion mass. In what follows, we set $B_0(x) = 1$ for demonstrative purposes; the proof holds if no terms linear in $\epsilon$ can appear in the equations of motion.

\subsection*{Intrinsic brane curvature operators}
Either brane is identically located at one value of the $y$-coordinate. Here we will discuss only the SM brane, taken to be at $y=0$ without loss of generality, but similar conclusions would follow for the other brane. The induced metric on the brane is just
\begin{align}
g_{\mu \nu}(x) = g_{M N}(x,0) \, \delta_\mu^M \delta_\nu^N,
\end{align}
where the Greek alphabet is used for 4D space and time components.

In the presence of branes, the gravitational equations of motion take on the form
\begin{align}
^{(5)}{\mathbb{R}^M}_N + {}^{(4)}{\mathbb{R}^\mu}_\nu \delta(y) \delta_\mu^M \delta^\nu_N = -\frac{\Lambda_5}{M_5^3} \delta^M_N - \frac{\sigma_1}{M_5^3} \delta(y) \delta^\mu_\nu \delta^M_\mu \delta^\nu_N. \label{eq:eomgrav5b}
\end{align}
We assume that 4D Poincar\'{e} symmetry is preserved by the vacuum on the brane {\bf (2)}, so that the vacuum stress-energy is $\langle{T^\mu}_\nu\rangle_0 \equiv \sigma_1 \, \delta^\mu_\nu$ with $\sigma_1$ the renormalized tension, and any currents directly coupled to gravitational operators take the form $\langle {J^\mu}_\nu(x) \rangle_0 = \text{constant}\times\delta^\mu_\nu$. The induced metric will again be maximally symmetric (from the 4D point of view), with Riemann tensor
\begin{align}
^{(4)}{{\mathcal{R}}_{\mu \nu}}^{\rho \sigma} = \frac{1}{12} {}^{(4)}\mathcal{R} \left( \delta_\mu^\rho \delta_\nu^\sigma - \delta_\mu^\sigma \delta_\nu^\rho \right). \label{eq:maxsym4}
\end{align}
Diffeomorphism invariance of the underlying theory implies that the operator ${}^{(4)}{\mathbb{R}^\mu}_\nu$ must be built out of the maximally-symmetric induced 4D Riemann tensors in eq.~\ref{eq:maxsym4} and Kronecker deltas for $\epsilon = 0$, leading to:
\begin{align}
f_{(5)}\left({}^{(5)}\mathcal{R}\right) \delta^M_N + f_{(4)}\left({}^{(4)}\mathcal{R}\right) \delta^\mu_\nu \delta^M_\mu \delta^\nu_N = -\frac{\Lambda_5}{M_5^3} \delta^M_N - \frac{\sigma_1}{M_5^3} \delta(y) \delta^\mu_\nu \delta^M_\mu \delta^\nu_N
\end{align}
for some functions $f_{(5)}$ and $f_{(4)}$. For constant $B_0(x)$, $\epsilon$ still parametrizes a diffeomorphism as before and will thus leave the LHS invariant, while the RHS does not depend on the metric. Again, no nonderivative terms in $B_0(x)$ appear, so intrinsic brane curvature operators cannot generate an additive radion mass. At most, they can modify by a small amount the kinetic term of $B_0(x)$.

\subsection*{Extrinsic brane curvature operators}

Operators involving factors of extrinsic brane curvature
\begin{align}
K_{MN} \equiv \nabla_M n_N,
\end{align}
where $n_N = \left\lbrace 0,0,0,0,g_{55}^{1/2} \right\rbrace$ is the unit normal vector to the brane, will also not contribute to the radion mass, as we will now show.

The components $K_{5 \mu}$ and $K_{\mu 5}$ are zero identically, while the others can be computed from the metric as:
\begin{align}
{K^\mu}_\nu = \nabla^\mu n_\nu = g^{\mu \alpha} \Gamma^5_{\alpha\nu} n_5 =  g^{\mu \alpha} \left( g^{55} \partial_5 g_{\alpha \nu}\right) n_5 = \left(g^{\mu \alpha} \partial_5 g_{\alpha \nu}\right) g_{55}^{-1/2}.
\end{align}
Setting $B_0(x) = 1$, we find that
\begin{align}
g_{55}^{-1/2} = 1 + \epsilon b(y), \qquad
g^{\mu \alpha} = a(y)^{-2} [1 - \epsilon b(y)] \delta^{\mu \alpha}, \qquad
\partial_5 g_{\alpha \nu} = -2k a(y)^2 \delta_{\alpha \nu},
\end{align}
all up to $\mathcal{O}(\epsilon^2)$. Putting this together, we thus conclude that the extrinsic curvature
\begin{align}
{K^\mu}_\nu = -2k \delta^\mu_\nu + \mathcal{O} \left[\epsilon \partial_\mu \partial^\mu B_0(x) \right] + \mathcal{O}\left(\epsilon^2\right)
\end{align}
also does not contain nonderivative factors of $B_0(x)$ to leading order in $\epsilon$.

Any operator ${}^{(4)}{\mathbb{R}^\mu}_\nu$ built out of intrinsic brane curvature factors $^{(4)}{{\mathcal{R}}_{\mu \nu}}^{\rho \sigma}$ and extrinsic brane curvature factors ${K^\mu}_\nu$ will thus not induce mass terms for $B_0(x)$ in its equation of motion---the expansion of eq.~\ref{eq:eomgrav5b} to first order in $\epsilon$. This completes the proof.

\section{Vacua in the ultraviolet}\label{sec:branetop}

As an example of a mechanism that provides multiple values of brane tensions only when the size of the fifth dimension is parametrically larger than the cutoff scale, we consider the following action of a 5D scalar $\Phi$ with bulk mass $M_\Phi$:
\begin{align}
S_\Phi = - \int \mathrm{d}^4 x \int_{-R}^{+R} \mathrm{d} y  \sqrt{-g^{(5)}} \left[\frac{\left( \partial_M \Phi\right)^2}{2} + \frac{M_{\Phi}^2 \Phi^2}{2} \right]
+ \sqrt{-g^{(4)}} \delta (y-R) \left[
M_b F_{a}^3 \cos \left(\frac{\Phi}{F_{a}^{3/2}} \right)\right]. 
\label{eq:Sphi}
\end{align}
For demonstrative purposes, we endow $\Phi$ with an axion-like periodic brane potential at $y = R$ with period $2 \pi F_a^{3/2}$ and barrier height $M_b F_a^3$, and with Dirichlet boundary conditions $\Phi(0) = \Phi_0$ from e.g.~a large boundary mass on the SM brane. The exact form of the boundary potentials is not important: any other boundary potential with multiple minima of sufficient depth and finite barrier height between them would suffice, and the minima of the potentials in the bulk and on the boundaries need not coincide.
Analogously, we could have fields with fixed boundary condition on the non-SM brane and a multi-valued potential on the SM brane. We assume the following hierarchy between the parameters in eq.~\ref{eq:Sphi}, the bulk curvature $k$ and the brane-bulk mixing term $\mu$ (which sets the scale of the stable extra dimension in Class I vacua):
\be
 M_{\rm UV} \gg M_b  \gtrsim M_{\Phi} \gg k > \mu,
\ee
such that the physics of $\Phi$ is well above the scales $\mu$ and $k$, but still within the regime of validity of our effective theory.

\begin{figure}[t]
\centering
\includegraphics[width = 0.8\textwidth]{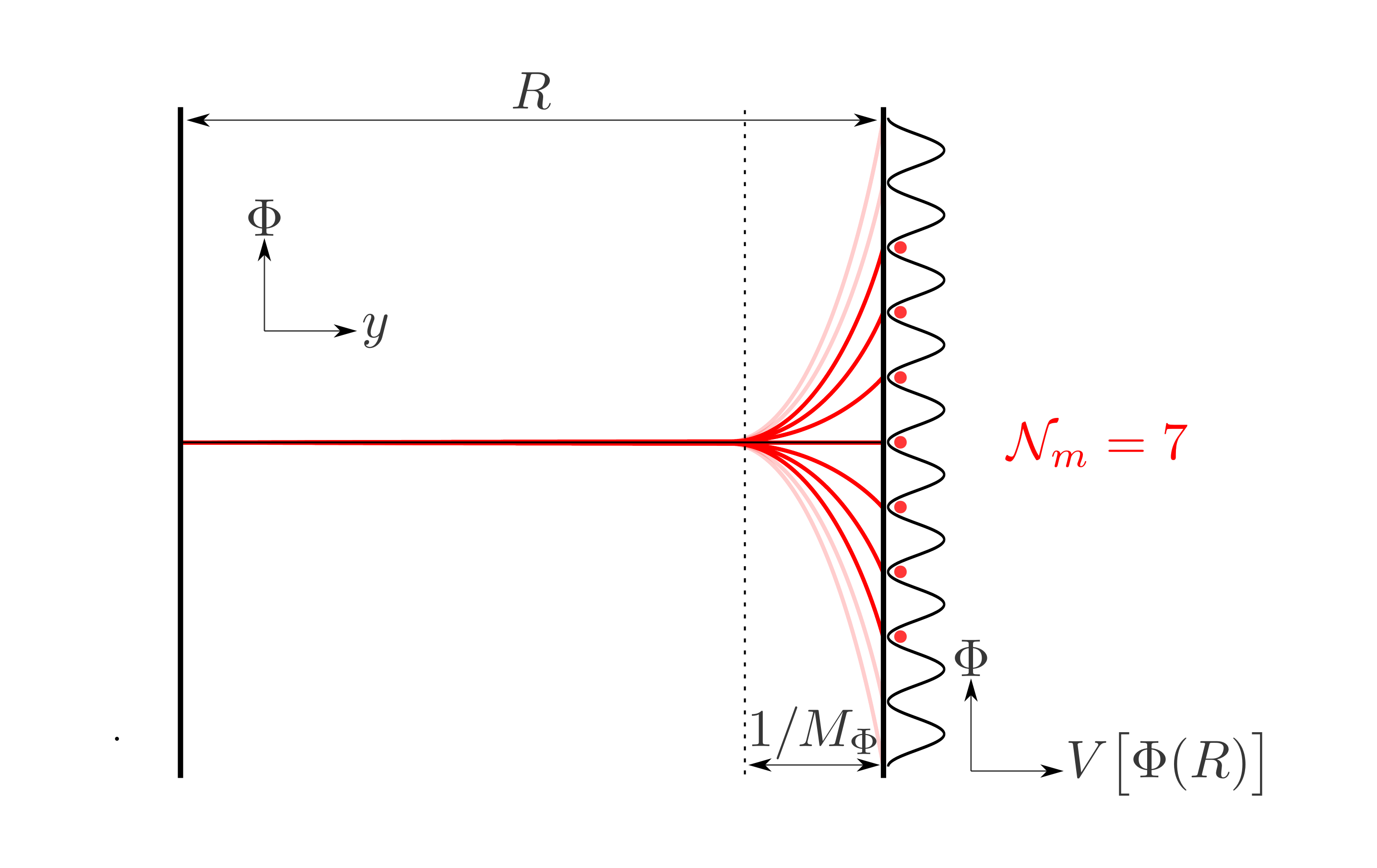}
\caption{Diagram of possible metastable field profiles $\Phi(y)$ outlined in red, and associated boundary values $\Phi(R)$ indicated by red dots. Due to the finite maximum derivative of the brane potential $V[\Phi(R)]$, the potentials can only accommodate $\mathcal{N}_m = 7$ metastable vacuum configurations. The stress-energy of the field profile is localized in a small range of width $\mathcal{O}(1/M_\Phi)$ near one of the branes, and at low energies manifests itself as a brane tension.}\label{fig:fields}
\end{figure}

For inter-brane separations $R$ much larger than $1/M_\Phi$, the classical profile of $\Phi$ is exponentially close to $\Phi=0$ in the bulk but can reside in one of the many minima on the boundary, as illustrated in figure~\ref{fig:fields}. The scalar field $\Phi$ will acquire a nontrivial profile---and thus contribute stress-energy---only in a small region of order $1/M_\phi$ near the non-SM brane. From a low-energy perspective, the many possible metastable $\Phi$ profiles simply amount to different values of the non-SM brane tension $\sigma_2$.

The number $\mathcal{N}_m$ of brane tension vacua generated in this setup can be estimated by $\mathcal{N}_m \sim \text{max}\lbrace \Phi(R) \rbrace/ F_a^{3/2}$, where $\text{max}\lbrace \Phi(R) \rbrace$ is the maximum field displacement on the brane that will still lead to a metastable solution. The maximum $\Phi(R)$ occurs at the point where the field derivative of the bulk potential and gradient energy density integrated over the Compton wavelength $1/M_\Phi$ equals the maximum field derivative on the brane potential: 
\begin{align}
M_\Phi \, \text{max}\lbrace \Phi(R) \rbrace \sim M_b F_a^{3/2},
\end{align}
giving a parametric estimate on the number of minima:
\begin{align}
\mathcal{N}_m \sim \frac{M_b}{M_{\Phi}}.
\end{align}

When the fifth dimension shrinks to a size smaller than $\Phi$'s Compton wavelength ($R < 1/M_\Phi$), the five-dimensional gradient energy density grows, reducing the number of minima until only a single stable profile remains once
\begin{align}
R \lesssim \frac{1}{M_b}.
\end{align}
This implies that in the region 
\be
M_b^{-1}\lesssim R \lesssim M_{\Phi}^{-1},
\label{eq:rangeR}
\ee
all but one of the $\mathcal{N}_m$ effective tension vacua are lost. With multiple scalar fields, the number of tension vacua at low energies can easily become exponentially large. For example, $\mathcal{N}_\Phi$ copies of scalar fields with the same action as in eq.~\ref{eq:Sphi} would give rise to a number of possible field profiles as large as $\mathcal{N}_m^{\mathcal{N}_\Phi}$, which could give rise to a significant part of the $\mathcal{N}_{\sigma_1}$  and $\mathcal{N}_{\sigma_2}$ tension vacua.

The above construction comes with one potential downside: in the range of radii given by eq.~\ref{eq:rangeR}, the $\Phi$ fields could act like Goldberger-Wise fields~\cite{Goldberger:1999uk} that can possibly stabilize the fifth dimension. This situation can be prevented by adding $\mathcal{N}_f$ extra fermions in the bulk with masses anywhere in the wide range between $\mu$ and $M_\Phi$, and whose Casimir force contribution is always attractive. (These fermions do not need to couple to SM fields.) The quantum Casimir stress scales linearly in the fermion multiplicity $\mathcal{N}_f$, while the classical Goldberger-Wise stress scales linearly in $\mathcal{N}_\Phi$ and polynomially in $\mathcal{N}_m$. Hence, for polynomially large fermion multiplicities ($\mathcal{N}_f$), there can be exponentially many ($\mathcal{N}_m^{\mathcal{N}_\Phi}$) low-energy tension vacua that evaporate before cutoff energy scales but without risking stabilization at any intermediate scale.

\bibliography{reference}

\providecommand{\href}[2]{#2}\begingroup\raggedright\begin{thebibliography}{100}

\bibitem{Agrawal:1997gf}
V.~Agrawal, S.~M. Barr, J.~F. Donoghue, and D.~Seckel, {\it {The Anthropic
  principle and the mass scale of the standard model}},  {\em Phys. Rev.} {\bf
  D57} (1998) 5480--5492, [\href{http://arxiv.org/abs/hep-ph/9707380}{{\tt
  hep-ph/9707380}}].

\bibitem{Weinberg:1987dv}
S.~Weinberg, {\it {Anthropic Bound on the Cosmological Constant}},  {\em Phys.
  Rev. Lett.} {\bf 59} (1987) 2607.

\bibitem{Weinberg:1988cp}
S.~Weinberg, {\it {The Cosmological Constant Problem}},  {\em Rev. Mod. Phys.}
  {\bf 61} (1989) 1--23.

\bibitem{Csaki:2003sh}
C.~Csaki, C.~Grojean, J.~Hubisz, Y.~Shirman, and J.~Terning, {\it {Fermions on
  an interval: Quark and lepton masses without a Higgs}},  {\em Phys. Rev.}
  {\bf D70} (2004) 015012, [\href{http://arxiv.org/abs/hep-ph/0310355}{{\tt
  hep-ph/0310355}}].

\bibitem{Romeo:2000wt}
A.~Romeo and A.~A. Saharian, {\it {Casimir effect for scalar fields under Robin
  boundary conditions on plates}},  {\em J. Phys.} {\bf A35} (2002) 1297--1320,
  [\href{http://arxiv.org/abs/hep-th/0007242}{{\tt hep-th/0007242}}].

\bibitem{birrell1984quantum}
N.~Birrell and P.~Davies, {\em Quantum Fields in Curved Space}.
\newblock Cambridge Monographs on Mathematical Physics. Cambridge University
  Press, 1984.

\bibitem{Hofmann:2000cj}
R.~Hofmann, P.~Kanti, and M.~Pospelov, {\it {(De)stabilization of an extra
  dimension due to a Casimir force}},  {\em Phys. Rev.} {\bf D63} (2001)
  124020, [\href{http://arxiv.org/abs/hep-ph/0012213}{{\tt hep-ph/0012213}}].

\bibitem{Kanti:2002zr}
P.~Kanti, K.~A. Olive, and M.~Pospelov, {\it {On the stabilization of the size
  of extra dimensions}},  {\em Phys. Lett.} {\bf B538} (2002) 146--158,
  [\href{http://arxiv.org/abs/hep-ph/0204202}{{\tt hep-ph/0204202}}].

\bibitem{Garriga:2000jb}
J.~Garriga, O.~Pujolas, and T.~Tanaka, {\it {Radion effective potential in the
  brane world}},  {\em Nucl. Phys.} {\bf B605} (2001) 192--214,
  [\href{http://arxiv.org/abs/hep-th/0004109}{{\tt hep-th/0004109}}].

\bibitem{rsi}
L.~Randall and R.~Sundrum, {\it {A Large mass hierarchy from a small extra
  dimension}},  {\em Phys. Rev. Lett.} {\bf 83} (1999) 3370--3373,
  [\href{http://arxiv.org/abs/hep-ph/9905221}{{\tt hep-ph/9905221}}].

\bibitem{Appelquist:1983vs}
T.~Appelquist and A.~Chodos, {\it {The Quantum Dynamics of Kaluza-Klein
  Theories}},  {\em Phys. Rev.} {\bf D28} (1983) 772.

\bibitem{Ponton:2001hq}
E.~Ponton and E.~Poppitz, {\it {Casimir energy and radius stabilization in
  five-dimensional orbifolds and six-dimensional orbifolds}},  {\em JHEP} {\bf
  06} (2001) 019, [\href{http://arxiv.org/abs/hep-ph/0105021}{{\tt
  hep-ph/0105021}}].

\bibitem{vonGersdorff:2005ce}
G.~von Gersdorff and A.~Hebecker, {\it {Radius stabilization by two-loop
  Casimir energy}},  {\em Nucl. Phys.} {\bf B720} (2005) 211--227,
  [\href{http://arxiv.org/abs/hep-th/0504002}{{\tt hep-th/0504002}}].

\bibitem{Kaloper:1999sm}
N.~Kaloper, {\it {Bent domain walls as brane worlds}},  {\em Phys. Rev.} {\bf
  D60} (1999) 123506, [\href{http://arxiv.org/abs/hep-th/9905210}{{\tt
  hep-th/9905210}}].

\bibitem{Randall:1998uk}
L.~Randall and R.~Sundrum, {\it {Out of this world supersymmetry breaking}},
  {\em Nucl.Phys.} {\bf B557} (1999) 79--118,
  [\href{http://arxiv.org/abs/hep-th/9810155}{{\tt hep-th/9810155}}].

\bibitem{Meszaros}
P.~Meszaros, {\it {The behaviour of point masses in an expanding cosmological
  substratum}},  {\em Astronomy and Astrophysics} {\bf 37} (1974) 225--228.

\bibitem{Weinberg:2002kg}
S.~Weinberg, {\it {Cosmological fluctuations of short wavelength}},  {\em
  Astrophys. J.} {\bf 581} (2002) 810--816,
  [\href{http://arxiv.org/abs/astro-ph/0207375}{{\tt astro-ph/0207375}}].

\bibitem{Craig:2015pha}
N.~Craig, A.~Katz, M.~Strassler, and R.~Sundrum, {\it {Naturalness in the Dark
  at the LHC}},  {\em JHEP} {\bf 07} (2015) 105,
  [\href{http://arxiv.org/abs/1501.05310}{{\tt arXiv:1501.05310}}].

\bibitem{Hook:2014cda}
A.~Hook, {\it {Anomalous solutions to the strong CP problem}},  {\em Phys. Rev.
  Lett.} {\bf 114} (2015), no.~14 141801,
  [\href{http://arxiv.org/abs/1411.3325}{{\tt arXiv:1411.3325}}].

\bibitem{Beenakker:1999xh}
W.~Beenakker, M.~Klasen, M.~Kramer, T.~Plehn, M.~Spira, et~al., {\it {The
  Production of charginos / neutralinos and sleptons at hadron colliders}},
  {\em Phys.Rev.Lett.} {\bf 83} (1999) 3780--3783,
  [\href{http://arxiv.org/abs/hep-ph/9906298}{{\tt hep-ph/9906298}}].

\bibitem{Kramer:2012bx}
M.~Kramer, A.~Kulesza, R.~van~der Leeuw, M.~Mangano, S.~Padhi, T.~Plehn, and
  X.~Portell, {\it {Supersymmetry production cross sections in $pp$ collisions
  at $\sqrt{s}=7$ TeV}},  \href{http://arxiv.org/abs/1206.2892}{{\tt
  arXiv:1206.2892}}.

\bibitem{Aad:2014nua}
{\bf ATLAS} Collaboration, G.~Aad et~al., {\it {Search for direct production of
  charginos and neutralinos in events with three leptons and missing transverse
  momentum in $\sqrt{s} =$ 8TeV $pp$ collisions with the ATLAS detector}},
  {\em JHEP} {\bf 04} (2014) 169, [\href{http://arxiv.org/abs/1402.7029}{{\tt
  arXiv:1402.7029}}].

\bibitem{Khachatryan:2014qwa}
{\bf CMS} Collaboration, V.~Khachatryan et~al., {\it {Searches for electroweak
  production of charginos, neutralinos, and sleptons decaying to leptons and W,
  Z, and Higgs bosons in pp collisions at 8 TeV}},  {\em Eur. Phys. J.} {\bf
  C74} (2014), no.~9 3036, [\href{http://arxiv.org/abs/1405.7570}{{\tt
  arXiv:1405.7570}}].

\bibitem{Chatrchyan:2013wxa}
{\bf CMS} Collaboration, S.~Chatrchyan et~al., {\it {Search for gluino mediated
  bottom- and top-squark production in multijet final states in pp collisions
  at 8 TeV}},  {\em Phys. Lett.} {\bf B725} (2013) 243--270,
  [\href{http://arxiv.org/abs/1305.2390}{{\tt arXiv:1305.2390}}].

\bibitem{Chatrchyan:2013mys}
{\bf CMS} Collaboration, S.~Chatrchyan et~al., {\it {Search for supersymmetry
  in hadronic final states with missing transverse energy using the variables
  $\alpha_T$ and b-quark multiplicity in pp collisions at $\sqrt s=8$ TeV}},
  {\em Eur. Phys. J.} {\bf C73} (2013), no.~9 2568,
  [\href{http://arxiv.org/abs/1303.2985}{{\tt arXiv:1303.2985}}].

\bibitem{Aad:2014vma}
{\bf ATLAS} Collaboration, G.~Aad et~al., {\it {Search for direct production of
  charginos, neutralinos and sleptons in final states with two leptons and
  missing transverse momentum in $pp$ collisions at $\sqrt{s} =$ 8 TeV with the
  ATLAS detector}},  {\em JHEP} {\bf 05} (2014) 071,
  [\href{http://arxiv.org/abs/1403.5294}{{\tt arXiv:1403.5294}}].

\bibitem{Khachatryan:2014mma}
{\bf CMS} Collaboration, V.~Khachatryan et~al., {\it {Searches for electroweak
  neutralino and chargino production in channels with Higgs, Z, and W bosons in
  pp collisions at 8 TeV}},  {\em Phys. Rev.} {\bf D90} (2014), no.~9 092007,
  [\href{http://arxiv.org/abs/1409.3168}{{\tt arXiv:1409.3168}}].

\bibitem{CMS:2016gvu}
{\bf CMS} Collaboration, C.~Collaboration, {\it {Search for electroweak SUSY
  production in multilepton final states in pp collisions at sqrt(s)=13 TeV
  with 12.9/fb}}, .

\bibitem{Han:2013kza}
T.~Han, S.~Padhi, and S.~Su, {\it {Electroweakinos in the Light of the Higgs
  Boson}},  {\em Phys. Rev.} {\bf D88} (2013), no.~11 115010,
  [\href{http://arxiv.org/abs/1309.5966}{{\tt arXiv:1309.5966}}].

\bibitem{Berggren:2013bua}
M.~Berggren, T.~Han, J.~List, S.~Padhi, S.~Su, and T.~Tanabe, {\it
  {Electroweakino Searches: A Comparative Study for LHC and ILC (A Snowmass
  White Paper)}},  in {\em {Community Summer Study 2013: Snowmass on the
  Mississippi (CSS2013) Minneapolis, MN, USA, July 29-August 6, 2013}}, 2013.
\newblock \href{http://arxiv.org/abs/1309.7342}{{\tt arXiv:1309.7342}}.

\bibitem{Gori:2014oua}
S.~Gori, S.~Jung, L.-T. Wang, and J.~D. Wells, {\it {Prospects for
  Electroweakino Discovery at a 100 TeV Hadron Collider}},  {\em JHEP} {\bf 12}
  (2014) 108, [\href{http://arxiv.org/abs/1410.6287}{{\tt arXiv:1410.6287}}].

\bibitem{CMS:2013xfa}
{\bf CMS} Collaboration, C.~Collaboration, {\it {Projected Performance of an
  Upgraded CMS Detector at the LHC and HL-LHC: Contribution to the Snowmass
  Process}},  in {\em {Proceedings, Community Summer Study 2013: Snowmass on
  the Mississippi (CSS2013): Minneapolis, MN, USA, July 29-August 6, 2013}},
  2013.
\newblock \href{http://arxiv.org/abs/1307.7135}{{\tt arXiv:1307.7135}}.

\bibitem{ATLAS:2013hta}
{\bf ATLAS} Collaboration, A.~Collaboration, {\it {Physics at a High-Luminosity
  LHC with ATLAS}},  in {\em {Community Summer Study 2013: Snowmass on the
  Mississippi (CSS2013) Minneapolis, MN, USA, July 29-August 6, 2013}}, 2013.
\newblock \href{http://arxiv.org/abs/1307.7292}{{\tt arXiv:1307.7292}}.

\bibitem{Arvanitaki:2012ps}
A.~Arvanitaki, N.~Craig, S.~Dimopoulos, and G.~Villadoro, {\it {Mini-Split}},
  {\em JHEP} {\bf 02} (2013) 126, [\href{http://arxiv.org/abs/1210.0555}{{\tt
  arXiv:1210.0555}}].

\bibitem{CEPC-SPPCStudyGroup:2015csa}
C.-S.~S. Group, {\it {CEPC-SPPC Preliminary Conceptual Design Report. 1.
  Physics and Detector}}, .

\bibitem{Baer:2013cma}
H.~Baer, T.~Barklow, K.~Fujii, Y.~Gao, A.~Hoang, S.~Kanemura, J.~List, H.~E.
  Logan, A.~Nomerotski, M.~Perelstein, et~al., {\it {The International Linear
  Collider Technical Design Report - Volume 2: Physics}},
  \href{http://arxiv.org/abs/1306.6352}{{\tt arXiv:1306.6352}}.

\bibitem{Aad:2015pla}
{\bf ATLAS} Collaboration, G.~Aad et~al., {\it {Constraints on new phenomena
  via Higgs boson couplings and invisible decays with the ATLAS detector}},
  {\em JHEP} {\bf 11} (2015) 206, [\href{http://arxiv.org/abs/1509.00672}{{\tt
  arXiv:1509.00672}}].

\bibitem{CMS:2015naa}
{\bf CMS} Collaboration, C.~Collaboration, {\it {A combination of searches for
  the invisible decays of the Higgs boson using the CMS detector}}, .

\bibitem{CMS:2016rfr}
{\bf CMS} Collaboration, C.~Collaboration, {\it {Searches for invisible Higgs
  boson decays with the CMS detector.}}, .

\bibitem{CMS:2016pod}
{\bf CMS} Collaboration, C.~Collaboration, {\it {Search for dark matter in
  final states with an energetic jet, or a hadronically decaying W or Z boson
  using $12.9~\mathrm{fb}^{-1}$ of data at $\sqrt{s} = 13~\mathrm{TeV}$}}, .

\bibitem{CMS:2016hmx}
{\bf CMS} Collaboration, C.~Collaboration, {\it {Search for dark matter in
  $\mathrm{Z}+E_\mathrm{T}^\mathrm{miss}$ events using $12.9~\mathrm{fb}^{-1}$
  of 2016 data}}, .

\bibitem{ATLAS:2016inv}
{\bf CMS} Collaboration, A.~Collaboration, {\it {Search for new phenomena in
  the $Z(\rightarrow ll) + E_\mathrm{T}^\mathrm{miss}$ final state at $\sqrt{s}
  = 13~\mathrm{TeV}$ with the ATLAS detector}}, .

\bibitem{Glaysher:2015bff}
{\bf ATLAS} Collaboration, P.~Glaysher, {\it {ATLAS Higgs physics prospects at
  the high luminosity LHC}},  {\em PoS} {\bf EPS-HEP2015} (2015) 160.

\bibitem{Martin:2009bg}
S.~P. Martin, {\it {Extra vector-like matter and the lightest Higgs scalar
  boson mass in low-energy supersymmetry}},  {\em Phys. Rev.} {\bf D81} (2010)
  035004, [\href{http://arxiv.org/abs/0910.2732}{{\tt arXiv:0910.2732}}].

\bibitem{Peskin:1991sw}
M.~E. Peskin and T.~Takeuchi, {\it {Estimation of oblique electroweak
  corrections}},  {\em Phys. Rev.} {\bf D46} (1992) 381--409.

\bibitem{Agashe:2014kda}
{\bf Particle Data Group} Collaboration, K.~A. Olive et~al., {\it {Review of
  Particle Physics}},  {\em Chin. Phys.} {\bf C38} (2014) 090001.

\bibitem{CMS:2016com}
{\bf CMS} Collaboration, C.~Collaboration, {\it {Search for new physics in the
  compressed mass spectra scenario using events with two soft opposite-sign
  leptons and missing momentum energy at 13 TeV}}, .

\bibitem{ArkaniHamed:2012kq}
N.~Arkani-Hamed, K.~Blum, R.~T. D'Agnolo, and J.~Fan, {\it {2:1 for Naturalness
  at the LHC?}},  {\em JHEP} {\bf 01} (2013) 149,
  [\href{http://arxiv.org/abs/1207.4482}{{\tt arXiv:1207.4482}}].

\bibitem{Kearney:2012zi}
J.~Kearney, A.~Pierce, and N.~Weiner, {\it {Vectorlike Fermions and Higgs
  Couplings}},  {\em Phys. Rev.} {\bf D86} (2012) 113005,
  [\href{http://arxiv.org/abs/1207.7062}{{\tt arXiv:1207.7062}}].

\bibitem{Carena:2012xa}
M.~Carena, I.~Low, and C.~E.~M. Wagner, {\it {Implications of a Modified Higgs
  to Diphoton Decay Width}},  {\em JHEP} {\bf 08} (2012) 060,
  [\href{http://arxiv.org/abs/1206.1082}{{\tt arXiv:1206.1082}}].

\bibitem{Khachatryan:2016vau}
{\bf ATLAS, CMS} Collaboration, G.~Aad et~al., {\it {Measurements of the Higgs
  boson production and decay rates and constraints on its couplings from a
  combined ATLAS and CMS analysis of the LHC pp collision data at $ \sqrt{s}=7
  $ and 8 TeV}},  {\em JHEP} {\bf 08} (2016) 045,
  [\href{http://arxiv.org/abs/1606.02266}{{\tt arXiv:1606.02266}}].

\bibitem{Fan:2014vta}
J.~Fan, M.~Reece, and L.-T. Wang, {\it {Possible Futures of Electroweak
  Precision: ILC, FCC-ee, and CEPC}},  {\em JHEP} {\bf 09} (2015) 196,
  [\href{http://arxiv.org/abs/1411.1054}{{\tt arXiv:1411.1054}}].

\bibitem{Garriga:1999yh}
J.~Garriga and T.~Tanaka, {\it {Gravity in the brane world}},  {\em Phys. Rev.
  Lett.} {\bf 84} (2000) 2778--2781,
  [\href{http://arxiv.org/abs/hep-th/9911055}{{\tt hep-th/9911055}}].

\bibitem{Csaki:1999mp}
C.~Csaki, M.~Graesser, L.~Randall, and J.~Terning, {\it {Cosmology of brane
  models with radion stabilization}},  {\em Phys. Rev.} {\bf D62} (2000)
  045015, [\href{http://arxiv.org/abs/hep-ph/9911406}{{\tt hep-ph/9911406}}].

\bibitem{Hoskins:1985tn}
J.~K. Hoskins, R.~D. Newman, R.~Spero, and J.~Schultz, {\it {Experimental tests
  of the gravitational inverse square law for mass separations from 2-cm to
  105-cm}},  {\em Phys. Rev.} {\bf D32} (1985) 3084--3095.

\bibitem{Kapner:2006si}
D.~J. Kapner, T.~S. Cook, E.~G. Adelberger, J.~H. Gundlach, B.~R. Heckel, C.~D.
  Hoyle, and H.~E. Swanson, {\it {Tests of the gravitational inverse-square law
  below the dark-energy length scale}},  {\em Phys. Rev. Lett.} {\bf 98} (2007)
  021101, [\href{http://arxiv.org/abs/hep-ph/0611184}{{\tt hep-ph/0611184}}].

\bibitem{Geraci:2008hb}
A.~A. Geraci, S.~J. Smullin, D.~M. Weld, J.~Chiaverini, and A.~Kapitulnik, {\it
  {Improved constraints on non-Newtonian forces at 10 microns}},  {\em Phys.
  Rev.} {\bf D78} (2008) 022002, [\href{http://arxiv.org/abs/0802.2350}{{\tt
  arXiv:0802.2350}}].

\bibitem{Fischbach:1996eq}
E.~Fischbach and C.~Talmadge, {\it {Ten years of the fifth force}},  in {\em
  {Dark matter in cosmology, quantum measurements, experimental gravitation.
  Proceedings, 31st Rencontres de Moriond, 16th Moriond Workshop, Les Arcs,
  France, January 2-27, 1996}}, pp.~443--451, 1996.
\newblock \href{http://arxiv.org/abs/hep-ph/9606249}{{\tt hep-ph/9606249}}.

\bibitem{ArkaniHamed:1999za}
N.~Arkani-Hamed, Y.~Grossman, and M.~Schmaltz, {\it {Split fermions in extra
  dimensions and exponentially small cross-sections at future colliders}},
  {\em Phys. Rev.} {\bf D61} (2000) 115004,
  [\href{http://arxiv.org/abs/hep-ph/9909411}{{\tt hep-ph/9909411}}].

\bibitem{Wagner:2012ui}
T.~A. Wagner, S.~Schlamminger, J.~H. Gundlach, and E.~G. Adelberger, {\it
  {Torsion-balance tests of the weak equivalence principle}},  {\em Class.
  Quant. Grav.} {\bf 29} (2012) 184002,
  [\href{http://arxiv.org/abs/1207.2442}{{\tt arXiv:1207.2442}}].

\bibitem{Smith:1999cr}
G.~L. Smith, C.~D. Hoyle, J.~H. Gundlach, E.~G. Adelberger, B.~R. Heckel, and
  H.~E. Swanson, {\it {Short range tests of the equivalence principle}},  {\em
  Phys. Rev.} {\bf D61} (2000) 022001.

\bibitem{Dimopoulos:2006nk}
S.~Dimopoulos, P.~W. Graham, J.~M. Hogan, and M.~A. Kasevich, {\it {Testing
  general relativity with atom interferometry}},  {\em Phys. Rev. Lett.} {\bf
  98} (2007) 111102, [\href{http://arxiv.org/abs/gr-qc/0610047}{{\tt
  gr-qc/0610047}}].

\bibitem{bertotti2003test}
B.~Bertotti, L.~Iess, and P.~Tortora, {\it A test of general relativity using
  radio links with the cassini spacecraft},  {\em Nature} {\bf 425} (2003),
  no.~6956 374--376.

\bibitem{Blas:2016ddr}
D.~Blas, D.~L. Nacir, and S.~Sibiryakov, {\it {Ultra-Light Dark Matter
  Resonates with Binary Pulsars}},  \href{http://arxiv.org/abs/1612.06789}{{\tt
  arXiv:1612.06789}}.

\bibitem{Arvanitaki:2010sy}
A.~Arvanitaki and S.~Dubovsky, {\it {Exploring the String Axiverse with
  Precision Black Hole Physics}},  {\em Phys. Rev.} {\bf D83} (2011) 044026,
  [\href{http://arxiv.org/abs/1004.3558}{{\tt arXiv:1004.3558}}].

\bibitem{Arvanitaki:2014wva}
A.~Arvanitaki, M.~Baryakhtar, and X.~Huang, {\it {Discovering the QCD Axion
  with Black Holes and Gravitational Waves}},  {\em Phys. Rev.} {\bf D91}
  (2015), no.~8 084011, [\href{http://arxiv.org/abs/1411.2263}{{\tt
  arXiv:1411.2263}}].

\bibitem{Gruzinov:2016hcq}
A.~Gruzinov, {\it {Black Hole Spindown by Light Bosons}},
  \href{http://arxiv.org/abs/1604.06422}{{\tt arXiv:1604.06422}}.

\bibitem{Arvanitaki:2014faa}
A.~Arvanitaki, J.~Huang, and K.~Van~Tilburg, {\it {Searching for dilaton dark
  matter with atomic clocks}},  {\em Phys. Rev.} {\bf D91} (2015), no.~1
  015015, [\href{http://arxiv.org/abs/1405.2925}{{\tt arXiv:1405.2925}}].

\bibitem{Graham:2015ifn}
P.~W. Graham, D.~E. Kaplan, J.~Mardon, S.~Rajendran, and W.~A. Terrano, {\it
  {Dark Matter Direct Detection with Accelerometers}},  {\em Phys. Rev.} {\bf
  D93} (2016), no.~7 075029, [\href{http://arxiv.org/abs/1512.06165}{{\tt
  arXiv:1512.06165}}].

\bibitem{Arvanitaki:2015iga}
A.~Arvanitaki, S.~Dimopoulos, and K.~Van~Tilburg, {\it {Sound of Dark Matter:
  Searching for Light Scalars with Resonant-Mass Detectors}},  {\em Phys. Rev.
  Lett.} {\bf 116} (2016), no.~3 031102,
  [\href{http://arxiv.org/abs/1508.01798}{{\tt arXiv:1508.01798}}].

\bibitem{Arvanitaki:2016fyj}
A.~Arvanitaki, P.~W. Graham, J.~M. Hogan, S.~Rajendran, and K.~Van~Tilburg,
  {\it {Search for light scalar dark matter with atomic gravitational wave
  detectors}},  \href{http://arxiv.org/abs/1606.04541}{{\tt arXiv:1606.04541}}.

\bibitem{VanTilburg:2015oza}
K.~Van~Tilburg, N.~Leefer, L.~Bougas, and D.~Budker, {\it {Search for
  ultralight scalar dark matter with atomic spectroscopy}},  {\em Phys. Rev.
  Lett.} {\bf 115} (2015), no.~1 011802,
  [\href{http://arxiv.org/abs/1503.06886}{{\tt arXiv:1503.06886}}].

\bibitem{Hees:2016gop}
A.~Hees, J.~Guéna, M.~Abgrall, S.~Bize, and P.~Wolf, {\it {Searching for an
  oscillating massive scalar field as a dark matter candidate using atomic
  hyperfine frequency comparisons}},  {\em Phys. Rev. Lett.} {\bf 117} (2016),
  no.~6 061301, [\href{http://arxiv.org/abs/1604.08514}{{\tt
  arXiv:1604.08514}}].

\bibitem{2013Sci...341.1215H}
N.~{Hinkley}, J.~A. {Sherman}, N.~B. {Phillips}, M.~{Schioppo}, N.~D. {Lemke},
  K.~{Beloy}, M.~{Pizzocaro}, C.~W. {Oates}, and A.~D. {Ludlow}, {\it {An
  Atomic Clock with $10^{-18}$ Instability}},  {\em Science} {\bf 341} (Sept.,
  2013) 1215--1218, [\href{http://arxiv.org/abs/1305.5869}{{\tt
  arXiv:1305.5869}}].

\bibitem{2015NatCo...6E6896N}
T.~L. {Nicholson}, S.~L. {Campbell}, R.~B. {Hutson}, G.~E. {Marti}, B.~J.
  {Bloom}, R.~L. {McNally}, W.~{Zhang}, M.~D. {Barrett}, M.~S. {Safronova},
  G.~F. {Strouse}, W.~L. {Tew}, and J.~{Ye}, {\it {Systematic evaluation of an
  atomic clock at $2 {\times} 10^{-18}$ total uncertainty}},  {\em Nature
  Communications} {\bf 6} (Apr., 2015) 6896,
  [\href{http://arxiv.org/abs/1412.8261}{{\tt arXiv:1412.8261}}].

\bibitem{2016NaPho..10..258N}
N.~{Nemitz}, T.~{Ohkubo}, M.~{Takamoto}, I.~{Ushijima}, M.~{Das}, N.~{Ohmae},
  and H.~{Katori}, {\it {Frequency ratio of Yb and Sr clocks with $5 \times
  10^{‑17}$ uncertainty at 150 seconds averaging time}},  {\em Nature
  Photonics} {\bf 10} (Apr., 2016) 258--261,
  [\href{http://arxiv.org/abs/1601.04582}{{\tt arXiv:1601.04582}}].

\bibitem{PhysRevLett.116.063001}
N.~Huntemann, C.~Sanner, B.~Lipphardt, C.~Tamm, and E.~Peik, {\it Single-ion
  atomic clock with $3 \times {10}^{-18}$ systematic uncertainty},  {\em Phys.
  Rev. Lett.} {\bf 116} (Feb, 2016) 063001.

\bibitem{2016arXiv160706867S}
M.~{Schioppo}, R.~C. {Brown}, W.~F. {McGrew}, N.~{Hinkley}, R.~J. {Fasano},
  K.~{Beloy}, T.~H. {Yoon}, G.~{Milani}, D.~{Nicolodi}, J.~A. {Sherman}, N.~B.
  {Phillips}, C.~W. {Oates}, and A.~D. {Ludlow}, {\it {Ultra-stable optical
  clock with two cold-atom ensembles}},  {\em ArXiv e-prints} (July, 2016)
  [\href{http://arxiv.org/abs/1607.06867}{{\tt arXiv:1607.06867}}].

\bibitem{peik2003nuclear}
E.~Peik and C.~Tamm, {\it Nuclear laser spectroscopy of the 3.5 ev transition
  in th-229},  {\em EPL (Europhysics Letters)} {\bf 61} (2003), no.~2 181.

\bibitem{campbell2012single}
C.~J. Campbell, A.~G. Radnaev, A.~Kuzmich, V.~A. Dzuba, V.~V. Flambaum, and
  A.~Derevianko, {\it Single-ion nuclear clock for metrology at the 19th
  decimal place},  {\em Physical review letters} {\bf 108} (2012), no.~12
  120802.

\bibitem{tkalya2015radiative}
E.~Tkalya, C.~Schneider, J.~Jeet, and E.~R. Hudson, {\it Radiative lifetime and
  energy of the low-energy isomeric level in th 229},  {\em Physical Review C}
  {\bf 92} (2015), no.~5 054324.

\bibitem{von2016direct}
L.~von~der Wense, B.~Seiferle, M.~Laatiaoui, J.~B. Neumayr, H.-J. Maier, H.-F.
  Wirth, C.~Mokry, J.~Runke, K.~Eberhardt, C.~E. D{\"u}llmann, et~al., {\it
  Direct detection of the 229th nuclear clock transition},  {\em Nature} {\bf
  533} (2016), no.~7601 47--51.

\bibitem{flambaum2006enhanced}
V.~Flambaum, {\it Enhanced effect of temporal variation of the fine structure
  constant and the strong interaction in th 229},  {\em Physical review
  letters} {\bf 97} (2006), no.~9 092502.

\bibitem{Branca:2016rez}
A.~Branca et~al., {\it {Search for light scalar Dark Matter candidate with
  AURIGA detector}},  \href{http://arxiv.org/abs/1607.07327}{{\tt
  arXiv:1607.07327}}.

\bibitem{Leaci:2008zza}
P.~Leaci, A.~Vinante, M.~Bonaldi, P.~Falferi, A.~Pontin, G.~A. Prodi, and J.~P.
  Zendri, {\it {Design of wideband acoustic detectors of gravitational waves
  equipped with displacement concentrators}},  {\em Phys. Rev.} {\bf D77}
  (2008) 062001.

\bibitem{Goryachev:2014yra}
M.~Goryachev and M.~E. Tobar, {\it {Gravitational Wave Detection with High
  Frequency Phonon Trapping Acoustic Cavities}},  {\em Phys. Rev.} {\bf D90}
  (2014), no.~10 102005, [\href{http://arxiv.org/abs/1410.2334}{{\tt
  arXiv:1410.2334}}].

\bibitem{Hall:2009bx}
L.~J. Hall, K.~Jedamzik, J.~March-Russell, and S.~M. West, {\it {Freeze-In
  Production of FIMP Dark Matter}},  {\em JHEP} {\bf 03} (2010) 080,
  [\href{http://arxiv.org/abs/0911.1120}{{\tt arXiv:0911.1120}}].

\bibitem{Graham:2015cka}
P.~W. Graham, D.~E. Kaplan, and S.~Rajendran, {\it {Cosmological Relaxation of
  the Electroweak Scale}},  {\em Phys. Rev. Lett.} {\bf 115} (2015), no.~22
  221801, [\href{http://arxiv.org/abs/1504.07551}{{\tt arXiv:1504.07551}}].

\bibitem{Abbott:1984qf}
L.~F. Abbott, {\it {A Mechanism for Reducing the Value of the Cosmological
  Constant}},  {\em Phys. Lett.} {\bf B150} (1985) 427--430.

\bibitem{Graham:2017abc}
P.~W. Graham, D.~E. Kaplan, and S.~Rajendran, {\it {to appear}}, .

\bibitem{Alberte:2016izw}
L.~Alberte, P.~Creminelli, A.~Khmelnitsky, D.~Pirtskhalava, and E.~Trincherini,
  {\it {Relaxing the Cosmological Constant: a Proof of Concept}},
  \href{http://arxiv.org/abs/1608.05715}{{\tt arXiv:1608.05715}}.

\bibitem{Susskind:2012xf}
L.~Susskind, {\it {Is Eternal Inflation Past-Eternal? And What if It Is?}},
  \href{http://arxiv.org/abs/1205.0589}{{\tt arXiv:1205.0589}}.

\bibitem{Linde:2006nw}
A.~D. Linde, {\it {Sinks in the Landscape, Boltzmann Brains, and the
  Cosmological Constant Problem}},  {\em JCAP} {\bf 0701} (2007) 022,
  [\href{http://arxiv.org/abs/hep-th/0611043}{{\tt hep-th/0611043}}].

\bibitem{Polchinski:2006gy}
J.~Polchinski, {\it {The Cosmological Constant and the String Landscape}},  in
  {\em {The Quantum Structure of Space and Time: Proceedings of the 23rd Solvay
  Conference on Physics. Brussels, Belgium. 1 - 3 December 2005}},
  pp.~216--236, 2006.
\newblock \href{http://arxiv.org/abs/hep-th/0603249}{{\tt hep-th/0603249}}.

\bibitem{Bousso:2007gp}
R.~Bousso, {\it {TASI Lectures on the Cosmological Constant}},  {\em Gen. Rel.
  Grav.} {\bf 40} (2008) 607--637, [\href{http://arxiv.org/abs/0708.4231}{{\tt
  arXiv:0708.4231}}].

\bibitem{Dvali:2003br}
G.~Dvali and A.~Vilenkin, {\it {Cosmic attractors and gauge hierarchy}},  {\em
  Phys. Rev.} {\bf D70} (2004) 063501,
  [\href{http://arxiv.org/abs/hep-th/0304043}{{\tt hep-th/0304043}}].

\bibitem{Polyakov:1993tp}
A.~M. Polyakov, {\it {A Few projects in string theory}},  in {\em {Gravitation
  and quantizations. Proceedings, 57th Session of the Les Houches Summer School
  in Theoretical Physics, NATO Advanced Study Institute, Les Houches, France,
  July 5 - August 1, 1992}}, pp.~0783--804, 1993.
\newblock \href{http://arxiv.org/abs/hep-th/9304146}{{\tt hep-th/9304146}}.

\bibitem{Damour:1994zq}
T.~Damour and A.~M. Polyakov, {\it {The String dilaton and a least coupling
  principle}},  {\em Nucl. Phys.} {\bf B423} (1994) 532--558,
  [\href{http://arxiv.org/abs/hep-th/9401069}{{\tt hep-th/9401069}}].

\bibitem{Arvanitaki:4D}
A.~Arvanitaki, S.~Dimopoulos, V.~Gorbenko, J.~Huang, and K.~Van~Tilburg, {\it
  {to appear}}, .

\bibitem{Bousso:2000xa}
R.~Bousso and J.~Polchinski, {\it {Quantization of four form fluxes and
  dynamical neutralization of the cosmological constant}},  {\em JHEP} {\bf 06}
  (2000) 006, [\href{http://arxiv.org/abs/hep-th/0004134}{{\tt
  hep-th/0004134}}].

\bibitem{Csaki:2004ay}
C.~Csaki, {\it {TASI lectures on extra dimensions and branes}},  in {\em {From
  fields to strings: Circumnavigating theoretical physics. Ian Kogan memorial
  collection (3 volume set)}}, pp.~605--698, 2004.
\newblock \href{http://arxiv.org/abs/hep-ph/0404096}{{\tt hep-ph/0404096}}.
\newblock [,967(2004)].

\bibitem{Csaki:2000zn}
C.~Csaki, M.~L. Graesser, and G.~D. Kribs, {\it {Radion dynamics and
  electroweak physics}},  {\em Phys. Rev.} {\bf D63} (2001) 065002,
  [\href{http://arxiv.org/abs/hep-th/0008151}{{\tt hep-th/0008151}}].

\bibitem{Charmousis:1999rg}
C.~Charmousis, R.~Gregory, and V.~A. Rubakov, {\it {Wave function of the radion
  in a brane world}},  {\em Phys. Rev.} {\bf D62} (2000) 067505,
  [\href{http://arxiv.org/abs/hep-th/9912160}{{\tt hep-th/9912160}}].

\bibitem{Goldberger:1999uk}
W.~D. Goldberger and M.~B. Wise, {\it {Modulus stabilization with bulk
  fields}},  {\em Phys. Rev. Lett.} {\bf 83} (1999) 4922--4925,
  [\href{http://arxiv.org/abs/hep-ph/9907447}{{\tt hep-ph/9907447}}].

\end{thebibliography}\endgroup
\bibliographystyle{JHEP}

\end{document}